\documentclass[longauth]{aa}  
\usepackage{graphicx}

\usepackage{txfonts}
\usepackage{amsmath}
\usepackage{textcomp, gensymb}
\usepackage{hyperref}
\usepackage{threeparttable}
\usepackage{booktabs}
\usepackage{arydshln}
\usepackage{color}
\usepackage{placeins}
\usepackage{multirow}
\usepackage{lastpage}
\usepackage{flafter} 
\usepackage{subcaption}
\usepackage{ftnxtra}
\usepackage{orcidlink}

\newcommand{\arcs}{$^{\prime\prime}$}
\defcitealias{CastroGonzalez2024}{CG24}
\defcitealias{Berger2023}{B23}

\begin{document} 

\title{Confirmation of the hot super-Neptune TOI-672 b with NIRPS and HARPS}
\titlerunning{A hot super-Neptune in the M dwarf Neptunian desert}
\subtitle{Insights into the Neptunian desert around M dwarfs}

\author{
Ares Osborn\inst{1,2,3,*}\orcidlink{0000-0002-5899-7750},
Ryan Cloutier\inst{2}\orcidlink{0000-0001-5383-9393},
Vincent Bourrier\inst{4}\orcidlink{0000-0002-9148-034X},
Bennett Skinner\inst{2}\orcidlink{0009-0000-9731-2462},
Nicole Gromek\inst{2}\orcidlink{0009-0000-1424-7694},
Avidaan Srivastava\inst{5}\orcidlink{0009-0009-7136-1528},
Fran\c{c}ois Bouchy\inst{4}\orcidlink{0000-0002-7613-393X},
Marion Cointepas\inst{4,1}\orcidlink{0009-0001-6168-2178},
Neil J. Cook\inst{5}\orcidlink{0000-0003-4166-4121},
Nicola Nari\inst{6,7,8},
Jose Manuel Almenara\inst{1}\orcidlink{0000-0003-3208-9815},
\'Etienne Artigau\inst{5,9}\orcidlink{0000-0003-3506-5667},
Xavier Bonfils\inst{1}\orcidlink{0000-0001-9003-8894},
Charles Cadieux\inst{5}\orcidlink{0000-0001-9291-5555},
Patrick Eggenberger\inst{4}\orcidlink{0000-0001-6319-9297},
Alexandrine L'Heureux\inst{5}\orcidlink{0009-0005-6135-6769},
Fr\'ed\'erique Baron\inst{5,9}\orcidlink{0000-0002-5074-1128},
Susana C. C. Barros\inst{10,11}\orcidlink{0000-0003-2434-3625},
Bj\"orn Benneke\inst{12,5}\orcidlink{0000-0001-5578-1498},
Marta Bryan\inst{13},
Bruno L. Canto Martins\inst{14}\orcidlink{0000-0001-5578-7400},
Nicolas B. Cowan\inst{15,16}\orcidlink{0000-0001-6129-5699},
Eduardo Cristo\inst{10},
Xavier Delfosse\inst{1}\orcidlink{0000-0001-5099-7978},
Jose Renan De Medeiros\inst{14}\orcidlink{0000-0001-8218-1586},
Ren\'e Doyon\inst{5,9}\orcidlink{0000-0001-5485-4675},
Xavier Dumusque\inst{4}\orcidlink{0000-0002-9332-2011},
David Ehrenreich\inst{4,17},
Jonay I. Gonz\'alez Hern\'andez\inst{7,8}\orcidlink{0000-0002-0264-7356},
David Lafreni\`ere\inst{5}\orcidlink{0000-0002-6780-4252},
Izan de Castro Le\~ao\inst{14}\orcidlink{0000-0001-5845-947X},
Christophe Lovis\inst{4}\orcidlink{0000-0001-7120-5837},
Lison Malo\inst{5,9}\orcidlink{0000-0002-8786-8499},
Claudio Melo\inst{18},
Lucile Mignon\inst{4,1},
Christoph Mordasini\inst{19}\orcidlink{0000-0002-1013-2811},
Francesco Pepe\inst{4}\orcidlink{0000-0002-9815-773X},
Rafael Rebolo\inst{7,8,20}\orcidlink{0000-0003-3767-7085},
Jason Rowe\inst{21},
Nuno C. Santos\inst{10,11}\orcidlink{0000-0003-4422-2919},
Damien S\'egransan\inst{4},
Alejandro Su\'arez Mascare\~no\inst{7,8}\orcidlink{0000-0002-3814-5323},
St\'ephane Udry\inst{4}\orcidlink{0000-0001-7576-6236},
Diana Valencia\inst{13}\orcidlink{0000-0003-3993-4030},
Gregg Wade\inst{22,23},
Jos\'e Luan A. Aguiar\inst{14}\orcidlink{0009-0006-6577-9571},
Romain Allart\inst{5}\orcidlink{0000-0002-1199-9759},
Khaled Al Moulla\inst{10,4}\orcidlink{0000-0002-3212-5778},
Andres Carmona\inst{1}\orcidlink{0000-0003-2471-1299},
Karen A. Collins\inst{24}\orcidlink{0000-0001-6588-9574},
Elisa Delgado-Mena\inst{25,10}\orcidlink{0000-0003-4434-2195},
Roseane de Lima Gomes\inst{5,14}\orcidlink{0000-0002-2023-7641},
George Dixon\inst{26},
Phil Evans\inst{27}\orcidlink{0000-0002-5674-2404},
Yolanda G. C. Frensch\inst{4,28,}\orcidlink{0000-0003-4009-0330},
Dasaev O. Fontinele\inst{14}\orcidlink{0000-0002-3916-6441},
Thierry Forveille\inst{1}\orcidlink{0000-0003-0536-4607},
Tianjun Gan\inst{29}\orcidlink{0000-0002-4503-9705},
Melissa J. Hobson\inst{4}\orcidlink{0000-0002-5945-7975},
Yuri S. Messias\inst{5,14}\orcidlink{0000-0002-2425-801X},
Louise D. Nielsen\inst{4,18,30}\orcidlink{0000-0002-5254-2499},
L\'ena Parc\inst{4}\orcidlink{0000-0002-7382-1913},
Ying Shu\inst{31}\orcidlink{0009-0002-5701-6276},
Atanas K. Stefanov\inst{7,8}\orcidlink{0000-0002-6059-1178},
Thiam-Gun Tan\inst{32}\orcidlink{0000-0001-5603-6895},
Jean-Pascal Vignes\inst{33},
Joost P. Wardenier\inst{19,5}\orcidlink{0000-0003-3191-2486},
Drew Weisserman\inst{2}\orcidlink{0000-0002-7992-469X}
}

\authorrunning{Ares Osborn}
\institute{
\inst{1}Univ. Grenoble Alpes, CNRS, IPAG, F-38000 Grenoble, France\\
\inst{2}Department of Physics \& Astronomy, McMaster University, 1280 Main St W, Hamilton, ON, L8S 4L8, Canada\\
\inst{3}Department of Physics, The University of Warwick, Gibbet Hill Road, Coventry, CV4 7AL, UK\\
\inst{4}Observatoire de Gen\`eve, D\'epartement d’Astronomie, Universit\'e de Gen\`eve, Chemin Pegasi 51, 1290 Versoix, Switzerland\\
\inst{5}Institut Trottier de recherche sur les exoplan\`etes, D\'epartement de Physique, Universit\'e de Montr\'eal, Montr\'eal, Qu\'ebec, Canada\\
\inst{6}Light Bridges S.L., Observatorio del Teide, Carretera del Observatorio, s/n Guimar, 38500, Tenerife, Canarias, Spain\\
\inst{7}Instituto de Astrof\'isica de Canarias (IAC), Calle V\'ia L\'actea s/n, 38205 La Laguna, Tenerife, Spain\\
\inst{8}Departamento de Astrof\'isica, Universidad de La Laguna (ULL), 38206 La Laguna, Tenerife, Spain\\
\inst{9}Observatoire du Mont-M\'egantic, Qu\'ebec, Canada\\
\inst{10}Instituto de Astrof\'isica e Ci\^encias do Espa\c{c}o, Universidade do Porto, CAUP, Rua das Estrelas, 4150-762 Porto, Portugal\\
\inst{11}Departamento de F\'isica e Astronomia, Faculdade de Ci\^encias, Universidade do Porto, Rua do Campo Alegre, 4169-007 Porto, Portugal\\
\inst{12}Department of Earth, Planetary, and Space Sciences, University of California, Los Angeles, CA 90095, USA\\
\inst{13}Department of Physics, University of Toronto, Toronto, ON M5S 3H4, Canada\\
\inst{14}Departamento de F\'isica Te\'orica e Experimental, Universidade Federal do Rio Grande do Norte, Campus Universit\'ario, Natal, RN, 59072-970, Brazil\\
\inst{15}Department of Physics, McGill University, 3600 rue University, Montr\'eal, QC, H3A 2T8, Canada\\
\inst{16}Department of Earth \& Planetary Sciences, McGill University, 3450 rue University, Montr\'eal, QC, H3A 0E8, Canada\\
\inst{17}Centre Vie dans l’Univers, Facult\'e des sciences de l’Universit\'e de Gen\`eve, Quai Ernest-Ansermet 30, 1205 Geneva, Switzerland\\
\inst{18}European Southern Observatory (ESO), Karl-Schwarzschild-Str. 2, 85748 Garching bei M\"unchen, Germany\\
\inst{19}Space Research and Planetary Sciences, Physics Institute, University of Bern, Gesellschaftsstrasse 6, 3012 Bern, Switzerland\\
\inst{20}Consejo Superior de Investigaciones Cient\'ificas (CSIC), E-28006 Madrid, Spain\\
\inst{21}Bishop's University, Dept of Physics and Astronomy, Johnson-104E, 2600 College Street, Sherbrooke, QC, Canada, J1M 1Z7, Canada\\
\inst{22}Department of Physics, Engineering Physics, and Astronomy, Queen’s University, 99 University Avenue, Kingston, ON K7L 3N6, Canada\\
\inst{23}Department of Physics and Space Science, Royal Military College of Canada, 13 General Crerar Cres., Kingston, ON K7P 2M3, Canada\\
\inst{24}Center for astrophysics $\vert$ Harvard \& Smithsonian, 60 Garden Street, Cambridge, MA 02138, USA\\
\inst{25}Centro de Astrobiolog\'ia (CAB), CSIC-INTA, Camino Bajo del Castillo s/n, 28692, Villanueva de la Ca\~nada (Madrid), Spain\\
\inst{26}Boyce Research Initiatives and Education Foundation, 3540 Carleton St., San Diego, CA 92106, USA\\
\inst{27}El Sauce Observatory, Coquimbo Province, Chile\\
\inst{28}European Southern Observatory (ESO), Av. Alonso de Cordova 3107,  Casilla 19001, Santiago de Chile, Chile\\
\inst{29}Department of Astronomy, Westlake University,Hangzhou 310030, Zhejiang Province, People's Republic of China\\
\inst{30}University Observatory, Faculty of Physics, Ludwig-Maximilians-Universit\"at M\"unchen, Scheinerstr. 1, 81679 Munich, Germany\\
\inst{31}Department of Physics and Astronomy, University of Waterloo, 200 University W, Waterloo, ON N2L 3G1, Canada\\
\inst{32}Perth Exoplanet Survey Telescope, Perth, Western Australia, Australia\\
\inst{33}American Association of Variable Star Observers, Cambridge, USA\\
\inst{*}\email{dr.ares.osborn@gmail.com}
}

\date{Received September 15, 1996; accepted March 16, 1997}

\abstract{
The Neptunian desert is a distinct lack of Neptune-sized planets at short orbital periods, purportedly carved by photoevaporation and tidal circularization following high-eccentricity migration. Constraining these processes and how they vary across different host-star spectral types requires the detailed characterization of planets in the desert and around its boundaries. In this study, we confirm the planetary nature of a massive super-Neptune identified by TESS around the M0 dwarf TOI-672. We analyse photometry from TESS and ExTrA and precise radial velocity measurements taken with the recently commissioned Near-InfraRed Planet Searcher (NIRPS) and HARPS spectrographs. We measure the planetary orbital period, radius, and mass of 3.634 days, $5.31^{+0.24}_{-0.26}\,R_{\oplus}$, and $50.9^{+4.5}_{-4.4}\,M_{\oplus}$, respectively. Our findings place TOI-672 b within the Neptunian ridge, a pile-up of planets from 3--5 days at the Neptunian desert boundary. We then use a novel approach to determine the desert boundaries in period-radius space and instellation-radius space, and, for the first time, compare the Neptunian desert boundaries for planets orbiting FGK versus M dwarf stars. We determine that the boundary ridge shifts slightly inward from $3.3 \pm 1.4$\,days for FGK host stars to $2.2 \pm 1.0$\,days for M dwarf host stars; these values do not statistically significantly differ from each other, and the shift to shorter periods for M dwarf planets is smaller than theoretical photoevaporation models predict. We also find that TOI-672\,b is a single-planet system within the sensitivity limits of our RV and TTV datasets.
}


\maketitle

\nolinenumbers
   

\section{Introduction} \label{sec:intro}

\begin{figure}
    \centering
    \includegraphics[width=\linewidth]{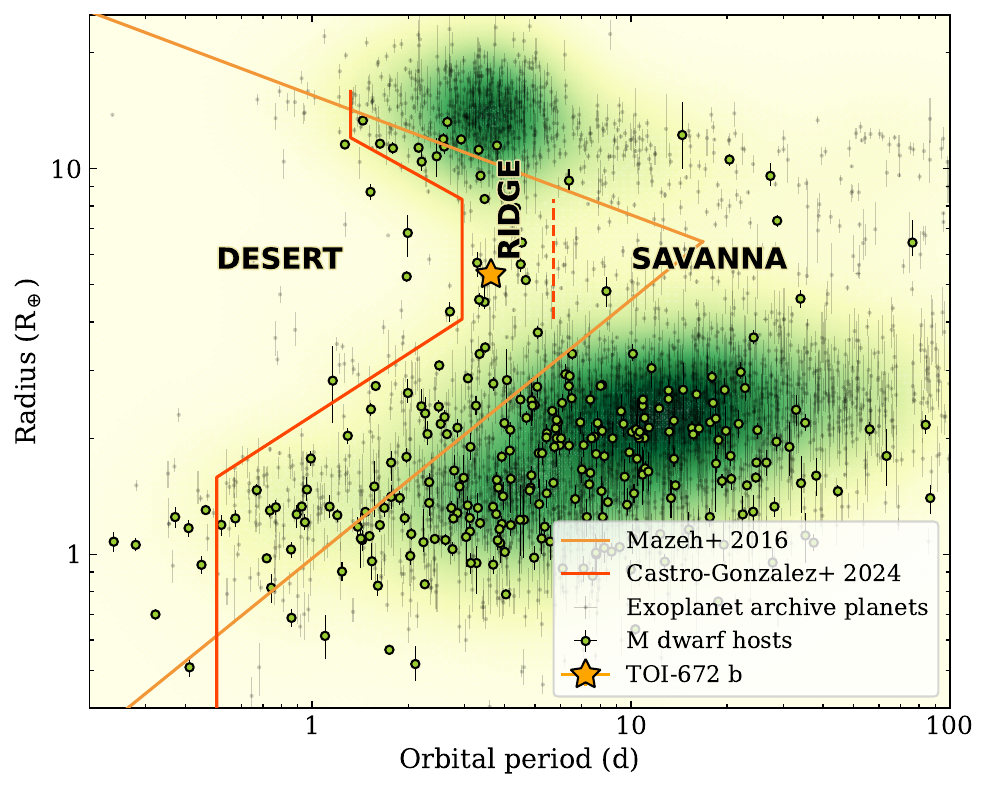}
    \caption{Period-radius diagram showing TOI-672\,b (yellow star) against the known planet population (light grey dots, taken from the NASA exoplanet archive\footnotemark, where only those with periods and radii determined to better than 4 $\sigma$ are shown; the background is also shaded according to this population). Planets around M dwarf hosts are highlighted (as green circles with a black outline, cut on $T_{\rm eff} < 3900$\,K, $M_{*}, R_{*} < 0.6$\,$M_{\odot}, R_{\odot}$). The Neptunian desert boundaries derived by \citet{Mazeh2016} (orange line) and \citet{CastroGonzalez2024} (red line) are shown. The desert, ridge, and savanna, as defined by \citet{CastroGonzalez2024}, are labelled, and the higher-period cut-off of the ridge is shown (red dashed line).}
    \label{fig:desert}
\end{figure}

A prominent feature in the exoplanet population is the ``Neptune desert'', a significant lack of Neptune-sized exoplanets (approximately $2-10\,R_{\oplus}$ and $6-250\,M_{\oplus}$) on short orbital periods ($\lesssim 3$\,days), shown in Fig.~\ref{fig:desert}. The existence of the Neptune desert was noted by many as the exoplanet population grew \citep[see e.g.,][]{Szabo2011,Beauge2013,Lundkvist2016,Mazeh2016} and given several names (e.g., the ``sub-Jovian Pampas'', a ``hot-super-Earth desert'', the ``short-period Neptunian desert''). 

The boundaries of the Neptunian desert and the planet populations that underpin it have been the subject of a number of studies. Perhaps the most prominent is the work by \citet{Mazeh2016}, which followed earlier work by \citet{Szabo2011}, but featured a larger sample of planets. \citet{Mazeh2016} defined empirical boundaries for the desert using linear slopes to designate the upper and lower boundaries in the period-radius and period-mass spaces, shown in Fig.~\ref{fig:desert}. The analysis used a mix of known planets with masses reported in the exoplanet encyclopedia\footnote{The exoplanet encyclopedia can be accessed at \url{https://exoplanet.eu}} and Kepler planet candidates, with no corrections for detection bias. These boundaries have become the de facto location of the desert in many works since, with extrapolation of the boundaries to intersect at an orbital period of $\sim 10$ days. 

There have now been several recent studies devoted to updating the desert boundaries and exploring nearby regions \citep[e.g.,][]{Deeg2023,Szabo2023,CastroGonzalez2024,Magliano2024,PelaezTorres2024}. It has been proposed that there is a Neptunian ``savanna'' \citep{Bourrier2023}, a somewhat milder deficit of Neptunian-sized planets ($\sim4-8\,R_{\oplus}$) at longer periods ($\gtrsim 5$\,days). \citet{CastroGonzalez2024} used the Kepler DR25 data to derive new boundaries with planets weighted according to transit and detection probabilities (i.e., correcting for observational biases). In period-radius space, they used a kernel density estimation on the weighted individual planets to define the desert boundaries as the region with no planets at the $3-\sigma$ level. Similarly to \citet{Mazeh2016}, \citet{CastroGonzalez2024} draw upper and lower boundaries but identify an over-density of planets between $\sim$$3-6$\,days, subsequently named the ``ridge'', which cuts off the tip of the desert and separates it from the savanna (see Fig.~\ref{fig:desert}).

There is a consensus that the Neptune desert is not the result of observational biases as transit and radial velocity surveys are highly sensitive to Neptune-like planets on short orbital periods. The prevailing theory is that the Neptune desert is sculpted by several processes. The upper boundary is predominantly shaped by high-eccentricity migration of large planets, where gravitational interactions kick the planet onto an eccentric orbit before being tidally circularized near periastron \citep{Matsakos2016}. Planets with larger masses can tidally circularize closer to their stars without experiencing tidal disruption, producing a negative slope for the upper boundary in period-radius and period-mass spaces \citep{Owen2018}. The lower boundary is thought to be predominantly shaped by photoevaporation wherein Neptune-sized planets are stripped of their H$_2$/He envelopes to become super-Earths due to XUV irradiation from their host stars \citep[e.g.,][]{Owen2018,Ionov2018,Owen2019,Vissapragada2022,Thorngren2023}. Mass-loss efficiency increases as orbital period decreases, producing a positive slope for the lower boundary in period-radius and period-mass spaces. Finally, the current understanding of planets within the ridge is that they may be a product of late dynamical migration, with elevated eccentricities and misaligned orbits \citep{CastroGonzalez2024}.

It follows from the physical mechanisms purported to carve out the Neptunian desert that its boundaries do not solely depend upon the properties of the planets, but also their host stars. Orbital period has been used as an important parameter in many Neptunian desert studies as it is readily available for many planets and with exceptionally high precision. However, a planet at a given orbital period around an F star, for example, is going to be more heavily irradiated than a planet at the same orbital period around an M dwarf. With this in mind, some studies have moved away from period space and instead use alternative dimensions such as instellation, planetary equilibrium temperature, and lifetime X-ray irradiation \citep[e.g.,][]{Lundkvist2016,McDonald2019,Burt2020,Kanodia2021,Persson2022,Powers2023,Magliano2024}. \citet{Szabo2019} and \citet{Szabo2023} showed the boundaries depend on stellar parameters including, in order of decreasing significance, effective temperature, metallicity, log~$g$, and stellar mass. In particular, \citet{Szabo2019} revealed, for Neptune-sized planets in the desert region, an increase in occurrence of close-in planets with decreasing stellar effective temperature. Approximately $60\%$ of planets with host stars colder than $5600$\,K have orbital periods less than 10\,days, but for host stars hotter than $5600$\,K, this drops to $\sim 10\%$. In short, in this particular region, cooler stars have a higher proportion of short period planets compared to hotter stars. Consequently, the Neptunian desert boundaries derived by e.g., \citet{Mazeh2016} and \citet{CastroGonzalez2024} in the period-radius space may not be the same when considering planets around different types of host stars. This is highlighted by \citet{Kanodia2021} and \citet{Powers2023}, where they translate the \citet{Mazeh2016} desert boundaries into insolation-radius space for different stellar types - the boundaries move towards lower instellations for M dwarf hosts compared to FGKs, leading to the question of whether your M dwarf planet is in the desert or not. Indeed, \citet{Hallatt2022} predict that the opening of the Neptunian desert (i.e., where the upper and lower boundaries intersect) in period-radius space should shift to shorter orbital periods around M dwarfs, beginning at $\sim0.7$\,days around $0.5\,M_\odot$ stars and $\sim 1.5$\,days around $0.8\,M_\odot$ stars, compared to $\sim 3$\,days around $1\,M_\odot$ stars. They predict that a shift in the opening and the width of the Neptunian desert would be most evident comparing the planet sample around M dwarfs versus solar-type stars. 

\footnotetext{The NASA exoplanet archive can be accessed at \url{https://exoplanetarchive.ipac.caltech.edu/}; accessed 24 Aug 2025.}

Investigation of the Neptunian desert thus far has been biased towards planetary systems around Sun-like stars as the Kepler primary mission, the source of a large proportion of exoplanet discoveries, was focused on the search for Earth-sized planets around FGK stars. Also, M dwarfs are typically harder to observe as they are fainter; this is especially true for precise radial velocity (RV) follow-up to confirm the nature of planetary candidates and obtain their masses, as observations of this kind are typically conducted at visible wavelengths. Consequently, samples used for studying the Neptunian desert have included comparatively few planets around M dwarfs compared to FGK stars. There are now, however, a number of spectrographs that are particularly suited to characterising planets around M dwarfs. Optical spectrographs that operate below $1\,\mu$m and are mounted on 8\,m-class telescopes, e.g., ESPRESSO \citep{Pepe2021} and MAROON-X \citep{Seifahrt2018}, are able to collect more light and potentially reach these fainter targets. Alternatively, M dwarfs are brighter in the red optical and near-infrared (nIR), thus motivating the push towards the commissioning of nIR spectrographs, such as CARMENES \citep{Quirrenbach2016}, IRD \citep{Tamura2012}, HPF \citep{Mahadevan2012}, SPIRou \citep{Donati2020}, and the Near-InfraRed Planet Searcher (NIRPS) \citep{Bouchy2025,Artigau2024SPIE}, the recently commissioned nIR arm of the High Accuracy Radial velocity Planet Searcher, HARPS \citep{Pepe2002,Mayor2003}. 

Here we present the confirmation of a hot Neptune orbiting an M0 dwarf star, TOI-672\,b, and the precise characterisation of its mass using the NIRPS and HARPS RV spectrographs. These data were obtained under the NIRPS Guaranteed Time Observations (GTO) program. We refine the planetary orbital and physical parameters and confirm its position in the Neptunian ridge, making it an interesting test case for formation and evolution models of Neptunian desert planets. 

Our paper is laid out as follows. In Section~\ref{sec:obs}, we describe our photometric and spectroscopic observations of TOI-672. We then report stellar parameters, including elemental abundances of the host star in Section~\ref{sec:stellar}. In Section~\ref{sec:fit}, we describe the joint fit model to the RV and transit data. In Section~\ref{sec:disc}, we discuss the results of our joint fit and the nature of TOI-672\,b, predicting its composition and looking at its photoevaporation history.  We also calculate the sensitivity of our RV data to detecting additional planets in the system, and search for transit timing variations. We finally present an analysis of the Neptunian desert boundaries around M dwarf planet hosts in comparison to FGK hosts in Section~\ref{sec:disc_desert}. We put forward our conclusions in Section~\ref{sec:conclusion}.


\section{Observations and previous validation}\label{sec:obs}

\begin{table}
    \small
    \centering
    \caption{Details for the TOI-672 system. }
    \label{tab:system}
    \begin{threeparttable}
    \begin{tabular}{llll}
    \toprule							
    \textbf{Property}   & \textbf{(Unit)}	& \textbf{Value}          & \textbf{Source}  \\
    \midrule							
    \multicolumn{4}{l}{\textbf{Identifiers}}                                             \\	
    TIC ID              &                   & 151825527               & TICv8            \\
    2MASS ID            &                   &	J11115769-3919400     &	2MASS            \\
    Gaia ID             &                   &	5396580575830873728   &	Gaia DR3   \\
    \midrule							
    \multicolumn{4}{l}{\textbf{Astrometric properties}}                                  \\
    R.A.                & (J2000.0)         & 11:11:57.696            &	Gaia DR3	 \\
    Dec.                & (J2000.0)	        & -39:19:40.08	          & Gaia DR3	 \\
    Parallax            & (mas)	            & 14.96 $\pm$ 0.02        & Gaia DR3	 \\
    Distance            & (pc)	            & 67.27$^{+0.07}_{-0.09}$ & Gaia DR3	 \\
    $\mu_{\rm{R.A.}}$	& (mas\,yr$^{-1}$)	& 85.07 $\pm$ 0.02        & Gaia DR3	 \\
    $\mu_{\rm{Dec}}$	& (mas\,yr$^{-1}$)	& -74.5 $\pm$ 0.01        & Gaia DR3	 \\
    Radial velocity     & (km\,s$^{-1}$)    & -5.07                   & Gaia DR3     \\
    Gaia RUWE           &                   & 1.16                    & Gaia DR3     \\
    \midrule							
    \multicolumn{4}{l}{\textbf{Photometric magnitudes}}							         \\
    TESS                         & (mag)	         & 11.674 $\pm$ 0.007	  & TICv8            \\
    \textit{B}	                 & \multirow[t]{18}{*}[-0.5em]{$\vdots$}	         & 15.094 $\pm$ 0.010	   & TICv8            \\
    \textit{V}                   &              & 13.576 $\pm$ 0.026      & TICv8            \\
    \textit{G}                   &              & 12.724 $\pm$ 0.003      & Gaia DR3   \\
    \textit{G$_{\rm BP}$}        &              & 13.820 $\pm$ 0.001      & Gaia DR3   \\
    \textit{G$_{\rm RP}$}        &              & 11.6800 $\pm$ 0.0007    & Gaia DR3   \\
    \textit{J}                   &              & 10.360 $\pm$ 0.026      & 2MASS            \\
    \textit{H}                   &              & 9.698 $\pm$ 0.022	   & 2MASS            \\
    \textit{K}                   &              & 9.507 $\pm$ 0.023	   & 2MASS            \\
    W1                           &              & 9.357 $\pm$ 0.022 & WISE \\
    W2                           &              & 9.337 $\pm$ 0.019 & WISE \\
    W3                           &              & 9.194 $\pm$ 0.033 & WISE \\
    W4                           &              & 8.789 $\pm$ 0.414 & WISE \\
    \textit{g'}                  &              & 14.388 $\pm$ 0.006 & CDS \\
    \textit{r'}                  &              & 12.987 $\pm$ 0.002 & CDS \\
    \textit{i'}                  &              & 12.078 $\pm$ 0.001 & CDS \\
    \textit{z'}                  &              & 11.606 $\pm$ 0.001 & CDS \\
    R                            &              & 12.597 $\pm$ 0.002 & CDS \\
    I                            &              & 11.485 $\pm$ 0.001 & CDS \\
    \bottomrule							
    \end{tabular}
    \begin{tablenotes}
    \item \textit{Sources:} TICv8 \citep{Stassun2019}, 2MASS \citep{Skrutskie2006}, WISE \citep{WISE2014}, Gaia DR3 \citep{GaiaDR3} and Gaia DR3 synthetic photometry (CDS) \citep{GaiaDR3Synth}.
    \end{tablenotes}
    \end{threeparttable}
\end{table}

\subsection{TESS photometry}\label{sec:obs_tess}

The TOI-672 system (Table~\ref{tab:system}) was observed in Transiting Exoplanet Survey Satellite (TESS) Sectors 9 (28 Feb -- 26 Mar 2019), 10 (26 Mar -- 22 Apr 2019), 36 (07 Mar -- 2 Apr 2021), and 63 (10 Mar -- 6 Apr 2023), all on Camera 2 with a 2-min cadence. TOI-672.01 (now TOI-672\,b) was detected by the Transiting Planet Searcher \citep[TPS,][]{Jenkins2002,Jenkins2010} using the light curves from the TESS Science Processing Operations Center (SPOC) pipeline at the NASA Ames Research Centre \citep{Jenkins2016,Caldwell2020}, as well as being detected by the MIT Quick-Look Pipeline \citep[QLP,][]{Huang2020}, and became a TESS Object of Interest (TOI) on 7 May 2019. The transiting planet parameters given in the TOI catalog\footnote{The TOI catalog can be accessed at \url{https://tess.mit.edu/toi-releases/}.} were updated after the Sector 36 then Sector 63 observations, and as of 3 Dec 2023 gave a reference mid-transit time of $2458546.4800 \pm 0.0003$ BJD, a period of $3.633574 \pm 0.000003$\,d, a transit duration of $1.78 \pm 0.05$\,hours, and a depth of $8668 \pm 124$\,ppm (parts per million). 

\begin{figure*}
    \centering
    \includegraphics[width=\linewidth]{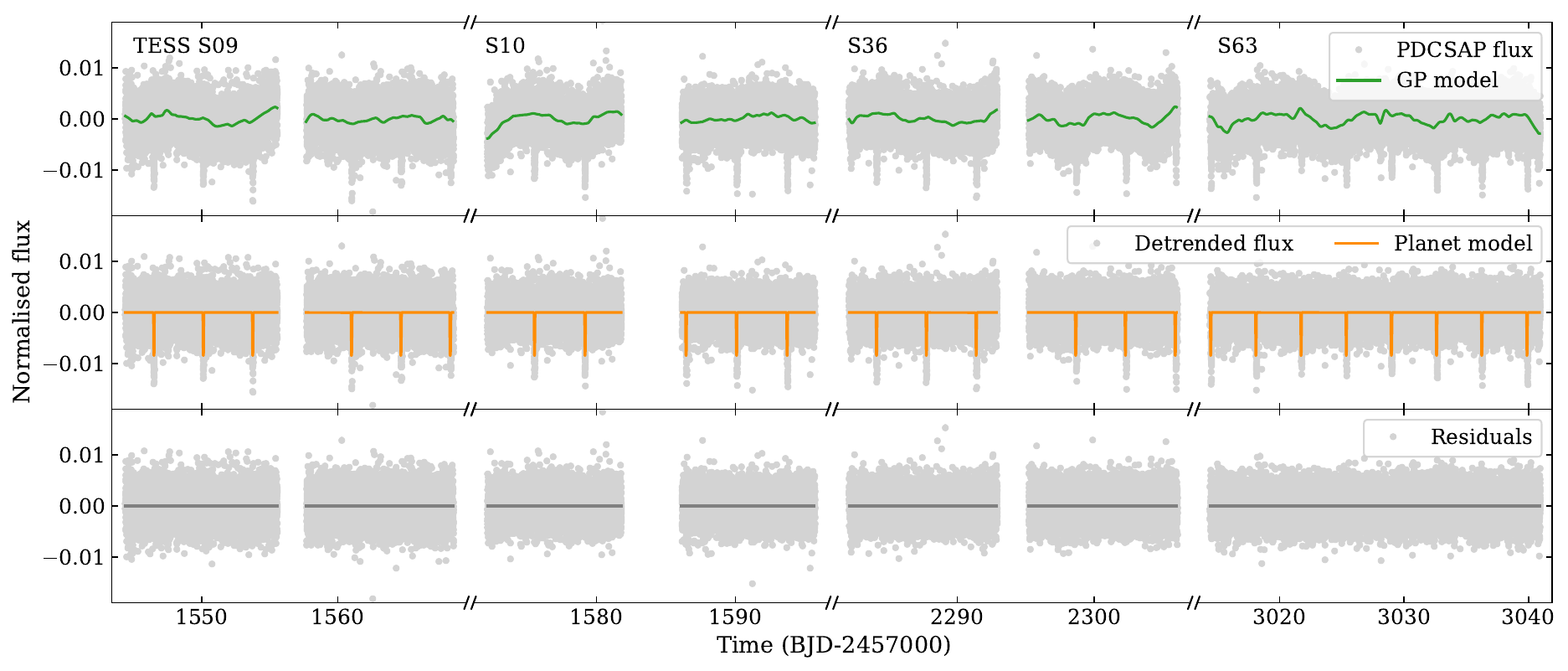}
    \caption{The TESS photometry for TOI-672 covering Sectors 9, 10, 36, and 63 (left to right, chronologically); the data are described in Section~\ref{sec:obs_tess} and the fit to these data is described in Section~\ref{sec:fit_tess}. \textit{Top panel:} the PDCSAP flux (grey circles) and the GP model (green line) used for detrending. \textit{Middle panel:} the detrended flux after the GP model is subtracted, showing the transit model (orange line). \textit{Bottom panel:} the residuals left after the GP and transit models are subtracted from the photometry.}
    \label{fig:tess}
\end{figure*}

The data products are available on the Mikulski Archive for Space Telescopes (MAST\footnote{TESS data products are accessible from MAST at \url{https://archive.stsci.edu/missions-and-data/tess}.}) and were produced by the TESS SPOC. We downloaded the publicly available photometry provided by the SPOC pipeline. In our main fit (Section~\ref{sec:fit}) we use the Presearch Data Conditioning Simple Aperture Photometry (PDCSAP), which is the product of removing common trends and artifacts from the Simple Aperture Photometry (SAP) by the SPOC Presearch Data Conditioning (PDC) algorithm \citep{Twicken2010,Smith2012,Stumpe2012,Stumpe2014}.  We remove all data points with a non-zero quality flag, and show the median-normalised PDCSAP flux in Fig.~\ref{fig:tess}. The PDCSAP photometry preserves stellar activity, which we will detrend as described in Section~\ref{sec:fit_tess}. 

\begin{figure}
    \centering
    \includegraphics[width=\columnwidth]{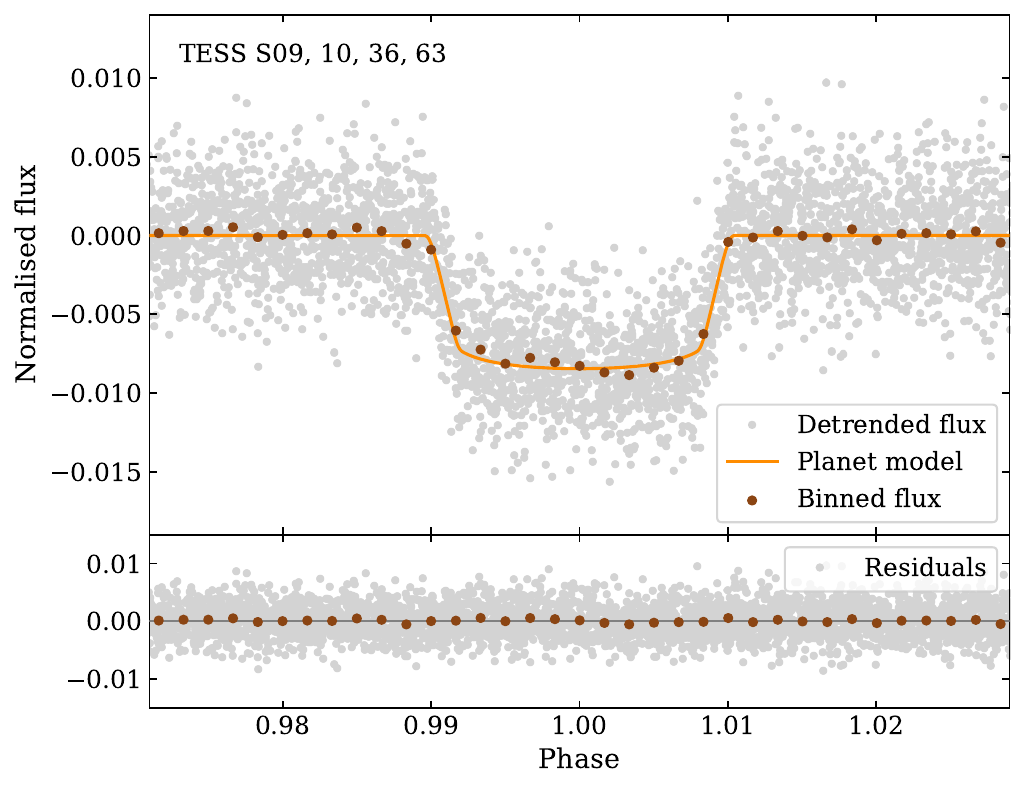}
    \caption{\textit{Top panel:} the detrended TESS photometry for TOI-672 from all sectors (grey circles, binned as dark brown circles), phase folded on the best-fit planetary period, with the transit model shown (orange line). \textit{Bottom panel:} the residuals after the GP and planet models are subtracted from the photometry.}
    \label{fig:tess_fold}
\end{figure}

\begin{figure}
    \centering
    \includegraphics[width=\columnwidth]{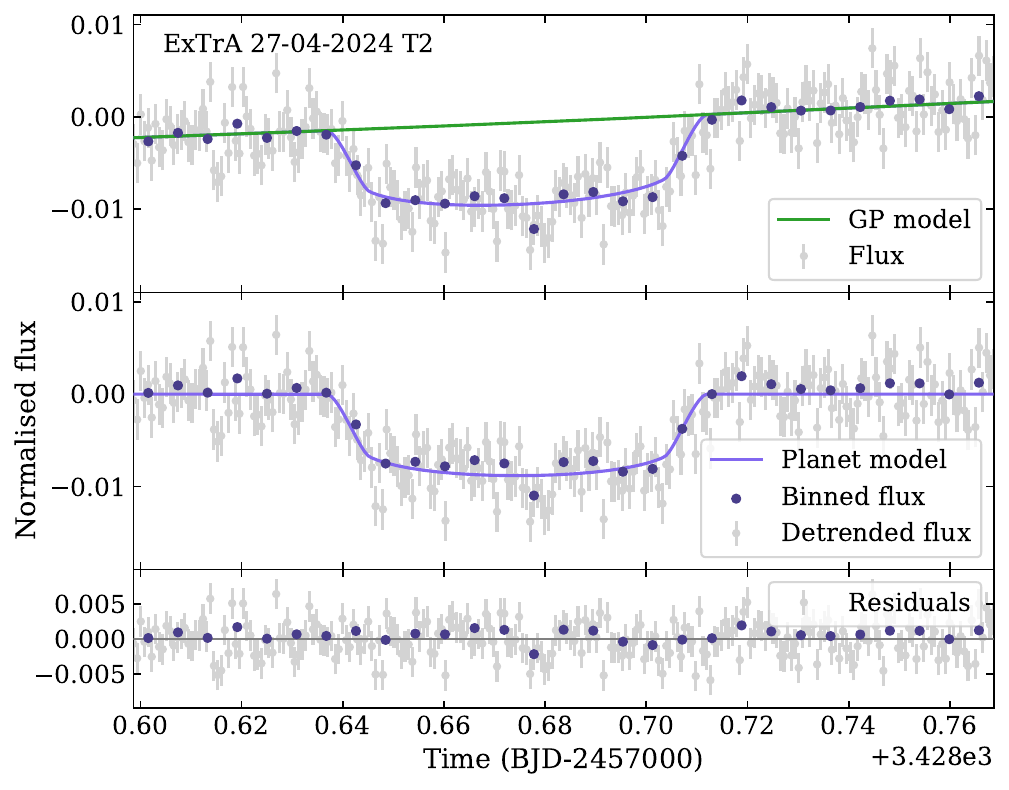}
    \caption{The ExTrA photometry for TOI-672 from the night of 27 April 2024 using Telescope 2 (T2); the data are described in Section~\ref{sec:obs_extra} and the fit to these data is described in Section~\ref{sec:fit_extra}. \textit{Top panel:} the extracted photometry (grey circles, binned as dark purple circles), and the GP model used to detrend it (green line). \textit{Middle panel:} the detrended photometry, showing the transit model (purple line). \textit{Bottom panel:} the residuals after the GP and transit models are subtracted from the photometry.}
    \label{fig:extra}
\end{figure}

\subsection{Ground-based photometry}\label{sec:obs_otherphotom}

The TESS pixel scale is $\sim$$21\arcsec$ per pixel, and while the SAP aperture for TOI-672 is variable, it is typically three pixels ($\sim 1\arcmin$) on a side. We show the apertures for each sector in Fig.~\ref{fig:tpfplotter}. Multiple stars can blend in an aperture of this size which can cause confusion in the source of the transit detection, and dilute the transit depth. Indeed, looking for other nearby stars within several magnitudes of TOI-672, there are two faint Gaia sources that fall within the aperture (with $\Delta G_{mag} = 3.21, 4.64$), and a bright neighbouring source ($\Delta G_{mag} = -0.34$) located $53.17 \arcsec$ to the north west (see Fig.~\ref{fig:tpfplotter}). The SAP photometry includes a dilution parameter (a ratio of the target flux to total flux in the aperture), which varies from 0.95 to 0.98 depending on sector.

All light curve data described below are available on the {\tt EXOFOP-TESS} website\footnote{The EXOFOP-TESS for TOI-672 can be accessed at \url{https://exofop.ipac.caltech.edu/tess/target.php?id=151825527}.}. We did not include these observations in our joint fit, with the exception of one transit from ExTrA (see Section~\ref{sec:fit}), but we do use their individual mid-transit times when conducting our Transit Timing Variations (TTV) search (see Section~\ref{sec:disc_ttvs}). The light curves are depicted in Fig.~\ref{fig:sg1}. A summary of all photometric transits, and whether they were used in the joint fit and for the TTV search, can be found in Table~\ref{tab:ttvs}.

\subsubsection{TFOP photometry}

To attempt to determine the true source of the TESS detection, ground-based time-series follow-up photometry of the field around TOI-672 were obtained as part of the TESS Follow-up Observing Program \citep[TFOP;][]{collins2019}\footnote{The TFOP webpage can be accessed at \url{https://tess.mit.edu/followup}.}. The majority of these observations are presented in the validation of this target \citep[detailed in Section \ref{sec:obs_validation};][]{Mistry2023}, so we do not describe them here except as a brief summary. 

Two full transits were observed from the Las Cumbres Observatory Global Telescope \citep[LCOGT;][]{Brown2013} network nodes on 2019 May 11 (at the Cerro Tololo Inter-American Observatory, CTIO, in Chile) and 2020 March 19 (at the Siding Spring Observatory, SSO, near Coonabarabran, Australia). A partial and a full transit were observed on UTC 2019 May 11 and 2020 February 01, respectively, from the Evans 0.36\,m telescope at El Sauce Observatory in Coquimbo Province, Chile. A full transit was observed on 2019 May 26 using the Perth Exoplanet Survey Telescope (PEST), located near Perth, Australia. One additional transit, not described by \citet{Mistry2023}, was made on June 14, 2022, using a 17$\arcsec$ telescope (CDK17) installed in Chile (Deep Sky Chile) with a Cousin R filter. The light curve was calculated using the HOPS software from the ExoClock collaboration \citep{Kokori2022}. 

\subsubsection{ExTrA photometry}\label{sec:obs_extra}

We supplemented the existing TFOP photometry with additional ground-based transit observations over many transit windows to aid in our TTV search, and to expand the baseline of transit observations to improve constraints on the orbital period of the planet. We used ExTrA \citep[Exoplanets in Transits and their Atmospheres,][]{Bonfils2015}, a low-resolution nIR ($0.85-1.55\,\mu$m) multi-object spectrograph fed by three 60-cm telescopes located at La Silla Observatory in Chile.  We used 8$\arcsec$ diameter aperture fibres and the low-resolution mode ($R\sim 20$) of the spectrograph, with an exposure time of 60~seconds. We positioned five fibres in the focal plane of each telescope to isolate light from the target and four comparison stars. The resulting ExTrA data were analysed using custom data reduction software \citep{Cointepas2021}.

We observed transits of TOI-672\,b using one, two, or all three of the ExTrA telescopes (T1, T2, and T3, respectively) on a total of 12 nights from 2019 to 2024: 22 Apr 2021 (T2, T3); 14 May 2021 (T2, T3); 3 Feb 2022 (T1, T2, T3); 4 Mar 2022 (T1, T2, T3); 6 Apr 2022 (T2, T3); 24 Apr 2022 (T2, T3); 16 Feb 2023 (T1, T2, T3); 28 Mar 2023 (T1, T2, T3); 8 Apr 2023 (T1, T2, T3); 18 Mar 2024 (T1, T2); 29 Mar 2024 (T1, T2); and 27 Apr 2024 (T2). Most ExTrA photometry was not used in the joint fit (see Section~\ref{sec:fit}), with the exception of the most recent observation from 27 April 2024 using T2 as this extended the overall baseline of the photometry. This particular observation is shown in Fig.~\ref{fig:extra}. The remaining photometry (a few nights excepted due to the lack of, or poor quality of, the transit ingress) was used in the aforementioned TTV search (see Section~\ref{sec:disc_ttvs}) and is shown in Fig.~\ref{fig:extra_extra}.

\subsection{Previous validation}\label{sec:obs_validation}

TOI-672\,b was previously validated by the Validation of Transiting Exoplanets using Statistical tools project \citep[VaTEST,][]{Mistry2023}, based on high-resolution imaging, ground-based photometry, and TESS Sectors 9, 10, and 36, but not the most recent Sector 63 data. In summary: TOI-672\,b passes multiple diagnostic tests and is statistically validated using the tool ${\rm TRICERATOPS}$ \citep{Giacalone2021}; the high-resolution imaging from the Zorro instrument on Gemini-S shows no close in contaminating stellar companions; and the ground-based follow-up (which includes the LCOGT, Evans, and PEST observations described above) shows the transit event occurring on the target star. The TESS data from Sectors 9, 10, and 36 were fit with a transit model using ${\rm Juliet}$ \citep{Espinoza2019} to obtain an orbital period of $3.633575 \pm 0.000001$\,days, a radius of $5.26^{+0.08}_{-0.10}$\,$R_{\oplus}$ assuming a stellar radius of $0.54 \pm 0.02$\,$R_{\odot}$, an impact parameter of $0.42^{+0.11}_{-0.21}$, and an inclination of $88.43^{+0.82}_{-0.52}\degree$. 

\subsection{NIRPS and HARPS spectroscopy}\label{sec:obs_nirpsharps}

\begin{figure*}
    \centering
    \includegraphics[width=\linewidth]{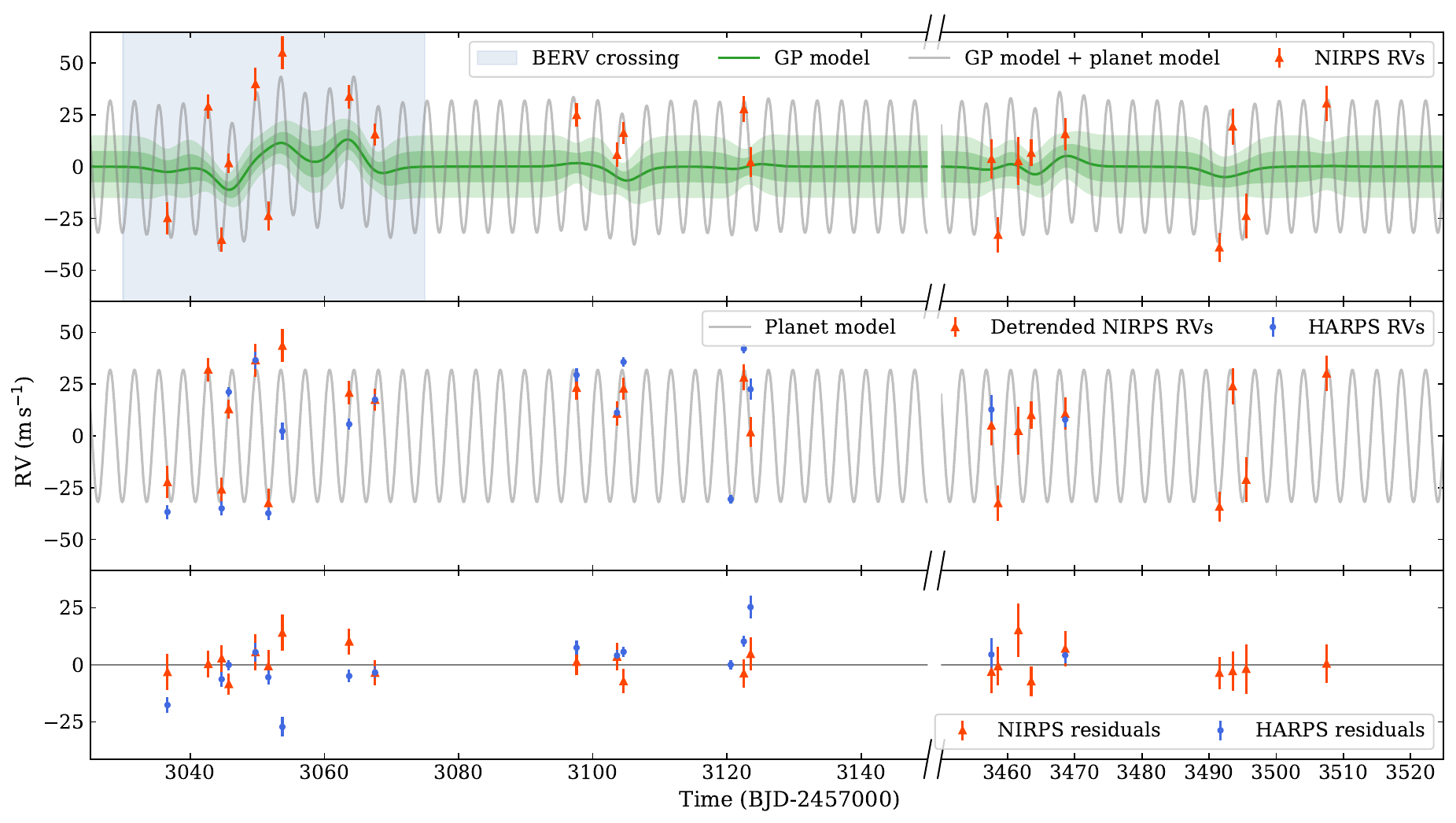}
    \caption{The NIRPS and HARPS data for TOI-672; the data are described in Section~\ref{sec:obs_nirpsharps} and the fit to these data are described in Section~\ref{sec:fit_nirpsharps}. \textit{Top panel:} the NIRPS RVs (red triangles), the GP model fit to the NIRPS RVs (green line, with the one and two standard deviations of the fit shaded), and the combined GP and planet model (grey line). The BERV crossing event described in Section~\ref{sec:obs_nirpsharps} is shaded in blue. \textit{Middle panel:} both sets of RVs, where NIRPS has been detrended with a GP and HARPS (blue circles) has no detrending, with the planet-only model is shown (grey line). \textit{Bottom panel:} the RV residuals after the GP and planet models are subtracted.}
    \label{fig:rvs}
\end{figure*}

\begin{figure}
    \centering
    \includegraphics[width=\columnwidth]{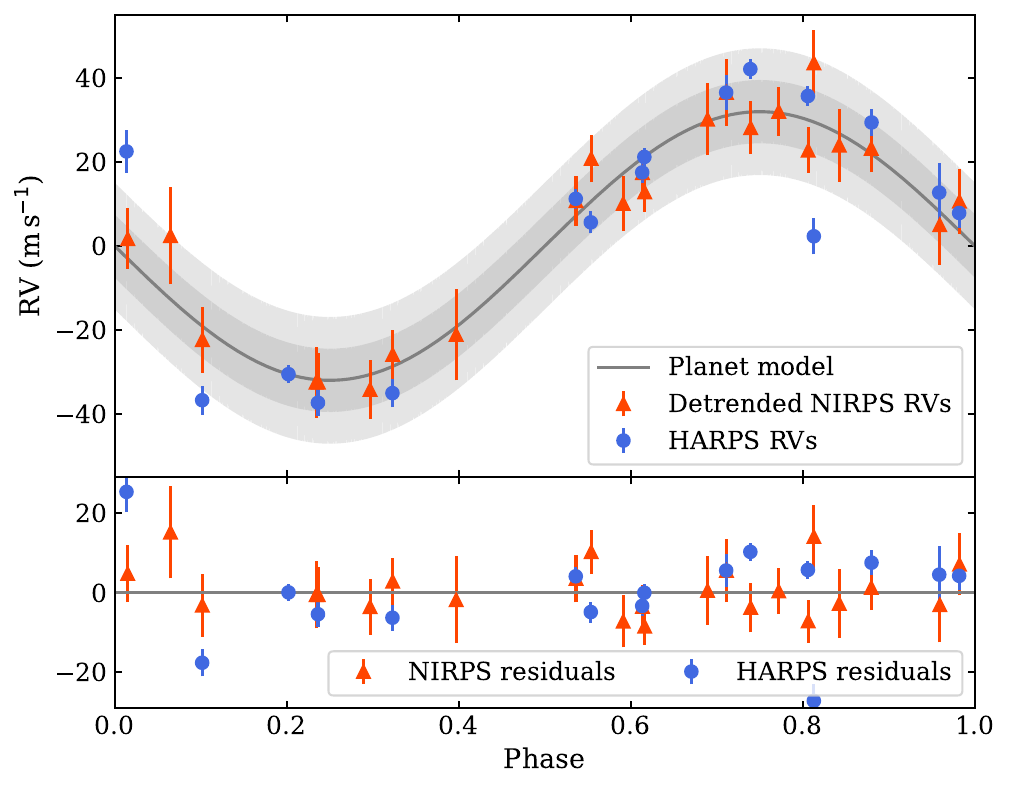}
    \caption{\textit{Top panel:} the detrended NIRPS and HARPS RVs (red triangles and blue circles, respectively), phase-folded on the best-fit planetary period, with the planet model shown (grey line, with the one and two standard deviations of the fit shaded). \textit{Bottom panel:} the RV residuals after the GP and planet model are subtracted.}
    \label{fig:rvs_fold}
\end{figure}

We obtained simultaneous radial velocity measurements of TOI-672 with the NIRPS \citep{Bouchy2025,Artigau2024SPIE} and HARPS \citep{Pepe2002,Mayor2003} spectrographs. Both instruments are mounted on the ESO 3.6-m telescope at the La Silla Observatory in Chile and can be operated simultaneously through the use of a VIS-nIR dichroic.

HARPS operates in the optical, covering a wavelength range of $0.38 - 0.69\,\mu$m. It has two modes corresponding to two different sets of fibres: a high accuracy mode (HAM, $R=115,000$) and a high efficiency mode (EGGS, $R=80,000$). While EGGS is lower in resolution, it uses a fibre with a projected aperture of 1.4\arcs\ on sky (a fibre diameter of 100\,$\mu$m), bigger than the 1.0\arcs\ aperture (70\,$\mu$m diameter) of the HAM fibre, and therefore gains in flux collection, important when observing fainter M dwarf targets. HARPS was used in EGGS mode with an exposure time of 1200\,s.

NIRPS is the recently commissioned nIR, adaptive optics-assisted spectrograph intended to complement HARPS as its ``red arm'', and began science operations on 1 April 2023. It covers a wavelength of $0.97-1.92\,\mu$m across the $YJH$ bands. NIRPS also has two observing modes: a high accuracy (HA) mode using a 0.4\arcs\ fibre with a resolution of 88,000, and a high efficiency (HE) mode using a bigger fibre of $0.9\arcsec$ at the expense of a lower spectral resolution of 75,200. NIRPS was used in HE mode with two back-to-back exposures of 600\,s, which are then binned per night. This is due to a practical limit on the integration time for NIRPS; there is no benefit to integrating longer than the point at where dark current surpasses the readout noise, which occurs at an integration time of approximately 900\,s. Longer total integrations are instead achieved by taking multiple exposures and combining them \citep{Artigau2024}.

Our observations of TOI-672 were obtained through the NIRPS Guaranteed Time Observations (GTO) program, which began in April 2023 and was allocated 725 nights over five years. The program is split into three primary sub-programs, each with their own distinct science cases \citep{Bouchy2025}. This work was performed under Work Package 2 (WP2), which is performing mass measurements of known planets around M dwarfs \citep{Parc2025,Frensch2025,Weisserman2026}. We specifically observed TOI-672 as part of the ``Deep Search'' programme. RV follow-up makes us sensitive to more planets than those known from lightcurves alone. Deep Search prioritises the follow-up of targets with an a priori higher likelihood of harbouring these extra planets, and additionally prioritises systems where there is a higher chance of later seeing these planets in transit too. 

We observed TOI-672 with NIRPS and HARPS simultaneously over a total of 24 nights between April 2023 and July 2024. There were nights when, for various reasons, only one instrument managed to successfully observe the target, namely, 8 Apr 2023 (NIRPS only); 24 Jun 2023 (HARPS only); 27 and 30 May, 1 and 29 Jun, and 1 and 3 Jul 2024 (NIRPS only). On the night of 10 Apr 2023, the target was observed twice by both HARPS and NIRPS (i.e., 2 HARPS spectra and 4 NIRPS spectra were obtained). Additionally, two observations were spuriously made in HARPS HAM on the nights of 29 Jun 2024 and 1 Jul 2024; we do not include these in our analysis as they would require treatment as a separate instrument and the RV precision is comparatively worse. No observations were obtained when the planet was in transit, so our RVs are not affected by the Rossiter-McLaughlin effect. In total, we obtained 48 usable spectra from NIRPS, which are binned nightly to produce 23 RV measurements. 18 HARPS spectra were obtained, but one from the night of 14 Jul 2024 only had a signal-to-noise ratio (S/N) of 2 at 550\,nm and was thus discarded, resulting in 17 HARPS RV measurements. The median S/N for the NIRPS at 1600\,nm is 40.0; it is 9.7 for HARPS at 550\,nm. 

The NIRPS GTO team currently uses two independent pipelines to reduce NIRPS data and produce 2D spectra. The first is an ESO pipeline based on the ESPRESSO pipeline \citep{Pepe2021}, which has been adapted to correct for the telluric absorption lines and OH emission lines \citep{Allart2022,Bouchy2025,Srivastava2026}. The second is the APERO pipeline initially created to reduce SPIRou data \citep{Cook2022}, another nIR spectrograph mounted on the Canada-France-Hawaii Telescope and with similar specifications to NIRPS. Our NIRPS data presented herein were reduced using the ESO pipeline. We reduced our HARPS data using the standard offline HARPS data reduction pipeline (ESO DRS 3.5). 

While the pipelines above try and correct for contamination from telluric absorption and in the nIR spectra, this is not always perfect, particularly at times where the systematic velocity of the star is close to the barycentric Earth radial velocity (BERV). Here, the stellar and telluric lines overlap, adversely affecting the correction. For this system, this ``BERV crossing'' occurs between approximately 3030 to 3075 BJD-2457000 (highlighted in Fig~\ref{fig:rvs}). This event reoccurs every 365 days, but our second year of observations do not start until after the BERV crossing for that year has already occurred. We attempt to mitigate the effect of the BERV crossing following the method of \citet{Srivastava2026}, where we mask pixels across all observations that exhibit variations in flux beyond a certain $\sigma$ threshold relative to the median stellar spectrum. This, however, does not produce a meaningful difference to either the individual RV values, nor in the RMS of the RVs, at several attempted sigma thresholds (down to 2.5$\sigma$). Therefore, we do not use these masked spectra, and instead remove the effects of the BERV crossing event with a GP on the extracted RVs, which is described in Section~\ref{sec:fit_nirpsharps}. We note other methods for correcting for the BERV crossing have been explored in \citet{Parc2025} and \citet{Frensch2025}.

After the 2D spectra is produced, RVs are then extracted. Again, precision radial velocities in the near infrared are affected by telluric absorption residuals that produce time-dependent effects on line profiles. The data-driven Line-By-Line (LBL) method \citep{Artigau2022} was specifically developed to provide resilience to this by operating over individual spectral lines, allowing for identification and removal of these spectral outliers that would otherwise bias the measurement of the radial velocity. We thus extract both NIRPS and HARPS RVs using LBL (version 0.65.006). Due to the relatively few observations of \hbox{TOI-672}, rather than constructing the template spectra from the combined spectra of this target, we instead use ``friend'' templates. The ``friend'' is a different, bright star with similar spectral type and rotational velocity to the target (here TOI-672), but which has had many observations using the same spectrograph, and thus has a template of sufficient S/N and excellent rejection of telluric features. For NIRPS RV extraction, we use NIRPS observations of GJ\,2066 ($T_{\rm eff} = 3550 \pm 157$\,K and \hbox{$\log g = 4.788 \pm 0.006$\,cgs} from TICv8, \citet{Stassun2019}) to construct the friend template. For HARPS RV extraction, we use HARPS observations of TOI-776 (\hbox{$T_{\rm eff} = 3725 \pm 60$\,K}; \hbox{$\log g = 4.8 \pm 0.1$\,cgs}; \hbox{[Fe/H] = $-0.21 \pm 0.08$\,dex}; and \hbox{$v \sin i = 2.2 \pm 1.0$\,km\,s$^{-1}$}, \citet{Fridlund2024}) to construct the friend template. 

The LBL method also measures a number of activity indicators, including the Full-Width at Half-Maximum (FWHM) and contrast, which are both also commonly measured using the Cross Correlation Function (CCF) method. It also provides two additional indicators, the differential line width (${\rm dLW}$) and differential temperature of the star ($d{\rm Temp}$), which are explained by \citet{Artigau2022} and \citet{Artigau2024}, respectively. The activity indicators and stellar rotation are discussed in Section~\ref{sec:rotation}. 

The resultant NIRPS and HARPS LBL time-series data can be found in Table~\ref{tab:nirpsrvs} and Table~\ref{tab:harpsrvs_lbl}, respectively. The RVs are shown in Fig.~\ref{fig:rvs} and Fig.~\ref{fig:rvs_fold}. Lomb-Scargle periodograms \citep{Lomb1976,Scargle1982,Press1989} of the RVs and activity indicators are shown in Fig.~\ref{fig:activityindicators_ls}.


\section{Stellar characterisation}\label{sec:stellar}

\begin{table*}
    \centering
    \small
    \caption{Stellar parameters and chemical abundances of the host star TOI-672.}
    \label{tab:stellarparams}
    \begin{threeparttable}
    \begin{tabular}{lllll}
    \toprule
    \textbf{Stellar parameters} & & \textbf{(Unit)} & \textbf{Value} & \textbf{Source} \\
	\midrule
    \multicolumn{5}{l}{\textbf{Physical parameters}} \\
    Effective temperature        & $T_{\rm eff}$        & (K)	          & $3783 \pm 52$             & Section~\ref{sec:stellparamrelations}, \citet{Mann2015} \\
                                 &                      &                 & $\mathbf{3810 \pm 33}$             & Section~\ref{sec:stellparamnirps}, \citet{Jahandar2024,Jahandar2025} \\
                                 &                      &                 & $3751 \pm 52$             & Section~\ref{sec:stellparamharps}, \citet{Antoniadis20,Antoniadis24} \\
    Spectral type                &                      &                 & M0V                       & \citet{Pecaut2013} \\
    Stellar radius               & $R_\star$            & (R$_{\odot}$)   & $\mathbf{0.54 \pm 0.03}$           & Section~\ref{sec:stellparamrelations}, \citet{Mann2015,Tayar2022} \\
                                 &                      &                 & $0.563 \pm 0.016$         & Section~\ref{sec:stellparamsed}, SED fit, derived \\
    Stellar mass                 & $M_\star$            & (M$_{\odot}$)   & $\mathbf{0.54 \pm 0.03}$           & Section~\ref{sec:stellparamrelations}, \citet{Mann2019,Tayar2022} \\
                                 &                      &                 & $0.570 \pm 0.021$         & Section~\ref{sec:stellparamsed}, SED fit, \citet{Schweitzer2019} \\
    Surface gravity              & $\log g$             & (cgs)           & 4.70 $\pm$ 0.03           & Section~\ref{sec:stellparamrelations}, derived \\
    Bolometric luminosity        & $L_{\rm bol}$        & ($L_{\rm bol, \odot}$) & 0.056$\pm$0.004    & Section~\ref{sec:stellparamrelations}, derived \\
                                 &                      &                 & $\mathbf{0.04799 \pm 0.00022}$     & Section~\ref{sec:stellparamsed}, SED fit \\
    Rotational velocity          & $v \sin i_\star$     & (km\,s$^{-1}$)  & < 3.75                    & Upper limit based on the width of a NIRPS resolution element \\
    Chromospheric activity index & $\log R'_{\rm HK}$   &                 & $-4.69^{+0.29}_{-0.34}$   & Section~\ref{sec:rotation}, \citet{Astudillo-Defru2017} \\
    Predicted rotation period    & $P_{\rm rot}$        & (days)          & $26^{+18}_{-9}$           & Section~\ref{sec:rotation}, \citet{Astudillo-Defru2017} \\
                                 &                      &                 & $19^{+35}_{-6}$           & Section~\ref{sec:rotation}, \citet{SuarezMascareno2018} \\
                                 &                      &                 & $\mathbf{\sim 18.5}$               & Section~\ref{sec:rotation}, ASAS-SN photometry \\
    \midrule
    \multicolumn{5}{l}{\textbf{Chemical abundances}} \\
    Overall metallicity     & $[$M/H$]$     & (dex)                               & $-0.3 \pm 0.1$     & Section \ref{sec:stellparamnirps}, \citet{Jahandar2024,Jahandar2025} \\
                            &           & \multirow[t]{14}{*}[-0.5em]{$\vdots$}   & $\mathbf{-0.25 \pm 0.05}$   & Section \ref{sec:abundances}, Gromek et al. (in prep) \\
    Alpha enhancement       & [$\alpha$/Fe] &                                     & $-0.13 \pm 0.09$   & Section \ref{sec:abundances}, Gromek et al. (in prep), \citet{Hinkel2022} \\
    Iron                    & $[$Fe/H$]$    &                                     & $0.09 \pm 0.11$    & Section \ref{sec:stellparamharps}, \citet{Antoniadis20,Antoniadis24} \\
                            &               &                                     & $\mathbf{-0.14 \pm 0.07}$   & Section \ref{sec:abundances}, Gromek et al. (in prep) \\
    Oxygen                  & $[$O/H$]$     &                                     & $-0.28 \pm 0.06$   & \multirow[t]{11}{*}[-0.5em]{$\vdots$} \\
    Sodium      & $[$Na/H$]$ &  & $0.17 \pm 0.04$ &  \\
    Magnesium   & $[$Mg/H$]$ &  & $-0.20 \pm 0.12$ & \\
    Aluminium   & $[$Al/H$]$ &  & $-0.15 \pm 0.06$ & \\
    Silicon     & $[$Si/H$]$ &  & $-0.15 \pm 0.10$ & \\
    Potassium   & $[$K/H$]$  &  & $0.16 \pm 0.09$ & \\
    Calcium     & $[$Ca/H$]$ &  & $0.07 \pm 0.04$ & \\
    Titanium    & $[$Ti/H$]$ &  & $-0.10 \pm 0.10$ & \\
    Chromium    & $[$Cr/H$]$ &  & $-0.07 \pm 0.04$ & \\
    Manganese   & $[$Mn/H$]$ &  & $0.15 \pm 0.07$ & \\
    Nickel      & $[$Ni/H$]$ &  & $-0.01 \pm 0.08$ & \\
    \midrule
    {\textbf{Elemental Ratios}}\\
    Iron to magnesium     & Fe/Mg   &  & $0.91^{+0.16}_{-0.14}$ & Section \ref{sec:abundances}, Gromek et al. (in prep) \\
    Magnesium to silicon  & Mg/Si   &  & $1.10^{+0.29}_{-0.23}$ & \multirow[t]{3}{*}[-0.5em]{$\vdots$} \\
    Silicon to oxygen     & Si/O    &  & $0.09 \pm 0.02$ \\
    \bottomrule
    \end{tabular}
    \begin{tablenotes}
    \item Methods for deriving parameters are described in full in Section~\ref{sec:stellar}. In the circumstance where multiple values are derived for the same parameter, we indicate in bold which value has been adopted.
    \end{tablenotes}
    \end{threeparttable}
\end{table*}

\begin{figure}[htbp]
    \centering
    \includegraphics[width=\linewidth]{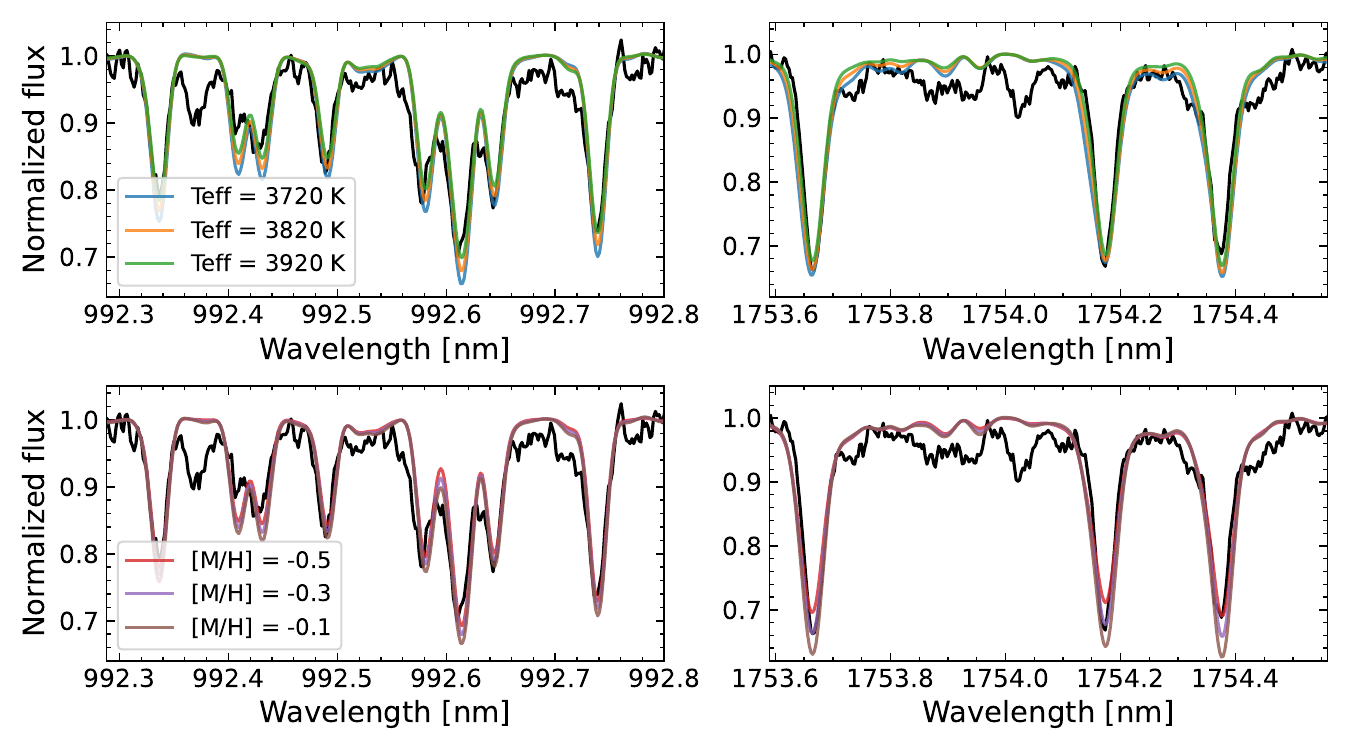}
    \caption{Normalized NIRPS spectrum of TOI-672 (black line) compared to PHOENIX ACES stellar models (coloured lines), described in Section~\ref{sec:stellparamnirps}. \textit{Top row:} we show models with a fixed metallicity of -0.3\,dex and $T_{\rm eff}$ values of 3720 (blue), 3820 (orange), and 3920 (green) K for two spectral regions. \textit{Bottom row:} for the same regions, we show models with a fixed $T_{\rm eff}$ of 3820\,K and metallicity values of -0.5 (red), -0.3 (purple), and -0.1 (brown) dex. For strong spectral lines, we perform $\chi^2$ minimization between the NIRPS spectrum and a grid of PHOENIX ACES models, finding $T_{\rm eff}=3810\pm33$\,K and [M/H]=$-0.3\pm0.1$\,dex.}
    \label{fig:phoenix}
\end{figure}

\subsection{Stellar parameters}\label{sec:stellarparams}

Here we derive parameters for the host star TOI-672, which are reported in Table~\ref{tab:stellarparams}.

\subsubsection{From the NIRPS spectra}\label{sec:stellparamnirps}

We first co-add the telluric-corrected NIRPS spectra from APERO into a single high-resolution template of S/N$\sim$280  to derive spectroscopic parameters following the methodology of \citet{Jahandar2024, Jahandar2025}. We derive the effective temperature, $T_\mathrm{eff}$, and overall metallicity, $\mathrm{[M/H]}$, by fitting a selection of strong spectral lines through $\chi^2$ minimization, using a grid of PHOENIX ACES stellar models \citep{Husser2013} interpolated to $\log g = 4.75$ and convolved to the spectral resolution of NIRPS (shown in Fig.~\ref{fig:phoenix}). We note that we use a slightly different $\log{g}$ than in \ref{sec:stellparamrelations}, but that a variation in $\log g$ of 0.2\,dex has negligible effect on the spectral analysis. We obtain $T_\mathrm{eff} = 3810 \pm 33$\,K and $\mathrm{[M/H]}=-0.3 \pm 0.1$\,dex. We use these values as starting points in our subsequent analysis of the stellar abundances (see Section~\ref{sec:abundances}), which we will ultimately use to measure the true overall metallicity, $\mathrm{[M/H]}$. To validate our spectroscopically-derived $T_\mathrm{eff}$, we also derive $T_\mathrm{eff}$ using the empirical M dwarf $T_\mathrm{eff}$-colour relation from \citet{Mann2015}. We use the {\it Gaia} $G_{\rm BP}$ and $G_{\rm RP}$ magnitudes to derive $T_\mathrm{eff}=3783 \pm 52$\,K, which agrees with the value measured from our NIRPS spectra.

\subsubsection{From the HARPS spectra}\label{sec:stellparamharps}
We then used the co-added spectra from HARPS to derive $T_\mathrm{eff}$ and [Fe/H] with the machine learning code {\tt ODUSSEAS}\footnote{\url{https://github.com/AlexandrosAntoniadis/ODUSSEAS}} \citep{Antoniadis20,Antoniadis24}. This code measures the pseudo-equivalent width of more than 4000 lines in the optical spectra which are then used as input in the machine learning model trained with a reference sample of 47 M dwarfs observed with HARPS. The reference values of $T_\mathrm{eff}$ for those stars come from interferometric calibrations \citep{Khata21}, whereas the [Fe/H] comes from photometric calibrations in binaries \citep{Neves12}. We obtained $T_\mathrm{eff}=3751 \pm 94$\,K, which is in good agreement with the above mentioned values, and \hbox{[Fe/H] = 0.09 $\pm$ 0.11\,dex}, which is higher than the value obtained from NIRPS spectra (see Section~\ref{sec:abundances}). Given that the chemical abundances are derived from the NIRPS spectra, we adopt the stellar parameters obtained with the NIRPS spectra as well. 

\subsubsection{From empirical M dwarf relations}\label{sec:stellparamrelations}
We use the empirical M dwarf radius-luminosity relation from \citet{Mann2015} to derive a stellar radius of \hbox{$R_{\star}= 0.54 \pm 0.02\,R_{\odot}$} from the star's absolute $K_s$-band magnitude. However, \citet{Tayar2022} argue that there is a systematic uncertainty floor that should be accounted for, with the recommendation to add a 4.2\% error in quadrature with the formal uncertainty on stellar radius. This results in \hbox{$R_{\star}= 0.54 \pm 0.03\,R_{\odot}$}. 
Similarly, we use the empirical M dwarf mass-luminosity relation from \citet{Mann2019} to derive a stellar mass of $M_{\star} = 0.54 \pm 0.01\,M_{\odot}$ based on the star's absolute $K_s$-band magnitude. Again, we add a systematic error in quadrature with this value as recommended by \citet{Tayar2022}; for stellar mass this is 5\%. This results in $M_{\star} = 0.54 \pm 0.03\,M_{\odot}$. We then use the stellar mass and radius to derive the stellar surface gravity, $\log g = 4.70 \pm 0.03$\,cgs. We can also use the stellar radius and effective temperature to derive the bolometric luminosity $L_{\rm bol} = 0.056 \pm 0.004$\,$L_{\rm bol, \odot}$. 

\subsubsection{From SED fitting}\label{sec:stellparamsed}

As an alternate approach to determining the bolometric luminosity and stellar radius and mass, we built the Spectral Energy Distribution (SED) using flux densities from several broadband photometric surveys: $G_{\rm BP}$, $G$, and $G_{\rm RP}$ from the Gaia mission \citep{GaiaDR3}; \textit{B} and \textit{V} from APASS \citep{Henden2015}; \textit{B} and \textit{I} from the Johnson UBVRI system \citep{Bessell1990}; \textit{J}, \textit{H}, and \textit{Ks} from 2MASS \citep{Skrutskie2006}; \textit{g'}, \textit{r'}, \textit{i'}, and \textit{z'} from the SDSS system \citep{Alam2015}; and \textit{W1}–\textit{W4} from WISE \citep{WISE2010}. All magnitude values are given in Table~\ref{tab:system}.
The SED fitting was carried out with the Virtual Observatory Spectral Analyzer (VOSA) \citep{Bayo2008}. Synthetic SEDs were generated from several atmospheric model grids, including BT-Settl \citep{Allard2012}, Kurucz \citep{Kurucz1993}, and Castelli \& Kurucz \citep{Castelli2003}. The best-fitting model was a BT-Settl atmosphere with $T_\text{eff}=3600$ K, [M/H] = –0.5 dex, and $\log g=5.0~\mathrm{cm}~\mathrm{s}^{-2}$.
VOSA applies a $\chi^2$ minimization approach to match the theoretical SED to the observed photometry, accounting for the observed and model fluxes, their associated uncertainties, the number of photometric points, the adopted input parameters, and the object's radius and distance. The resulting fit is presented in Fig.~\ref{fig:sed}.
We integrated the observed SED to obtain the bolometric luminosity, $L_{\rm bol} = 0.04799 \pm 0.00022 L_\odot$. The stellar radius, $R_\star = 0.563 \pm 0.016 R_\odot$, was then derived using the Stefan–Boltzmann relation $L_{bol} = 4\pi R^2 \sigma T_\mathrm{eff}^4$. Finally, the stellar mass was estimated as $M_\star = 0.570 \pm 0.021 M_\odot$ using Equation 6 from \citet{Schweitzer2019}. The stellar radius and mass agree with the values obtained using the empirical M dwarf relations. 

\subsection{Stellar elemental abundances} \label{sec:abundances}

The stellar elemental abundances for TOI-672 were determined from our NIRPS spectra following the methodology of \citet{Gromek2025} and Gromek et al. (in prep), based on the methodology of \citet{Hejazi2023}, and are presented in Table~\ref{tab:stellarparams}. Here we provide a brief summary of the framework to recover elemental abundances from our NIRPS spectrum. Elemental abundances are derived from spectral synthesis methods using MARCS stellar atmosphere models \citep{Gustafsson2008} and the Turbospectrum radiative transfer code \citep{Alvarez1998,Plez2012} within modified iSpec functions \citep{Blanco-Cuaresma2014,Blanco-Cuaresma2019}. We adopt the solar reference abundances from \citet{Asplund2009}. We identified prominent lines in our 1D telluric-corrected, NIRPS template spectrum that exceeded an absorption depth of 5\% from the continuum level and did not appear to be contaminated by an imperfect telluric correction. Our final line list followed from cross-matching these lines with atomic and molecular (i.e., OH) lines in the VALD linelist \citep{Kupka2011}, and is given in Table~\ref{tab:linelist}.

The initial stellar parameters for the spectral synthesis were fixed to {$T_{\rm eff} = 3810 \pm 33$\,K}, {${\rm [M/H]} = -0.3 \pm 0.1$}, {$\log g = 4.70 \pm 0.03$}, $v_{mic} = 1 \pm 1$\,km\,s$^{-1}$, and \hbox{$v_{mac} = 1.50 \pm 0.25$\,km\,s$^{-1}$}. While the micro- and macroturbulence values are treated as nuisance parameters within our spectral synthesis, their values and range follow from a $\chi^2$-minimization of the OH lines in the NIRPS spectrum and our synthetic spectra as a function of $v_{mic}$ and $v_{mac}$, respectively. We use the OH lines as they have been shown to be especially sensitive to changes in $v_{mic}$ and $v_{mac}$ \citep{Souto2017, Hejazi2023}. Then, on a line-by-line basis, the individual elemental abundances of the synthesized spectra were varied from $[\rm X/H]\pm 0.75$\,dex in increments of 0.25\,dex while keeping all other stellar parameters constant. The synthetic spectra were then interpolated (using a linear interpolation) to a finer grid with a resolution of 0.015 dex. We determined the best-fit abundance for each line based on a $\chi^2$-minimization over each line's core region, using the aforementioned abundance grid. For certain spectral lines where the continuum levels of the model and the observed spectra do not align, we use a psuedo-continuum, meaning we apply a uniform flux offset to the observed spectra to match the continuum of the model spectra. This offset is determined by identifying points in the continuum outside each line's core region within $\pm$0.5-nm and minimizing the $\chi^2$ values between the model and the observed data for these points in the continuum. The final elemental abundances are then computed as weighted averages of the individual line abundances. The weights are calculated as the RMSE between the best-fit model and the observed spectrum for each line, divided by the line depth. To compute error terms, we add in quadrature the random error $\sigma_{ran}=\mathrm{std([X/H])}/ \sqrt{N}$ among $N$ lines of the element X to the systematic uncertainties $\sigma_{T_{\rm eff}}$, $\sigma_{\rm [M/H]}$, $\sigma_{\log g}$, $\sigma_{mac}$, and $\sigma_{mic}$. The parameters $\sigma_{T_{\rm eff}}$, $\sigma_{\rm [M/H]}$ etc. indicate the systematic errors resulting from varying $T_{\rm eff}$, [M/H] etc. by their corresponding uncertainties (here, 70\,K and 0.09\,dex, respectively). We then calculate these error terms by resampling each stellar parameter individually from a Gaussian distribution and repeating our analysis with 15 iterations \citep{Hejazi2023} to quantity the abundance dispersion. The results from our full error analysis are presented in Table~\ref{tab:Abundance_table}.

We use the results from our abundance analysis to recompute the overall metallicity and $\alpha$ enhancement of TOI-672 following the procedure of \citet{Hinkel2022}. We assume O, Mg, Ca, Si, and Ti as the alpha elements in the alpha enhancement computation, where [$\alpha$/Fe] is computed as the number fraction of all alpha elements summed together, scaled to the number fraction of iron, relative to solar values. We find that $[\rm M/H] = -0.25\pm 0.05$, which agrees well with our initial value of $[\rm M/H] = -0.3\pm 0.1$ from our preliminary analysis. 

\subsection{Stellar rotation} \label{sec:rotation}

TOI-672 does not have a stellar rotation period $P_{\rm rot}$ reported in the literature, thus here we attempt to determine it. We begin by estimating the stellar rotation period from the empirical M dwarf rotation-activity relation from \citet{Astudillo-Defru2017}, which uses the $\log R'_{HK}$ index as an activity diagnostic. The S-index is calculated by the standard offline HARPS reduction pipeline (DRS 3.5, Table~\ref{tab:harpsrvs_drs}), which we use to calculate $\log R'_{\rm HK} = -4.69^{+0.29}_{-0.34}$. This value places TOI-672 in the unsaturated regime of magnetic activity and predicts $P_{\rm rot} = 26^{+18}_{-9}$\,days. We can also use the relation between $\log R'_{HK}$ and $P_{\rm rot}$ for M dwarf stars derived by \citet{SuarezMascareno2018}, which gives $P_{\rm rot} = 19^{+35}_{-6}$\,days, consistent with the previous value, though both with very large uncertainties. 

We then inspected the spectroscopic activity indicators shown in Fig.~\ref{fig:activityindicators_ls}. We do not identify any significant rotation-like signals in the HARPS data. There are signals that appear in multiple NIRPS activity indicators (i.e., ${\rm dLW}$, $d{\rm Temp}$, and FWHM) around $\sim20-30$\,days. This value is consistent with the $P_{\rm rot}$ value predicted above from the M dwarf activity-rotation relation; however, this could also be attributed to the BERV crossing event described in Section~\ref{sec:obs_nirpsharps}, as the timescale of this event is also on order of $20-30$ days, and is expected to affect the activity indicators also. This BERV crossing event only affects the RVs in the nIR, and so the lack of signal in the HARPS indicators could also corroborate the signal being due to this.

Shifting focus to photometric measurements of TOI-672, we note that \citet{Boyle2025} reported that rotation periods determined via the Lomb-Scargle periodogram from carefully extracted TESS light curves are reliable out to $\sim10$ days. This is due to the 27-day duration of single TESS sectors for which the reliability of $P_{\rm rot}$ detections drops off severely (to below 40\,\%) for $P_{\rm rot}>10$\,days. Beyond 15\,days, it is no more reliable than a randomly assigned period, even when stitching together multiple sectors \citep{Boyle2025}. The predicted rotation periods from both \citet{Astudillo-Defru2017} and \citet{SuarezMascareno2018} are $>15$\,days, and as a consequence we do not attempt to recover a rotation period from the TESS data.

\begin{figure}
    \centering
    \includegraphics[width=\linewidth]{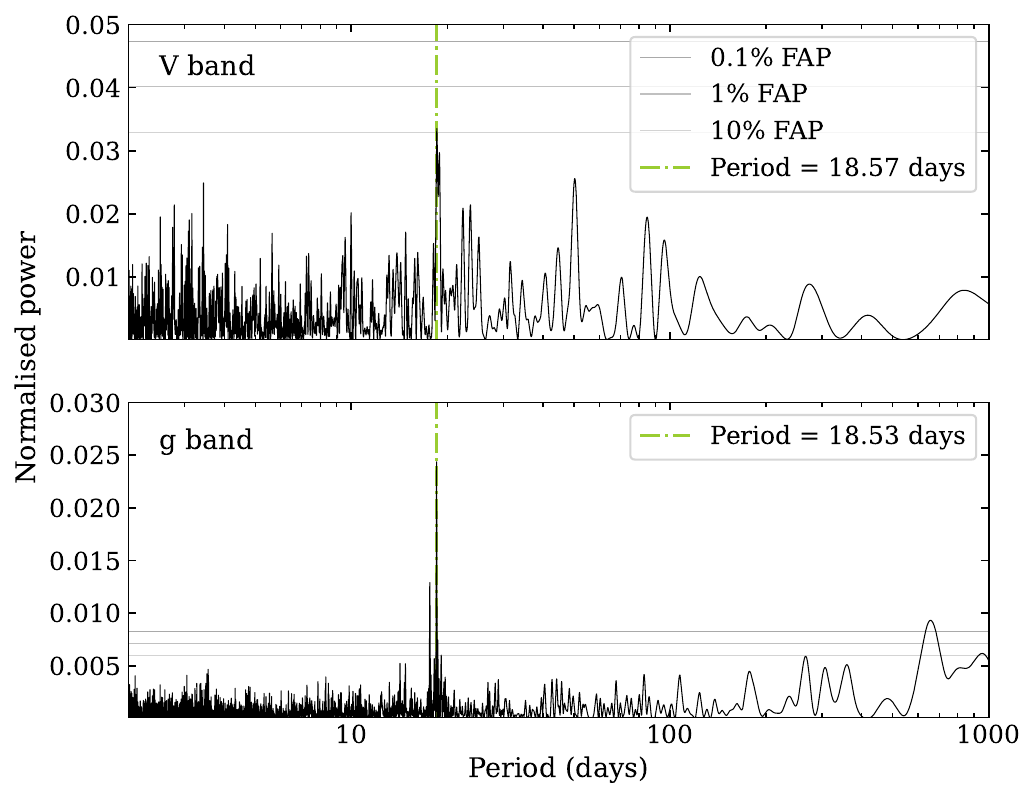}
    \caption{Lomb-Scargle periodograms of the V band (\textit{top panel}) and g' band (\textit{bottom panel}) photometry from ASAS-SN, described in Section~\ref{sec:rotation}. The period of the highest peak is highlighted by a green dash-dot line, and likely corresponds to the stellar rotation period.}
    \label{fig:asassn}
\end{figure}

Finally, we queried the  All-Sky Automated Survey for Supernovae \citep[ASAS-SN;][]{Shappee2014,Kochanek2017} data archive for long-term photometric monitoring of TOI-672. We specifically re-compute the light curves through the Sky Patrol portal going back to 2015, rather than using the pre-computed light curves. These data consist of 692 V-band observations from Nov 2015 to August 2018, and 4776 g'-band observations from Dec 2017 to August 2025 (we note there are usually multiple observations per night). We treat the light curve from each band separately, first performing a 3-$\sigma$  clip, then removing any long-term trends with a simple two-dimensional polynomial, and finally computing Lomb-Scargle periodograms which are displayed in Fig.~\ref{fig:asassn}. In the g' band, there is a very clear signal above the 0.1\% FAP at 18.53\,d. There is a corresponding signal at 18.57\,d in the V band, but less significant at just above the 10\% FAP. This is probably due to fewer data in this band. These signals are probably indicative of the stellar rotation period and match to the rotation periods predicted from the $\log R'_{\rm HK}$, so we can tentatively say we detect a 18.5\,d rotation signal. We also annotate this period on the Lomb-Scargle periodograms for the RVs and activity indicators in Fig.~\ref{fig:activityindicators_ls}. This aligns with the aforementioned signal in the activity indicators for NIRPS, but as previously stated, the BERV crossing event predicted at a similar period could also contribute. There are also very small signals in the HARPS RVs and indicators at this period, but they are not significant. 

\section{The joint transit and RV fit}\label{sec:fit}

We fit the photometry from TESS and a select light curve from ExTrA, plus the RVs from HARPS and NIRPS. As most of the other ground-based photometry from Section~\ref{sec:obs_otherphotom} were obtained during the timespan of the TESS observations, we elect to omit them from our joint fit model to avoid inflating the number of free parameters in return for little contribution towards improving the precision of the planet's ephemeris. While we do not include most ground-based photometric data here, we will ultimately use it to search for transit timing variations in Section~\ref{sec:disc_ttvs}. The exception to these omissions is the latest transit observation with ExTrA, which was taken on the night of 24 April, 2024, which occurred after the most recent TESS sector of data.

To perform the fit, we use the {\tt exoplanet} package \citep{Foreman-Mackey2020}. {\tt exoplanet} builds upon the light curve modelling package {\tt starry} to compute transit models, the probabilistic programming library {\tt PyMC} for MCMC sampling \citep{Salvatier2016}, and the fast Gaussian Process (GP) regression package {\tt celerite} \citep{Foreman-Mackey2017}. We describe the joint fit model set-up below.

\subsection{TESS photometry}\label{sec:fit_tess}

We normalise the light curves for each TESS sector by its median out-of-transit flux level and concatenate all available sectors. This light curve shows clear variability as noted in Section~\ref{sec:obs_tess}, which we detrend using a GP whose kernel is a stochastically-driven, damped harmonic oscillator with a power spectral density of

\begin{equation}
    S(\Omega) = \sqrt{\frac{2}{\pi}} \frac{S_0\Omega^4_0}{(\Omega^2 - \Omega^2_0)^2 + \Omega^2_0 \Omega^2 (\frac{1}{Q^2})},
\label{equ:sho}
\end{equation}

\noindent where $\Omega_0$ is the undamped angular frequency, $Q$ is the quality factor, and $S_0$ is proportional to the power at $\Omega=\Omega_0$ \cite{Foreman-Mackey2017}. We fit the maximum power, $S_0w^4_0$, rather than $S_0$ directly. Finally, we add a log jitter term, $\log(J_{\rm TESS})$, as an additional error term to diagonal elements of the GP covariance matrix. Priors and posteriors on these parameters are reported in Table\,~\ref{tab:priors}. 

We fit the planetary model components with a single Keplerian orbit model. The Keplerian model is parameterised by the planetary orbital period $P$, time of midtransit $t_0$, orbital eccentricity $e$, and the argument of periastron $\omega$. We discuss the eccentricity in Section~\ref{sec:fit_nirpsharps}.
These parameters are used in light curve models created with {\tt starry}, alongside the planetary radius $R_p$ and stellar parameters. We use values from the TESS SPOC pipeline \citep{Li2019} to place priors on $P$, $t_0$, and $R_p$ (see Table~\ref{tab:priors}). We use a limb-darkened transit model with the quadratic limb-darkening parameterisation from \citet{Kipping2013}. We use priors informed by the values on the stellar radius $R_\star$ and the stellar mass $M_\star$ derived in Section~\ref{sec:stellar}.

\subsection{ExTrA photometry}\label{sec:fit_extra}

The ExTrA light curves are extracted using the methodology established in \citet{Cointepas2021}. These light curves exhibit systematic trends that we detrend using a GP with a Matern-3/2 kernel of the form

\begin{equation}
    \kappa(\tau) = \sigma^2 [(1+1/\epsilon)e^{-(1-\epsilon)\sqrt{3}\tau/\rho}(1-1/\epsilon)e^{-(1+\epsilon)\sqrt{3}\tau/\rho}],
\end{equation}

\noindent where $\tau =\,\mid t_i-t_j \mid$ is the time lag between two observations taken at times $t_i$ and $t_j$, and the hyperparameters are the amplitude scale $\sigma$ and correlation timescale $\rho$. We adopt log-uniform priors on both hyperparameters (see Table~\,\ref{tab:priors}). We fit distinct limb-darkening coefficients from the TESS photometry as \hbox{ExTrA's} wavelength coverage spans a unique wavelength range (i.e., $0.8-1.55\,\mu$m with ExTrA compared to $0.6-1.0\,\mu$m with TESS). Again, we use a quadratic limb-darkening parameterisation.

\subsection{NIRPS and HARPS RVs}\label{sec:fit_nirpsharps}

We try two different methods for fitting the NIRPS and HARPS RVs. The first incorporates no detrending and is solely a single-planet Keplerian. The second incorporates an additional quasi-periodic GP kernel to detrend the NIRPS RVs, in order to remove the effect of the BERV crossing event described in Section~\ref{sec:obs_nirpsharps}. We do not include a GP on the HARPS data. We compare the evidence for the two models in Section~\ref{sec:fit_sampling}.

We use DACE\footnote{The DACE platform is available at \url{https://dace.unige.ch}} with a simple Keplerian model to estimate values for the systematic RV offsets and the semi-amplitude of the planetary signal in the RVs, $K$. The semi-amplitude of the planetary signal should be the same across the different wavelengths of NIRPS and HARPS, so we fit for one value between them. The RV offsets, however, are different, so we fit these separately (see Table~\ref{tab:priors}). We incorporate separate jitter terms for NIRPS and HARPS, which encapsulate any uncharacterised signal or noise that is perceived as white noise in the RV data. This could be, for example, instrumental effects, which will be different across the two instruments, and/or short-scale stellar activity. We use priors informed by the error on the NIRPS and HARPS data (see Table~\ref{tab:priors}), and the jitter is added in quadrature with the error on the RVs. For the first model, this is all that is incorporated.

For the second model, we are removing the effect of the BERV crossing window, which appears as a larger scatter in the RVs (Fig.~\ref{fig:rvs}) and also shows in the activity indicators (Fig.~\ref{fig:activityindicators_ls}). There are $\sim20$-day signals above the 10\,\% FAP level in the NIRPS ${\rm dLW}$, $d{\rm Temp}$, and FWHM indicator. We detrend this signal using a quasi-periodic GP.

As in \citet{Osborn2021}, we create our own quasi-periodic GP kernel using {\tt PyMC}, as no such exact kernel is available in {\tt exoplanet}. {\tt PyMC} provides many simple covariance functions which can be combined. We use their {\tt Periodic} kernel

\begin{equation}
    k(x,x') = \exp \left(-\frac{\sin^2(\pi |x-x'| \frac{1}{\theta})}{2\gamma^2} \right),
\label{equ:periodic}
\end{equation}

\noindent and {\tt ExpQuad} (squared exponential) kernel

\begin{equation}
    k(x,x') = \exp \left(-\frac{(x-x')^2}{2\lambda^2} \right),
\label{equ:expquad}
\end{equation}

\noindent and multiply them together, along with an additional hyperparameter to describe the amplitude of the GP, to create the quasi-periodic kernel\footnote{Code available on request to the corresponding author.}

\begin{equation}
    k(x,x') = \eta^2 \exp \left(-\frac{\sin^2(\pi |x-x'| \frac{1}{\theta})}{2\gamma^2} - \frac{(x-x')^2}{2\lambda^2} \right).
\label{equ:qp}
\end{equation}

The hyperparameters are: $\eta$, the amplitude of the GP; $\theta$, the recurrence timescale, which is the period of the signal; $\gamma$, the smoothing parameter; and $\lambda$, the timescale for signal growth and decay \citep[e.g.][]{Rasmussen2006,Haywood2014,Grunblatt2015}. This is commonly applied to stellar activity signals but is more broadly used to remove any quasi-periodic signal, so is applicable regardless of whether our signal is due to the BERV crossing or instead stellar activity (as noted in Section~\ref{sec:rotation}, there is a potential stellar rotation period of 18.5\,d). We note that the {\tt Periodic} term implemented by {\tt PyMC} has a slightly different scaling than is commonly used \citep[e.g. in][]{Haywood2014}, where $1/2$ is used as the exponent rather than 2. $\gamma$ may be halved to recover the standard definition. We use a uniform prior on the recurrence timescale, looking to encompass the peaks present in the stellar activity indicator periodograms for NIRPS, the expected timescale of the BERV crossing event, and the potential stellar rotation period, therefore using a 10-40\,day range. As $\lambda$ is some multiple of the recurrence we cover a range from 1 to 10 times the bounds of our recurrence timescale prior. We bound the smoothing parameter between 0 and 1. The amplitude is given a wide prior.

As it is noted that warm Neptunes tend to present zero eccentricity \citep{Correia2020}, we run initial fits of both models (with and without the GP) with a uniform prior on eccentricity and argument of periastron. The posterior distributions in both cases indicate zero eccentricity, with the 95 per cent confidence interval of $e < 0.094$ for the model without the GP, and $e < 0.099$ with the GP. We therefore elect to fix $e$ to zero in the final model presented here. This is further justified by the expectation that the planet should be tidally circularised given its short orbital period of $\sim 3.6$ days. We also note that there is no change in fit values of any parameters within error whether eccentricity is fixed to zero or not.

\subsection{Sampling the joint model posterior} \label{sec:fit_sampling}

We fit two models as described in Section~\ref{sec:fit_nirpsharps}: model 1 incorporates no detrending of the RVs while model 2 incorporates a quasi-periodic GP to detrend the NIRPS RVs. We follow the same procedure to estimate each  model's joint posterior probability density (PDF) function as outlined below.

We first use {\tt exoplanet} to maximise the log-posterior probability of the model. We use the best-fit values from this optimisation as the starting point of the {\tt PyMC} sampler. {\tt PyMC} draws samples from the posterior using the No-U-Turn Sampler, a variant of Hamiltonian Monte Carlo. We examine chains from earlier test runs of the model to inform our choice of 4 chains of 60,000 steps, with an additional 1000 steps that are discarded as burn-in. We calculate the rank-normalised split-$\hat{R}$ statistic \citep{Vehtari2021} for each parameter to test for non-convergence. For both models, $\hat{R} \approx 1.0$ for all parameters, implying both models have converged. 

To identify which model is statistically favoured, we calculate and compare the Bayesian evidences $Z$ for each model using the Perrakis estimator \citep{Perrakis2014}. We calculate an evidence ratio (i.e., $Z_{\rm model2}/Z_{\rm model1}$) of 5.88. This ratio (or Bayes factor) is not large enough to conclude that model 2 is statistically favoured by our data \citep{Nelson2020}, but because we are physically motivated to remove the BERV crossing signal in the NIRPS RVs, we choose to go with model 2. The results from model 2 are thus what we report as our final model parameters in Table~\ref{tab:planetparams} using the median values from the marginalized posteriors and the 16th and 84th percentiles as the approximate $1\sigma$ errors. We also highlight that the RV semi-amplitudes recovered in both models are consistent, where $K_{\rm model1} = 32.7^{+2.6}_{-2.7}$\,m\,s$^{-1}$ and $K_{\rm model2} = 32.0^{+2.6}_{-2.5}$\,m\,s$^{-1}$, so the choice of model should not affect the conclusions we draw about this planet.

\begin{table*}
    \centering
    \small
    \caption{Transit, orbital, and physical parameters of the planet TOI-672\,b.}
    \label{tab:planetparams}
    \begin{threeparttable}
    \begin{tabular}{lllll}
    \toprule
    \textbf{Planet parameters} & & \textbf{(Unit)} & \textbf{Value} & \textbf{Source} \\
	\midrule
    Period                              & $P$               & (days)            & 3.633581 $\pm$ 0.000001           & Joint fit \\
    Full transit duration               & $T_{\rm dur}$     & (hours)           & 1.80 $\pm$ 0.02                   & Joint fit (derived) \\
    Reference time of midtransit        & $t_0$             & (BJD-2457000)     & 1546.4796 $\pm$ 0.0003            & Joint fit \\
    Radius                              & $R_p$             & (R$_{\oplus}$)    & 5.31$^{+0.24}_{-0.26}$            & Joint fit \\
    Planet-to-star radius ratio         & $R_p/R_\star$     &                   & 0.0893$^{+0.0009}_{-0.0010}$      & Joint fit (derived) \\
    Impact parameter                    & $b$               &                   & 0.51$^{+0.05}_{-0.07}$            & Joint fit \\
    Inclination                         & $i$               & ($^{\circ}$)      & 88.0$^{+0.4}_{-0.3}$              & Joint fit \\
    Eccentricity                        & $e$               &                   & 0 (fixed)                         & Joint fit \\
    Argument of periastron              & $\omega$          & ($^{\circ}$)      & 0 (fixed)	                        & Joint fit \\
    Radial velocity semi-amplitude      & $K$               & (ms$^{-1}$)       & 32.0$^{+2.6}_{-2.5}$              & Joint fit \\
    Mass                                & $M_p$             & (M$_{\oplus}$)    & 50.9$^{+4.5}_{-4.4}$              & Joint fit (derived) \\
    Bulk density                        & $\rho$            & (g\,cm$^{-3}$)    & 1.86$^{+0.34}_{-0.26}$            & Joint fit (derived) \\
    Semi-major axis                     & $a$               & (AU)              & 0.0376 $\pm$ 0.0007               & Joint fit (derived) \\
    System scale                        & $a/R_{\star}$     &                   & 14.8$^{+0.6}_{-0.5}$              & Joint fit (derived) \\
    Equilibrium temperature$^*$         & $T_{\rm eq}$      & (K)               & 699.5$^{+14.5}_{-15.5}$           & Joint fit (derived) \\
    Bolometric instellation             & $S_{\rm bol}$     & ($S_{\rm bol, \oplus}$) & 39.8$\pm 3.4$               & Joint fit (derived) \\
    \bottomrule
    \end{tabular}
    \begin{tablenotes}
    \item $^*$Equilibrium temperature is calculated assuming full heat redistribution and an albedo of 0. Further parameters from the joint fit model can be found in Table~\ref{tab:priors}.
    \end{tablenotes}
    \end{threeparttable}
\end{table*}


\begin{figure*}
    \centering
    \includegraphics[width=0.95\linewidth]{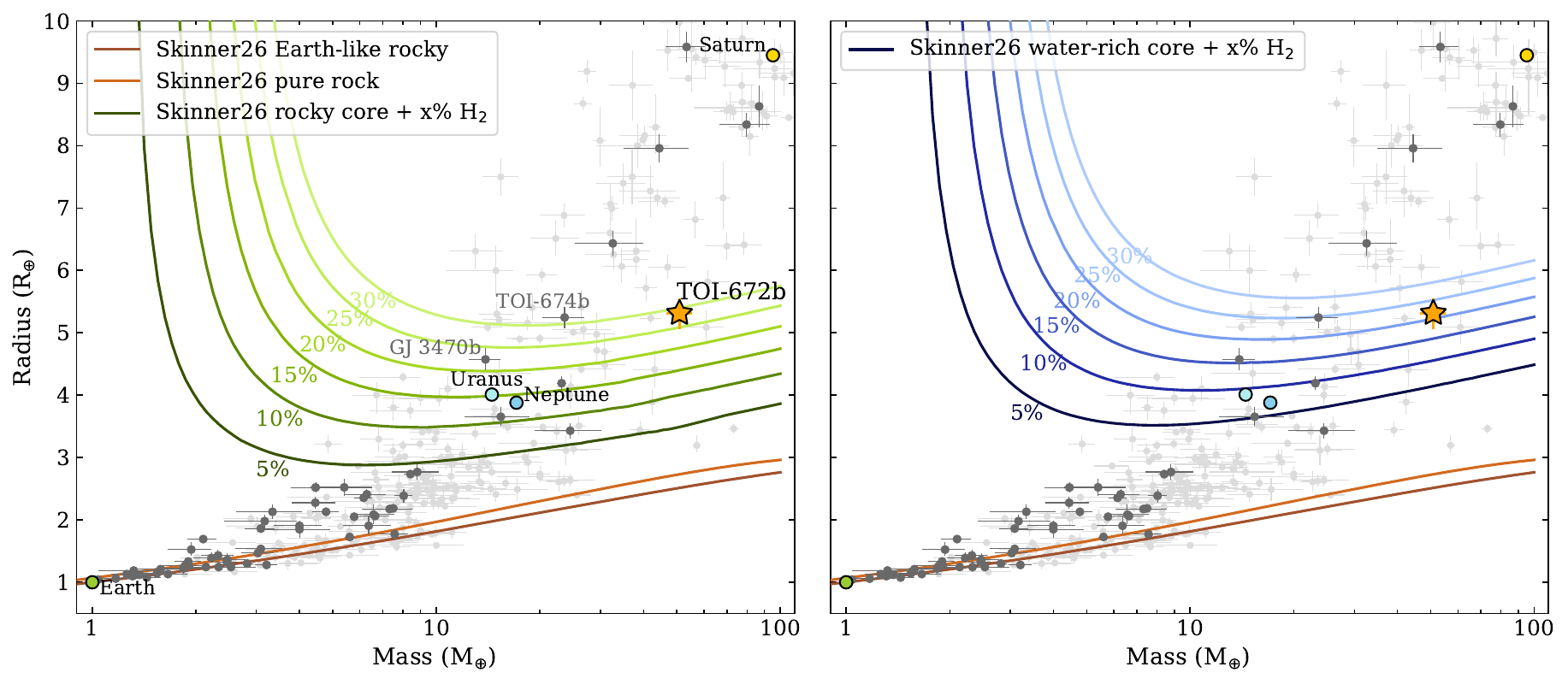}
    \caption{Mass-radius diagrams showing TOI-672\,b (yellow star) against the well-characterised planet population (light grey dots) from the PlanetS catalogue, available at \url{https://dace.unige.ch/exoplanets}. It contains planets with precisions on planet mass and radius of $< 25$\% and 8\%, respectively \citep{Otegi2020,Parc2024}. Planets around M dwarf hosts (i.e., $T_{\rm eff} < 3900$\,K and $M_{*}, R_{*} < 0.6$\,$M_{\odot}, R_{\odot}$) are highlighted as dark grey dots. Solar system planets are shown for comparison (outlined circles, labelled). Both panels show mass-radius relations from \citet{Skinner2026} for Earth-like rocky (32.5\% core + 67.5\% mantle) compositions (dark brown) and pure rock (100\% mantle) compositions (light brown). Both panels also show model mass-radius relations from \citet{Skinner2026} (solid lines), as described in Section~\ref{sec:disc_comp} for an equilibrium temperature of 700\,K, with annotated percentages reporting the hydrogen mass fraction of each curve. The mass-radius relations in the \textit{left panel} correspond to an Earth-like core composition (i.e., 32.5\% iron + 67.5\% mantle) while the \textit{right panel} assumes a 50\% water + 50\% Earth-like core composition.}
    \label{fig:mr}
\end{figure*}

\section{Results and discussion}\label{sec:disc}

\subsection{Planetary parameters}\label{sec:disc_params}

We measure the mass and radius of TOI-672\,b to be 50.9$^{+4.5}_{-4.4}$\,$M_{\oplus}$ and 5.31$^{+0.24}_{-0.26}$\,$R_{\oplus}$, respectively. TOI-672\,b is a massive super-Neptune orbiting an M0V star every 3.63\,days. Our measured radius and period agree with the previously measured values reported in the TOI release and validation paper \citep{Mistry2023}. However, the planet is much more massive than predicted by the \citet{Chen2017} mass-radius relations presented in \citet{Mistry2023} (i.e., $24.2 \pm 10.7$\,$M_{\oplus}$). Nonetheless, the planet's low density of 1.86$^{+0.34}_{-0.26}$\,g\,cm$^{-3}$ suggests that it has a significant H$_2$/He envelope, which we explore in Section~\ref{sec:disc_comp}. We compare the mass and radius of TOI-672\,b to the known, well-characterised planet population in Fig.\,\ref{fig:mr}. There are no planets around M dwarfs with similar masses and radii; TOI-1728\,b is the closest in radius at $5.05^{+0.16}_{-0.17}$\,$R_{\oplus}$ with a similar period of $3.49$\,d, but only half the mass, $26.78^{+5.43}_{-5.13}$\,$M_{\oplus}$ \citep{Kanodia2020}; it has a low precision on its mass measurement and so does not feature in Fig.~\ref{fig:mr}. It does, however, probably lie on a similar composition track to TOI-672\,b (see Section~\ref{sec:disc_comp}), as does TOI-674 \citep[$P = 1.98$\,d,][]{Murgas2021} and GJ\,3470\,b \citep[\textcolor{red}{$P = 3.34$\,d,}][]{Awiphan2016}. Kepler-101\,b has very similar properties ($3.63$\,d; $51.1^{+5.1}_{-4.7}$\,$M_{\oplus}$; 5.77$^{+0.85}_{-0.79}$\,$R_{\oplus}$), but it is orbiting a G3 star \citep{Bonomo2014} and so has a higher equilibrium temperature.

TOI-672\,b is moderately irradiated by its host star with an instellation of $39.8\times$ Earth's instellation and an equilibrium temperature of 699.5\,K (assuming an albedo of zero and full heat redistribution). It has a Transmission Spectroscopy Metric (TSM) and Emission Spectroscopy Metric (ESM) of $69^{+11}_{-10}$ and $26 \pm 3$, respectively \citep{Kempton2018}. With its Neptunian size and short orbital period, TOI-672\,b falls within the Neptune desert as defined by \citet{Mazeh2016}. That study, however, focuses on FGK planet hosts, and we revisit the desert boundaries for FGK versus M dwarf hosts in Section~\ref{sec:disc_desert}. 

\subsection{Planetary composition}\label{sec:disc_comp} 

\begin{figure*}
    \centering
    \includegraphics[width=0.95\linewidth]{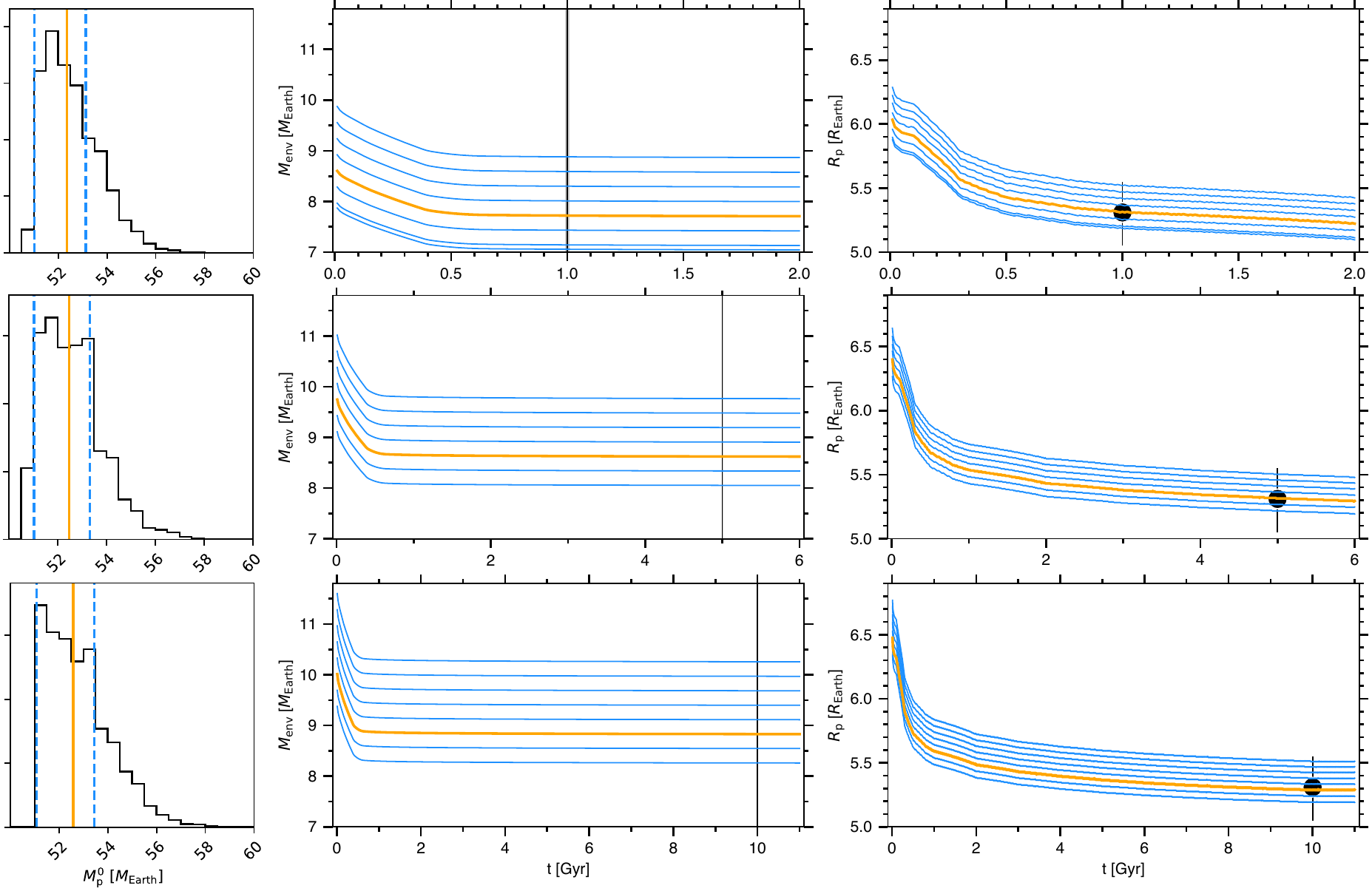}
    \caption{\textsc{jade} exploration of TOI-672~b atmospheric evolution, assuming ages of 1, 5, and 10 Gyr \textit{(top, middle, and bottom rows, respectively)}. \textit{Left column:} PDF of the initial planet mass, with the median highlighted as an orange line and 1$\sigma$ HDI interval as dashed blue lines. \textit{Middle column}: Temporal evolution of the planet envelope mass, in the best-fit simulation (orange curve) and a set of simulations from within the HDI interval (blue curves). A vertical black line with shaded band marks the assumed age and uncertainties. \textit{Right column}: Same as the middle panel, but for the planet radius. The black point indicates the measured value.}
    \label{fig:JADE_evol}
\end{figure*}

We add TOI-672\,b to the mass-radius diagram in Fig.~\ref{fig:mr}. We compare the known exoplanet population to a suite of compositional curves, including novel mass-radius relations for irradiated, H$_2$/He-dominated planets \citep{Skinner2026}. We use these novel compositional mass-radius relations because the somewhat unique mass and radius of TOI-672\,b place it in a region of the parameter space for which we are unaware of any existing mass-radius relations in the recent literature. 

Here we provide a brief summary of our interior structure models, which are detailed by \citet{Skinner2026}. Planets are assumed to be differentiated bodies in hydrostatic equilibrium with convective interiors and atmospheres following the radiative transfer-informed temperature profile of \citet{ParmentierGuillot2014,ParmentierGuillot2015}. Planets are composed of an Fe-rich core with mixed-in volatile O and S, a mantle with full mineralogy composed of FeO, SiO$_2$, MgO, and combinations thereof (e.g. MgSiO$_3$), a multi-phase layer of pure water that uses the \texttt{AQUA} models to calculate the non-isothermal H$_2$O equation of state \citep{Haldemann2020}, and a non-ideal H$_2$/He envelope with an opacity arising from a $50\times$ solar metallicity equilibrium mixture of metals. This metallicity is in line with the metallicity expected for a planet of this mass following mass-metallicity relationships and similar to Uranus and Neptunes' atmospheric metallicities \citep{SwainHasegawa2024}. As this model is for a static interior, it cannot directly capture the crucial physics of envelope contraction. This is accounted for by including internal luminosity in the outer temperature boundary condition following the models of \citet{Mordasini2020}. Without a constrained stellar age, we assume a planetary age of 4.5 Gyr, analogous to the solar system and the near average of the potential system ages considered in section \ref{sec:atmospheric_evolution}. Given an input planetary mass, instellation, age, and core, mantle, water, and envelope mass fractions, we solve the standard differential equations of planetary structure to calculate the planet's transit radius by integrating the atmospheric optical depth along a grazing chord as observed through a transit geometry \citep{Guillot2010}. Since the refractory abundance ratios Fe/Mg and Mg/Si of TOI-672 are consistent with solar (c.f. Table~\ref{tab:stellarparams}), we consider two fiducial core compositions in the mass-radius relations depicted in Fig.~\ref{fig:mr}. The first is an Earth-like composition (i.e., 32.5\% core + 67.5\% mantle, mantle molar Mg/Si=1.205 and Fe/Mg=0.123; left panel) and the other a water-rich core (i.e., 50\% water + 50\% Earth-like; right panel), as is expected for a water-rich core formed from solar metallicity material beyond the snowline \citep{Lodders2003}.

Our models imply that TOI-672\,b is consistent with having a H$_2$/He envelope mass fraction of 20-30\%, depending on the assumed water mass fraction of the core. In contrast, the mass-radius curves of \citet{Fortney2007}, which are computed using an evolutionary planetary interior model rather than a static one as considered here, imply a lower H$_2$/He envelope mass fraction of $\sim$11\% (see their fig. 7). \citet{Fortney2007} do not account for high-pressure phase transitions, resulting in a systematic underprediction of planetary densities leading to lower inferred H$_2$/He abundances. We note that the stellar refractory abundances are insufficient to constrain the core compositions because TOI-672\,b may have a core that is volatile-rich, which is largely set by its unknown formation location and evolutionary history. Additionally, the unconstrained age of TOI-672 means that it is impossible to determine how much envelope contraction has occurred, introducing a degeneracy between the inferred planetary radius and its age.

\subsection{Atmospheric evolution}\label{sec:atmospheric_evolution}

We explored the scenario in which TOI-672\,b migrated early-on within the protoplanetary disk and started evolving on its present-day orbit, performing atmospheric evolutionary simulations of TOI-672\,b with the \textsc{jade} code \citep{Attia2021,Attia2025}\footnote{We used version 1.0.0 of the code, publicly available at \url{https://gitlab.unige.ch/spice_dune/jade}.} according to the procedure detailed in \citet{Bourrier2025}. As the age of TOI-672 remains largely unconstrained, we performed simulations for a range of field ages that includes 1, 5, and 10\,Gyr.

First, rotating stellar models representative of TOI-672 were computed with the Geneva stellar evolution code \citep[GENEC,][]{Eggenberger2008} using stellar radius and effective temperature constraints given in Table~\ref{tab:stellarparams}. These models account for the internal transport of angular momentum by hydrodynamic and magnetic instabilities \citep[see e.g.,][]{Eggenberger2022}. Based on the structural and rotational evolution of these models, the evolution of the high energy fluxes emitted by the host star is then deduced \citep[see][]{Pezzotti2021}. The initial velocity of the star being unknown, we consider the case of a moderate rotator with an initial rotation rate of 5 $\Omega_{\sun}$ \citep[see][]{Eggenberger2019}. This rotational history then results in a given evolution of the X-ray luminosity of the host star and to a corresponding extreme ultraviolet (EUV) luminosity computed with the relation between the X-ray and EUV luminosity provided by \citet{Sanz-Forcada2011} \citep[see e.g.][]{Pezzotti2021}. The stellar luminosity was set to the GENEC evolutionary curve at the three assumed ages.

We then use the \textsc{jade} internal structure solution for TOI-672\,b, which we condition on our findings from Section~\ref{sec:disc_comp}. Namely, the mass and orbital properties of the planet were set to their nominal present-day values and the envelope mass fraction was fixed to 0.2, which is consistent with our findings in Section~\ref{sec:disc_comp}. Uniform priors ($\mathcal{U}$(0,1)) were set on the atmosphere ($f_\mathrm{Env/Pl}$) and mantle ($f_\mathrm{Ma/Pl}$) mass fractions relative to the total planet mass, and a narrower prior ($\mathcal{U}$(0,0.1)) on the trace atmospheric metallicity, although this value is not constrained by our data. The retrieval is constrained by the measured planet radius. We find that the derived $f_\mathrm{Env/Pl}$ increases with the assumed system age, from 15.1$\stackrel{+3.1}{_{-3.4}}$\% for 1\,Gyr to 17.9$\stackrel{+3.4}{_{-3.3}}$\% for 10\,Gyr, corresponding to absolute median solid and envelope masses of 43.2 and 7.7\,M$_\mathrm{\oplus}$ for 1\,Gyr and 41.8 and 9.1\,M$_\mathrm{\oplus}$ for 10\,Gyr. This is likely because the atmosphere cools and shrinks as stellar irradiation decreases over time, so that a larger atmospheric mass fraction and lower bulk planetary density are required to yield the same radius at a more advanced age. 

\begin{figure}[ht!]
    \centering
    \includegraphics[width=0.95\columnwidth]{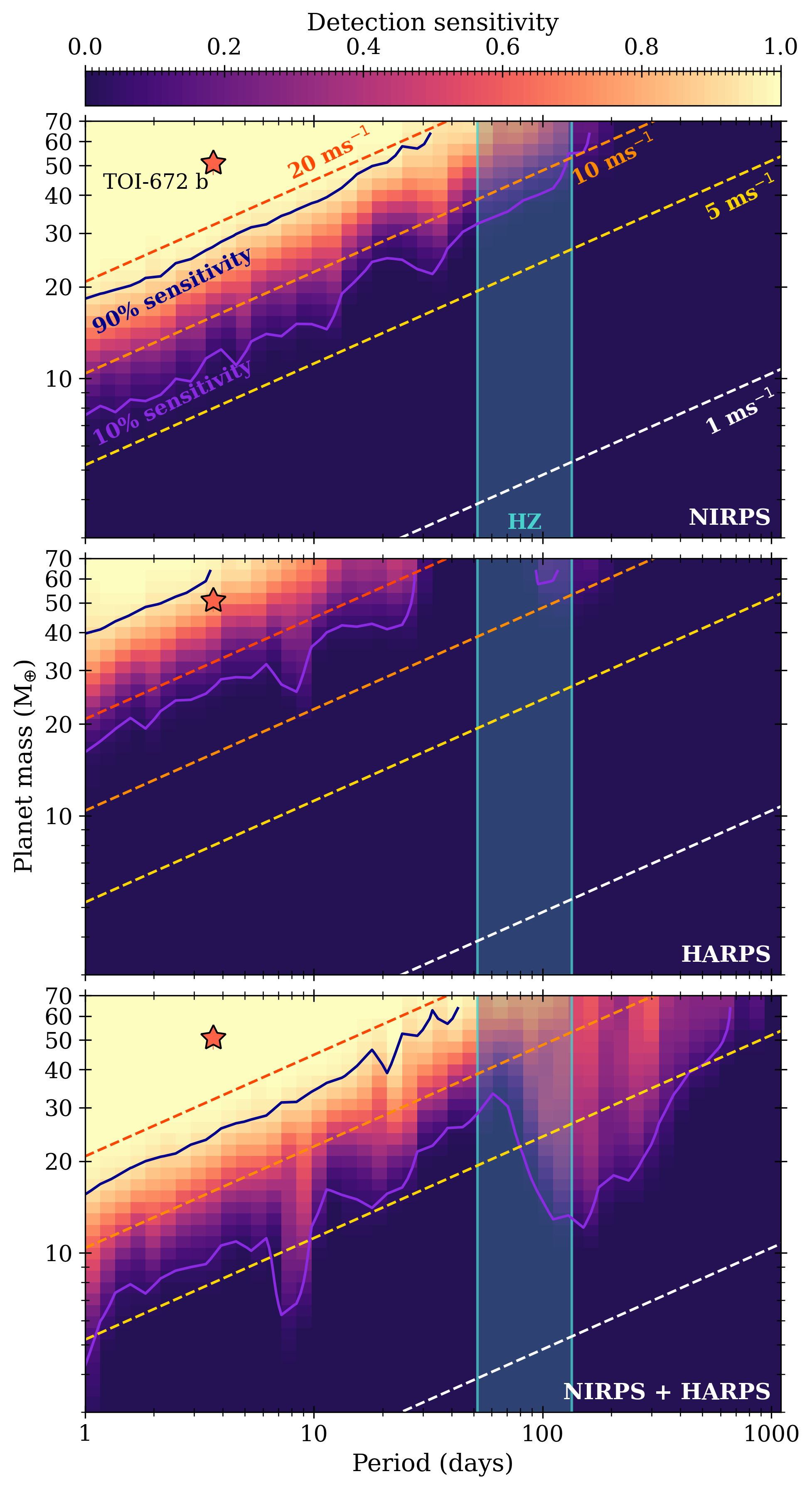}
    \caption{Sensitivity of the RV data, specifically the NIRPS data \textit{(top panel)}, HARPS data \textit{(middle panel)}, and combined NIRPS+HARPS data \textit{(bottom panel)}, to additional planets orbiting TOI-672 as a function of planet mass and orbital period (using $50\times50$ bins in this parameter space). The calculation is described in Section~\ref{sec:disc_sensitivity}. The sensitivity is shaded as in the colourbar, from 0 (dark purple) to 1 (light yellow). The 90$\%$ and 10$\%$ sensitivity limits are denoted by the dark purple and medium purple solid lines, respectively, and labelled in the top panel. RV semi-amplitudes of 20, 10, 5, and 1\,m\,s$^{-1}$ are denoted by the red, orange, yellow, and white dashed lines, respectively, and labelled in the top panel. TOI-672\,b is shown by the star marker (where error bars are too small to be seen). The habitable zone defined by \citet{Kopparapu2013}, specifically the moist greenhouse inner edge and the maximum greenhouse outer edge calculated for the specific host star TOI-672, is shaded in turquoise.}
    \label{fig:sensitivity}
\end{figure}

We then simulated the atmospheric evolution of TOI-672~b with \textsc{jade} over a grid of initial planet mass, running three sets of simulations from the expected dissipation of the disk (10\,Myr) up to the assumed present-day ages. System properties were fixed to their present-day values from Table~\ref{tab:planetparams}, and to the results of the internal structure retrieval for each age. The stellar luminosity curve was set to the model computed with \textsc{GENEC}. The evolutionary retrieval was performed as described in \citet{Bourrier2025}, constrained by the present-day planet radius and assumed system age. Results are shown in Fig.~\ref{fig:JADE_evol}. Most atmospheric erosion would have occurred within the first 300\,Myr, when the star was most active, although the planet may only have lost about 20\% of its primordial envelope mass during this phase. Indeed, atmospheric escape is made inefficient by the strong gravity of TOI-672~b. Even for an assumed system age of 10\,Gyr, the planet would subsequently have lost $<10$\% of its remaining envelope mass. Initial masses derived for TOI-672\,b are consistent within 1$\sigma$ across the three assumed ages, ranging from 52.4$\stackrel{+0.8}{_{-1.3}}$\,M$_{\oplus}$ to 52.6$\stackrel{+0.8}{_{-1.5}}$\,M$_{\oplus}$. The corresponding initial planet radius would be on the order of 6.1--6.5\,R$_{\oplus}$, so that the nature of TOI-672~b would not have changed substantially across its evolution.

\begin{figure*}[ht!]
    \centering
    \includegraphics[width=\linewidth]{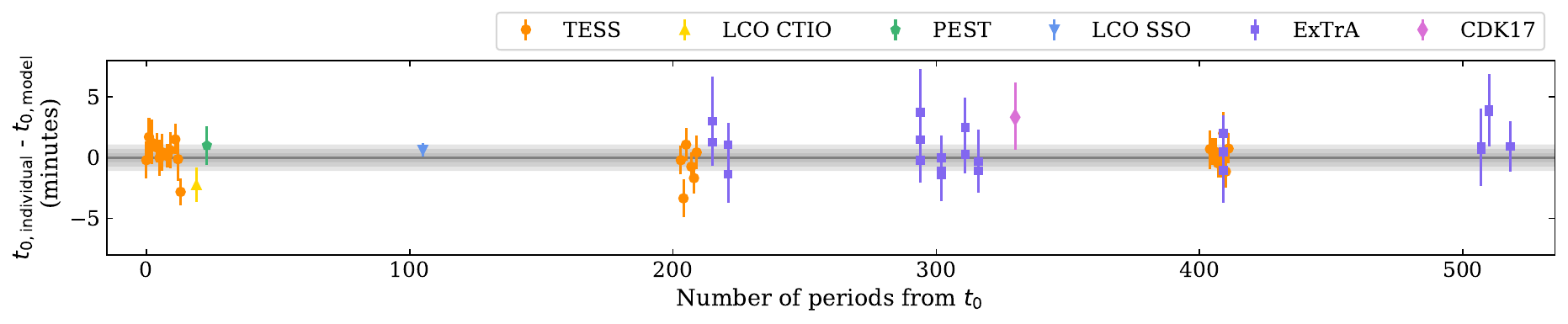}
    \caption{Searching for TTVs in the transits of TOI-672\,b. Mid-transit times from fits to each individual transit (described in Section~\ref{sec:disc_ttvs}) are compared to the linear ephemeris from the joint fit model (i.e., the dark grey line at $y=0$ in this plot, where the 1, 2 and 3 standard deviations from the value are shaded). The TESS transits are orange circles; the LCO CTIO transit is the yellow triangle; the PEST transit is the green pentagon; the LCO SSO transit is the blue upside down triangle; the ExTrA transits are purple squares; and the Deep Sky Chile CDK17 transit is the orchid diamond.}
    \label{fig:ttvs}
\end{figure*}

\subsection{RV sensitivity to additional planets}\label{sec:disc_sensitivity}

Multi-planet systems are common around M dwarfs \citep{Bonfils2013,Tuomi2014,Dressing2015,Gaidos2016,Cloutier2021b,Mignon2025}. In particular, \citet{Dressing2015} and \citet{Gaidos2016} found that early M dwarfs host an average of 2.5 small, short period planets per star while $90^{+5}_{-21}$\% of mid-to-late M dwarfs host multiple small planets \citep{Cloutier2021b}. It is therefore reasonable to expect additional planets around \hbox{TOI-672} that have insofar remained undetected by TESS and our RV analysis due to either a long orbital period, low inclination, or small size/mass. Here we quantify the sensitivity of our RV dataset to additional planets in the TOI-672 system.

We calculate the detection sensitivity of our RV time series as a function of orbital period and planet mass via a set of injection-recovery tests. Our procedure follows the methods of \citet{Cloutier2021} and \citet{Cherubim2023}. We inject synthetic Keplerian signals into the residuals of our RV time series. 
We produce $10^5$ realisations of Keplerian signals produced by a single planet with planet masses and orbital periods sampled uniformly in log space from 1-70\,M$_{\oplus}$ and 1-1000\,d, respectively. Orbital phases are sampled uniformly from 0 to 2$\pi$; we sample orbital inclinations as a Gaussian distribution centred on the median value of inclination from our joint fit (i.e., 88.0$\degree$) with a dispersion in mutual inclinations of 2$\degree$ following \citet{Ballard2016}. We note that this results in mostly coplanar planets being injected. We sample the stellar mass from its posterior and calculate the RV semi-amplitude due to the injected planet assuming a circular orbit. By injecting the Keplerian signal into the RV residuals, we preserve any residual noise that was not perfectly detrended and we maintain the individual measurement uncertainties and time stamps.

We then attempt to recover the injected planets as signals that satisfy two criteria. First, a recovered signal must appear in the Generalised Lomb-Scargle (GLS) periodogram \citep{Zechmeister2009} with a FAP $\leq 1$\% and whose period is within 5\,\% of the injected period. The GLS periodogram is created for each realisation and the FAP is calculated analytically \citep{Zechmeister2009}. Secondly, the Keplerian model must be statistically favoured over a flat line. We base the model comparison on the difference in Bayesian Information Criterion (BIC) between the two models; ${\rm BIC} = 2 \ln \mathcal{L} + \nu \ln N$, where $\mathcal{L}$ is the Gaussian likelihood of the RV data given the assumed model, $\nu$ is the number of model parameters (one or six for the null or Keplerian models, respectively), and $N$ is the number of RV measurements. Our second detection criterion is therefore $\Delta {\rm BIC} = {\rm BIC}_{Keplerian} - {\rm BIC}_{null} \geq 10$ \citep{Cloutier2021b,Cherubim2023}. We define the sensitivity of our RV data as the ratio of the number of recovered planets to the number of injected planets per bin, which is shown in Fig.~\ref{fig:sensitivity}.

Fig.~\ref{fig:sensitivity} highlights the different levels of sensitivity between our HARPS and NIRPS data. We note that there are fewer HARPS data than NIRPS, and only one HARPS point after the time span in which the star was unobservable (Fig.~\ref{fig:rvs}). We also note that the apparent improved sensitivity at approximately 9 and 200 days is likely due to aliasing in the periodogram of the RVs, i.e. it is artificial and a region of high false detection probability rather than enhanced sensitivity.

Considering orbital periods out to 200\,days, with our NIRPS data we expect to detect approximately 40\,\% of signals with a semi-amplitude of 10\,m\,s$^{-1}$. With the HARPS data, there is no sensitivity to signals of this semi-amplitude; we only start to become sensitive to semi-amplitudes above 20 \,m\,s$^{-1}$. Combining the two datasets slightly improves the sensitivity. It is also worth noting here that, despite the NIRPS sensitivity being better, using HARPS+NIRPS gives double the amount of RV measurements for the same time investment, as they operate simultaneously. 

We can reliably detect planets the mass of TOI-672\,b out to around 100\,d orbital periods. With the RV data in hand, we are unable to recover small terrestrial planets with $M_{p} \leq 5 M_{\oplus}$, even at very short orbital periods of 1 day. Our data are also insufficient to probe super-Earth mass planets ($M_p\lesssim 20\, M_\oplus$) within the habitable zone bounded by the moist greenhouse inner edge and the maximum greenhouse outer edge (shaded in Fig.~\ref{fig:sensitivity}, \citet{Kopparapu2013}).

\subsection{Searching for Transit Timing Variations}\label{sec:disc_ttvs}

Here we search for Transit Timing Variations (TTVs) by fitting the individual transits from TESS, ExTrA, and the other ground-based facilities. The baseline of these data is 518 days. We exclude several transits within these datasets due to excessive noise or a lack of clear ingress or egress. These excluded observations include two nights of ExTrA observations from all telescopes on the nights of 16 Feb 2023 and 8 Apr 2023, and both Evans 0.36\,m telescope observations. In total, we use 51 transits, listed in Table~\ref{tab:ttvs}.  We follow the method outlined in Sections\,\ref{sec:fit_tess} and \ref{sec:fit_extra} to fit transit models to each individual transit event. We adopt Gaussian priors on the stellar radius and mass, planetary period, and radius based on the results of our full joint fit model (Table\,\ref{tab:priors}). The mid-transit time is left free to vary with a uniform prior whose width spans the full duration of the transit observation. The TESS light curves are detrended with the SHO GP as described in Section\,\ref{sec:fit_tess}; the ExTrA light curves are detrended with the Matern 3/2 GP as described in Section\,\ref{sec:fit_extra}; the remaining ground-based light curves require no detrending further to what was presented in \citet{Mistry2023}. The resultant mid-transit times for each transit are provided in Table~\ref{tab:ttvs} and shown in Fig.\,\ref{fig:ttvs}. 

We find no evidence for TTVs in the transits of TOI-672\,b. The root mean square (rms) deviation across all transits is 1.7\,minutes. The TESS transits show the smallest deviation with a maximum of 2\,minutes and rms of 1.3\,minutes. The typical uncertainty on $t_0$ is 1.4\,minutes for TESS, 2.3\,minutes for ExTrA, and 1.6\,minutes for the other facilities. The $\chi^2$ statistic comparing the individual mid-transit times to the linear ephemeris is 0.76, indicating that the transit times are consistent with no detectable TTVs, and that the transit time uncertainties might be slightly overestimated. We perform least-squares minimisations to fit a straight-line model and then a quadratic model to the transit times, and compare these models to the linear ephemeris model. Neither of these more complex models are preferred as their $\Delta$ BIC values are $> 3$ and 5, respectively. We also examine a Lomb-Scargle periodogram of the mid-transit times and uncover no significant periodicities.

\section{M dwarf Neptunian desert}\label{sec:disc_desert} 

We place TOI-672\,b in the period-radius diagram in Fig~\ref{fig:desert} and overplot the Neptunian desert boundaries from \citet{Mazeh2016} and \citet{CastroGonzalez2024} (hereafter \citetalias{CastroGonzalez2024}). TOI-672\,b appears to lie in the newly-coined Neptunian ridge (\citetalias{CastroGonzalez2024}). However, planets orbiting M dwarfs are highly under-represented in these boundary studies due to observational biases.  The desert boundaries may not uniformly apply to all systems due to differences in the environments of planets around M dwarfs compared to earlier type stars. Indeed, \citet{Hallatt2022} predict that the boundaries of Neptunian desert in radius-period space should be altered around different types of host stars due to the planets' differing radiation environments and evolutionary histories. Here we explore for the first time the consistency of the Neptune desert boundaries between M versus FGK dwarfs in period-radius space, and also extend our analysis to instellation-radius space.

We build off the methodology presented in \citetalias{CastroGonzalez2024}. However, the number of M dwarf systems in the Kepler DR25 catalog is limited, which inhibits a meaningful statistical comparison between the M and FGK system properties. We instead use the catalogue of \citet{Berger2023} (hereafter \citetalias{Berger2023}), which presents homogeneous, precise stellar parameters for Kepler, K2 and TESS host stars, and consequent planetary parameters based on Gaia DR3 astrometry. Homogeneous stellar properties are important as the significant systematics between the different methodologies for determining these propagate into the determination of the physical parameters of the planets like radii and masses. Using a consistent method for determining parameters means comparing like-for-like. 

We apply several cuts to this catalogue to produce our final sample. We begin by restricting our sample to planets discovered by Kepler and TESS. We exclude K2 discoveries as the recovery fraction of planets from the K2 light curves is significantly lower than Kepler \citep{Zink2021}. We only include ``confirmed'' and ``known'' planets (dispositions of ``CP'' and ``KP'', respectively). We exclude planet candidates as the high incidence of false positives within the Neptunian desert \citep[see e.g.,][]{Sanchis-Ojeda2014,Lillo-Box2024,Doyle2025} would contaminate our sample if included. We also identify and remove post-main sequence stars that do not satisfy the temperature-dependent stellar radius filter given by Eq. 1 in \citet{Fulton2017}:

\begin{equation}
    \frac{R_{\star}}{R_\odot} > 10^{0.00025(T_{\rm eff}-5500)+0.20},
\end{equation}

\noindent where the stellar effective temperature $T_{\rm eff}$ is given in kelvin. We also eliminate stars earlier than F0V and later than M9.5V by restricting to stars with $ 2300 < T_{\rm eff}$\,(K)$ < 7300$\,K.

The \citetalias{Berger2023} catalogue uses $R_p/R_{\star}$ in combination with their computed stellar radii to recalculate the planet radii. For the Kepler planets, $R_p/R_{\star}$ comes from the Cumulative KOI table at the NASA Exoplanet Archive, which provides accurate planetary parameters. As such, we then use the planet radius values $R_p$ from \citetalias{Berger2023}. By contrast, the parameters for the TESS planets come from the TOI catalogue \citep{Guerrero2021}. These are the values provided when a TESS planet is released as a candidate and they are not updated to parameters from later refereed publications on the systems. The latter are generally considered more reliable, but are available for known and confirmed planets only. This is no issue for this study, as we are restricting our sample to only known and confirmed planets. Therefore, we perform a crossmatch with the NASA Exoplanet Archive\footnote{Accessed 24 Aug 2025.} to obtain published $R_p/R_{\star}$ values for our TESS planets, and then use $R_{\star}$ from \citetalias{Berger2023} to recalculate $R_p$. The \citetalias{Berger2023} catalogue also provides stellar luminosity values, which we can combine with their semi-major axis values to compute instellations, $S$, for all of our planets.

We account for the biases in our sample from geometric transit probability and detection sensitivity using the method of \citetalias{CastroGonzalez2024}. In short, we use an Inverse Detection Efficiency Method where the probability of each planet in our sample being detected around its host star is calculated. Each planet's detection probability is used to weight the planet in our sample according to

\begin{equation} \label{equ:weights}
    w = \frac{1}{P_{\rm transit}} \times \frac{1}{P_{\rm detection}}.
\end{equation}

\noindent $P_{\rm transit}$ is the geometric transit probability

\begin{equation}
    P_{\rm transit} = \frac{R_{\star} + R_{\rm P}}{a} \approx \frac{R_{\star}}{a},
\end{equation}

\noindent where $a$ is the planet's semi-major axis taken from \citetalias{Berger2023}. $P_{\rm detection}$ is the signal recoverability, i.e., the probability of detecting a planet given the S/N of its combined transit signal. For the Kepler pipeline, signal recoverability is well described by a gamma cumulative distribution of the form

\begin{equation}\label{equ:kepler}
    F(x | a,b,c) = \frac{c}{b^a \Gamma(a)} \int_0^x t^{a-1} e^{-t/b} dt
\end{equation}

\noindent where $x$ is the transit S/N and the best fit coefficients for the DR25 catalogue are $a = 33.54$, $b = 0.2478$, and $c = 0.9731$ \citep{Christiansen2020}, which we adopt here. Unfortunately, no similar calculation is publicly available yet for the TESS Transit Search Pipeline, and performing this analysis is beyond the scope of this work. As an ad hoc solution, we follow  \citet{Rodel2024} and apply the Kepler distribution in Eq.~\ref{equ:kepler} to our TESS targets. Of course, there are many differences between the TESS and Kepler missions (e.g., aperture size, precision, observation baseline, brightness of target stars), and their respective pipelines (with our confirmed planets from TESS being discovered by multiple pipelines). This work is therefore more of an illustrative first look at these Neptunian desert boundaries, and can be improved upon in the future.

The S/N of transit signals in the Kepler pipeline can be calculated as 

\begin{equation}\label{equ:keplersn}
    \rm S/N = \frac{\delta}{\sigma_{\rm CDPP}} \sqrt{N},
\end{equation}

\noindent where $\delta$ is the transit depth, $\sigma_{\rm CDPP}$ is the Root Mean Square (RMS) Combined Differential Photometric Precision (CDPP) computed at the transit duration \citep{Christiansen2012}, and $N$ is the number of transits observed. For the Kepler planets in our sample, we obtain $N$ and $\sigma_{\rm CDPP}$ values for each target from Kepler DR25, as well as the impact parameter. We use the impact parameter alongside the stellar radii, planetary radii, and orbital periods from \citetalias{Berger2023} to reconstruct the transit duration. We then use the value of the $\sigma_{\rm CDPP}$ calculated for the closest matching transit duration (the available transit durations range from 1.5 to 15 hours) for each target in our sample. 

The S/N of transit signals in the TESS pipeline can also be calculated using Eq.~\ref{equ:keplersn}, but this only holds for one sector of observations because the CDPP values for the same star can vary across sectors \citep{Twicken2025}. The S/N for multiple sectors is calculated as

\begin{equation}\label{equ:tesssn}
    {\rm S/N} = \frac{\delta}{\sqrt{\sum_i N_i \sigma_{{\rm CDPP},i}^2 }} N,
\end{equation}

\noindent such that $N_i$ transits are observed in sector $i$ \citep{Twicken2025}\footnote{RMS CDPP values for all TESS 2\,min cadence targets with SPOC light curves are delivered to MAST, and are available at  \href{https://archive.stsci.edu/doi/resolve/resolve.html?doi=10.17909/xx44-3n34}{doi:10.17909/xx44-3n34}}. Here, we use transit duration and depth from the TOI catalog and again use the value of the $\sigma_{\rm CDPP}$ calculated for the closest matching transit duration (in this case, the available transit durations range from 0.5 to 15 hours) for each target in our sample. Finally, we remove any planets with S/N $< 7.1$, which is the Kepler pipeline detection threshold.

To split the sample into FGK planet hosts versus M dwarf planet hosts, we define M dwarfs as $T_{\rm eff} < 3900$\,K. In summary, our sample is drawn from Kepler and TESS known and confirmed planets, and contains 146 planets around 105 M dwarf hosts, and 2637 planets around 1994 FGK hosts, whose parameters are provided in Table~\ref{tab:desert_sample} and displayed in Fig~\ref{fig:desert_sample}. 

Unfortunately, with so few M dwarf planets, the Gaussian Kernel Density Estimate (KDE) method of \citetalias{CastroGonzalez2024} results in uninformative desert boundaries. We therefore devised an alternative approach to calculating the desert boundaries, which we apply to the M dwarf and FGK samples separately, as well as to the combined sample for comparison. We do this in both period-radius space, as is used in \citetalias{CastroGonzalez2024}, and also instellation-radius space.

\begin{figure}
    \centering
    \includegraphics[width=0.9\linewidth]{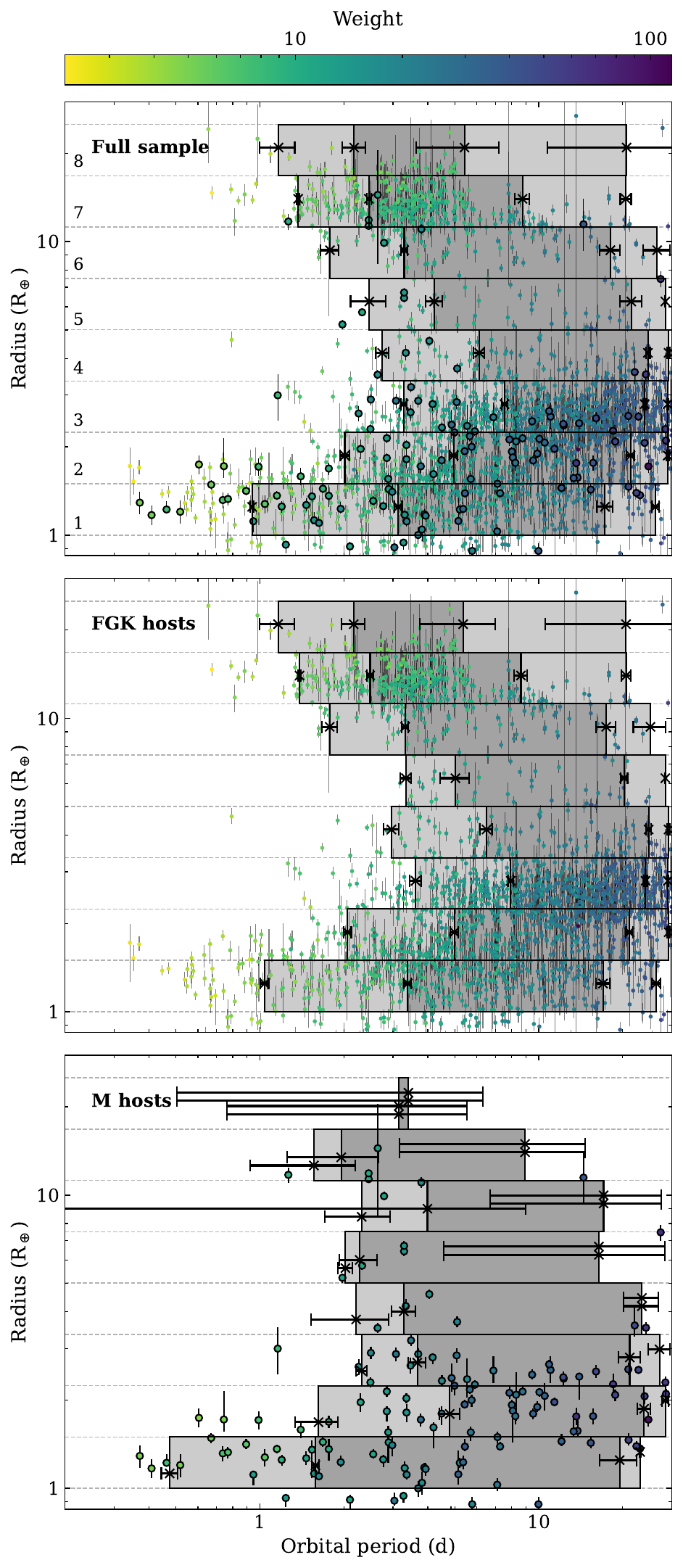}
    \caption{A study of the Neptunian desert boundaries for FGK planet hosts versus M dwarf hosts in period-radius space, as described in Section\,\ref{sec:disc_desert}. Our planet sample (Table~\ref{tab:desert_sample}; Fig.~\ref{fig:desert_sample}) is shown as coloured dots, according to the weighting of each planet. In the top panel, our full planet sample is shown (where M dwarf hosts are highlighted with black outlines); the middle panel shows planets around FGK hosts only; and the bottom panel shows planets around M dwarf hosts. The boundaries of the bins in radius space are depicted with dashed lines and the bins are numbered 1-8 in the top panel. The period percentile values calculated for each bin and their errors are shown, where the 2.3 to 97.7 percentiles are shaded in light grey while the 16 to 84 percentiles are shaded in dark grey.}
    \label{fig:desert_bins}
\end{figure}

\begin{figure}
    \centering
    \includegraphics[width=0.9\linewidth]{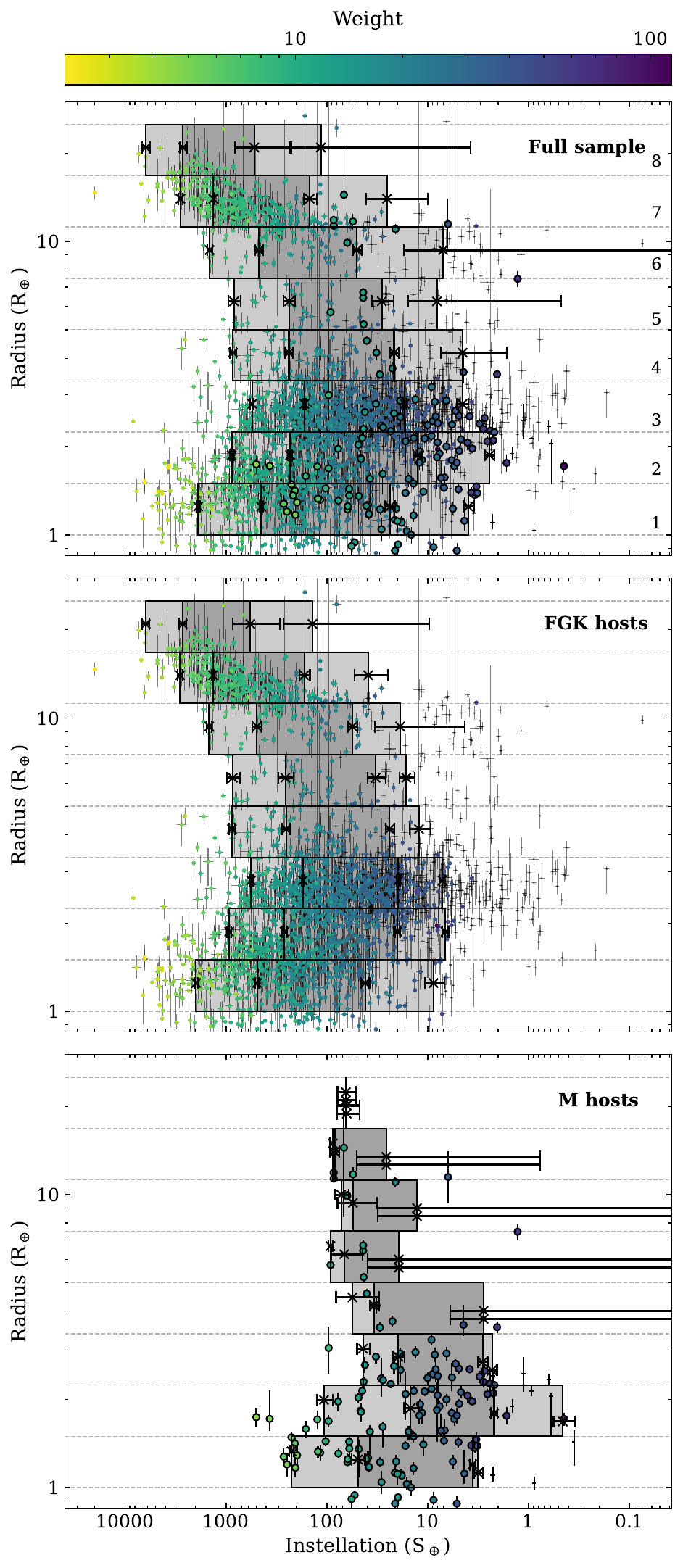}
    \caption{A study of the Neptunian desert boundaries for FGK planet hosts versus M dwarf hosts in instellation-radius space, as described in Section\,\ref{sec:disc_desert}. The details are the same as in Fig.~\ref{fig:desert_bins}.}
    \label{fig:desert_bins_instell}
\end{figure}

\begin{figure}
    \centering
    \includegraphics[width=0.95\linewidth]{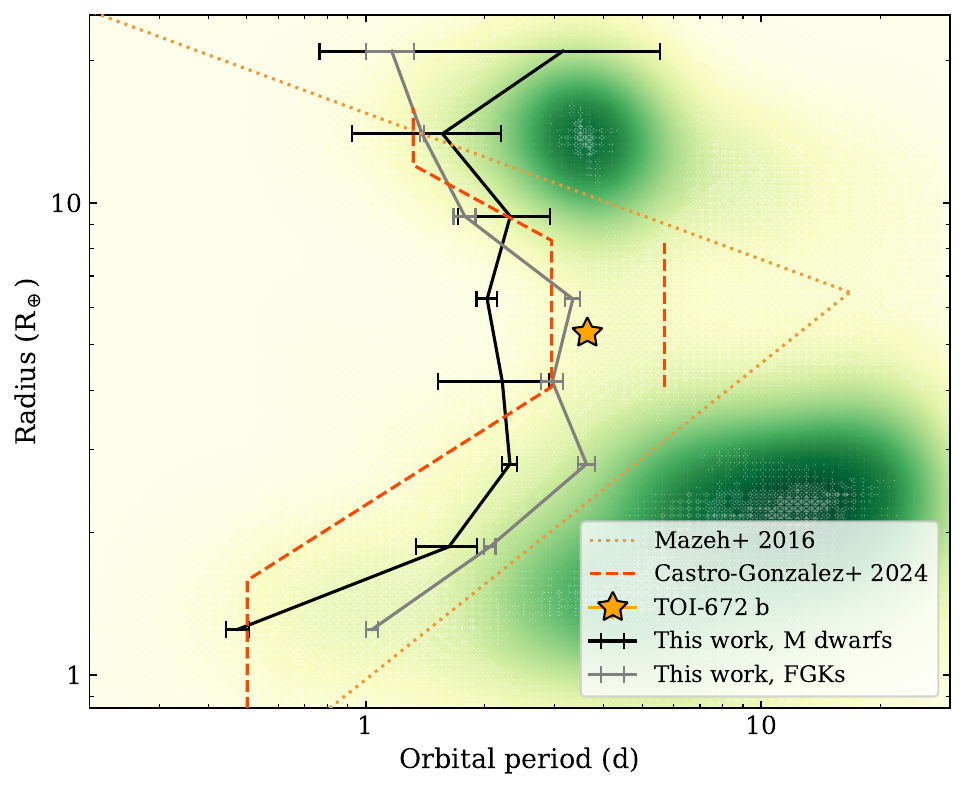}
    \caption{A comparison of the Neptunian desert boundaries in period-radius space as derived by \citet{Mazeh2016} (yellow dotted lines), \citet{CastroGonzalez2024} (orange dashed lines), and this work, where we calculate the desert boundaries for FGK planet hosts (grey solid line) and M dwarf hosts (black solid line) separately. The background is shaded according to the known planet population from the NASA exoplanet archive, as in Fig.~\ref{fig:desert}.}
    \label{fig:desert_lines}
\end{figure}

\begin{figure}
    \centering
    \includegraphics[width=0.95\linewidth]{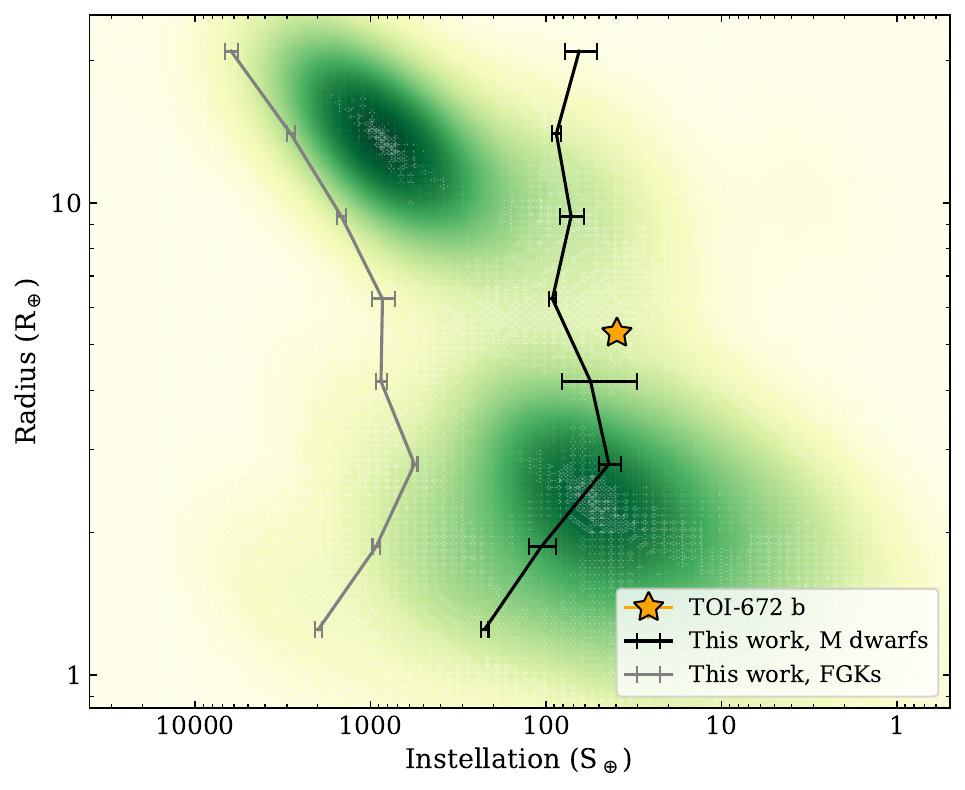}
    \caption{A comparison of the Neptunian desert boundaries in instellation-radius space as derived in this work, where we calculate the desert boundaries for FGK planet hosts (grey solid line) and M dwarf hosts (black solid line) separately. The background is shaded according to the known planet population from the NASA exoplanet archive, as in Fig.~\ref{fig:desert}.}
    \label{fig:desert_lines_instell}
\end{figure}

We split our samples into eight equal size bins in $\{\log{R_p},\log{P}\}$ (or $\{\log{R_p},\log{S_p}\}$) space. Following \citetalias{CastroGonzalez2024}, we restrict the parameter space the bins cover to $1 < R_p < 25$\,R$_{\oplus}$ and $P < 30$\,days, in order to avoid areas of parameter space that have uncorrectable incompleteness, i.e., where very few or no planets have been detected (we keep this period cut for the sample in instellation space also). The number of bins is restricted by our M dwarf sample, which has $18 \times$ fewer planets compared to those around FGK stars from Kepler and TESS. To account for uncertainties in the radii and period (or instellation) of each planet, we sample $10^5$ realizations from normal distributions of each planet's radius and period (or instellation) given its reported $1\sigma$ errors. For each realization, we calculated the 2.3, 16, 84, and 97.7 percentiles of the orbital periods (or instellations) in each radius bin. In these calculations, each planet is weighted by $w$ given in Eq.~\ref{equ:weights}. We then take the mean and standard deviation of these percentiles over the $10^5$ realizations as our final period (or instellation) distributions, which are displayed in Fig.~\ref{fig:desert_bins} for period-radius space, and in Fig.~\ref{fig:desert_bins_instell} for instellation-radius space. Finally, we adopt the 2.3 percentile values as the edges of the Neptunian desert, which are shown in Fig.~\ref{fig:desert_lines} for period-radius space, and in Fig.~\ref{fig:desert_lines_instell} for instellation-radius space. 

We first look at our boundaries in period-radius space. We reproduce a similar desert shape as \citealt{CastroGonzalez2024}, where the triangular tip as in \citet{Mazeh2016} is truncated. It is worth noting here that there are very few giant planets in our M dwarf sample, making the errors on the upper boundary, i.e., bins 7 and 8, very large. It could be the case that the upper boundary does not present in the same way for M dwarf planets, but this might change if we discover more planets in this regime. 

As introduced in Section~\ref{sec:intro}, \citet{Hallatt2022} predict that the opening of the Neptunian desert, where the upper and lower boundaries intersect, should shift to shorter orbital periods around M dwarfs. This begins at $\sim0.7$\,days around $0.5\,M_\odot$ stars and $\sim 1.5$\,days around $0.8\,M_\odot$ stars, compared to $\sim 3$\,days around $1\,M_\odot$ stars. By taking the weighted average of the 2.3 percentiles for bins 3--5 (labelled in Fig.~\ref{fig:desert_bins}), we can estimate the period at which the desert opens. The average stellar mass of our FGK sample is $0.99 \pm 0.2\,M_\odot$ and we find the opening at $3.3 \pm 1.4$\,days, matching the prediction. The average stellar mass of our M dwarf sample is $0.43 \pm 0.13\,M_\odot$, and while the opening shifts to a slightly lower period as predicted, we find the opening is located at $2.2 \pm 1.0$\,days. This is within $\sim$1~$\sigma$ of the period for the FGK hosts and in slight tension with the predicted 0.7\,days \citep{Hallatt2022}. 

The prediction of \citet{Hallatt2022} is predicated on atmospheric mass loss processes as the dominant mechanism in opening the desert. 
The discrepancy between our observation and the theoretical prediction might mean that they are not. \citetalias{CastroGonzalez2024} remark that the period range of the Neptunian ridge is similar to the hot Jupiter pileup and note that a large number of Jupiter and Neptune-sized planets with eccentric, misaligned orbits exists in these over-dense regions. Thus, they posit that high-eccentricity migration may be the main process bringing planets into the ridge. The mechanisms sculpting the desert and the ridge should be explored further once a larger sample of M dwarf planets will make comparisons of planets across stellar types more robust. This may be accomplished through looking at atmospheric escape efficiencies and spin-orbit angles of planets in and around the desert. 

Next we considered the desert boundaries in bolometric instellation-radius space. Unlike the boundaries in period-radius space, we find that the desert boundaries around FGK hosts versus M dwarf hosts are differ significantly with the latter boundaries occurring at much lower bolometric instellations. Repeating the same boundary calculation as in period-radius space, we estimate the instellation at which the desert opens to be $830.7 \pm 0.1\,S_{\oplus}$ around FGK stars compared to $57.8 \pm 0.2\,S_{\oplus}$ around M dwarfs. In both parameter spaces, TOI-672\,b is located just outside the desert.

Photoevaporation caused by X-ray and Extreme Ultra-Violet (together, XUV) radiation from host stars is believed to be a main driver of the Neptunian desert's lower boundary \citep{Owen2018}. However, the XUV luminosity histories of the host stars in our sample are unknown, which necessitated our consideration of the desert as a function of bolometric instellation. What is known is that the ratio of XUV luminosity to bolometric luminosity differs across stellar types, as does the evolutionary timescale for the saturated phase of high XUV emission, prior to the emission falling off as stars age and spin-down \citep{Jackson2012,Shkolnik2014,Pass2025}. In general, later type stars exhibit longer saturation timescales and higher XUV:bolometric luminosity ratios. As such, we expect that per bolometric instellation, planets orbiting field-age M dwarfs will experience higher levels of lifetime XUV irradiation compared to around FGK stars \citep{McDonald2019}. While present day bolometric instellation is not the main driver of the location of the desert, the location of the M dwarf boundary at lower bolometric instellations than around FGK stars in Figure~\ref{fig:desert_lines_instell} is consistent with predictions from XUV-driven photoevaporation. However, it is important to note that the boundary for M dwarf hosts in instellation space could not overlap the boundary for FGKs. At an instellation of $\sim 800$\,S$_{\oplus}$ (the position of the boundary for FGKs), a Neptune around an average M dwarf in our sample would experience tidal disruption due to its proximity to the Roche radius of its star. Nevertheless, the current placement of the boundary at $\sim 60$\,S$_{\oplus}$ for M dwarfs is far outside the Roche radius, so we do not expect its placement to be sculpted primarily by tidal disruption.

Future work should focus on calculating the lifetime XUV irradiation for all planets in our sample (noting that \citet{McDonald2019} only do this for lifetime \textit{X-ray} irradiation), and recalculating the desert boundaries. If XUV photoevaporation is the main driver of the desert boundaries, then the empirical boundaries in lifetime integrated XUV irradiation-radius space should be consistent across different spectral types.


\section{Conclusions}\label{sec:conclusion}

In this study we confirmed and characterised the previously validated planet \hbox{TOI-672\,b}. We used photometry from four TESS sectors and a single transit observation from ExTrA, and modelled these data jointly with RV data taken with the NIRPS and HARPS spectrographs. 

We use empirical M dwarf relations to measure the mass and radius of the M0V star TOI-672, and the spectra from NIRPS to obtain its effective temperature and individual elemental abundances from which we derive its overall metallicity [M/H] and $\alpha$-enhancement $[\alpha/Fe]$. The stellar rotation period is estimated as 18.5\,days from ASAS-SN photometry, within the range of predicted periods from the $\log R'_{\rm HK}$.

We show that TOI-672\,b is a massive super-Neptune with a period of 3.63\,days, a radius of $5.31^{+0.24}_{-0.26}$\,$R_{\oplus}$, and a mass of $50.9^{+4.5}_{-4.4}$\,$M_{\oplus}$. We present novel mass-radius relations for irradiated H$_2$/He-dominated planets and determine that TOI-672\,b has a H$_2$/He envelope mass fraction of $20-30$\% (depending on the assumed core composition). We perform atmospheric evolution simulations, assuming early migration and evolution at the present-day orbit of TOI-672\,b, and find that the nature of TOI-672\,b would not have changed substantially over its lifetime, atmospheric escape made inefficient by its strong gravity.

The period and radius of TOI-672\,b place it within the Neptunian desert or ridge. However, existing desert boundaries have been calculated from planets predominantly orbiting FGK stars. We perform the first study that directly compares the Neptunian desert boundaries in both period-radius space and instellation-radius space for planets hosted by FGK stars versus M dwarf stars. We devise a novel method for determining the desert boundary, and reproduce a truncation of the right-hand-side of the desert in period-radius space (the ridge, as in \citetalias{CastroGonzalez2024}), and find this to be at comparable periods of $2.2 \pm 1.0$\,days and $3.3 \pm 1.4$\,days for M dwarf hosts and FGK hosts, respectively. This shift is small compared to theoretical predictions based on atmospheric escape processes \citep{Hallatt2022}. We find that in instellation-radius space, the boundary changes significantly between M-dwarf and FGK hosts, shifting to lower instellations for the former. We encourage further study of the mechanisms sculpting the desert and ridge to better determine the predominant processes and whether the desert changes across stellar types.

Lastly, we explore the possibility of additional planets in the TOI-672 system. It was selected as a favourable target for the NIRPS GTO ``Deep Search'' sub-program, as a system with the potential for having additional planets not seen in transit.  We use the photometry from TESS, ExTrA, and other ground-based facilities to search for TTVs, for which we find no evidence. We quantify the sensitivity of our RV time series and find that, from 1--200\,d, we would expect to recover signals down to approximately 10\,m\,s$^{-1}$ with the combined NIRPS and HARPS data, and thus are unable to recover any small terrestrial planets below 5\,$M_{\oplus}$ that could be present.

\section*{Data Availability}

Tables \ref{tab:linelist} and \ref{tab:desert_sample} are available in full in electronic form at the CDS via anonymous ftp to \url{cdsarc.u-strasbg.fr} (130.79.128.5) or via \url{http://cdsweb.u-strasbg.fr/cgi-bin/qcat?J/A+A/}. Tables \ref{tab:nirpsrvs}, \ref{tab:harpsrvs_drs}, and \ref{tab:harpsrvs_lbl} are available on ExoFOP-TESS at \url{https://exofop.ipac.caltech.edu/tess/target.php?id=151825527}. The code underlying the joint fit is available on reasonable request to the corresponding author.


\begin{acknowledgements}

We first thank the anonymous referee for their comments which improved the quality of this manuscript. We would like to thank George W. King for helpful discussion regarding XUV-driven photoevaporation. 

AO acknowledges this research was part funded by the UKRI (Grants ST/X001121/1, EP/X027562/1).

This project has received funding from the European Research Council (ERC) under the European Union's Horizon 2020 research and innovation programme (project {\sc Spice Dune}, grant agreement No 947634). This material reflects only the authors' views and the Commission is not liable for any use that may be made of the information contained therein.

This work has been carried out within the framework of the NCCR PlanetS supported by the Swiss National Science Foundation under grants 51NF40\_182901 and 51NF40\_205606.

NJC, \'EA, CC, AL, FBa, BB, RD, LMa, RA \& JPW  acknowledge the financial support of the FRQ-NT through the Centre de recherche en astrophysique du Qu\'ebec as well as the support from the Trottier Family Foundation and the Trottier Institute for Research on Exoplanets.

NN, JIGH, RR, ASM \& AKS  acknowledge financial support from the Spanish Ministry of Science, Innovation and Universities (MICIU) projects PID2020-117493GB-I00 and PID2023-149982NB-I00.

NN  acknowledges financial support by Light Bridges S.L, Las Palmas de Gran Canaria.

NN acknowledges funding from Light Bridges for the Doctoral Thesis "Habitable Earth-like planets with ESPRESSO and NIRPS", in cooperation with the Instituto de Astrof\'isica de Canarias, and the use of Indefeasible Computer Rights (ICR) being commissioned at the ASTRO POC project in the Island of Tenerife, Canary Islands (Spain). The ICR-ASTRONOMY used for his research was provided by Light Bridges in cooperation with Hewlett Packard Enterprise (HPE).

We acknowledge funding from the European Research Council under the ERC Grant Agreement n. 337591-ExTrA.

\'EA, FBa, RD \& LMa  acknowledges support from Canada Foundation for Innovation (CFI) program, the Universit\'e de Montr\'eal and Universit\'e Laval, the Canada Economic Development (CED) program and the Ministere of Economy, Innovation and Energy (MEIE).

XB, XDe, AC \& TF  acknowledge funding from the French ANR under contract number ANR\-24\-CE49\-3397 (ORVET), and the French National Research Agency in the framework of the Investissements d'Avenir program (ANR-15-IDEX-02), through the funding of the ``Origin of Life" project of the Grenoble-Alpes University.

AL  acknowledges support from the Fonds de recherche du Qu\'ebec (FRQ) - Secteur Nature et technologies under file \#349961.

SCB, EC, NCS \& ED-M  acknowledge the support from FCT - Funda\c{c}\~ao para a Ci\^encia e a Tecnologia through national funds by these grants: UIDB/04434/2020, UIDP/04434/2020.

SCB acknowledges the support from Funda\c{c}\~ao para a Ci\^encia e Tecnologia (FCT) in the form of a work contract through the Scientific Employment Incentive program with reference 2023.06687.CEECIND and DOI \href{https://doi.org/10.54499/2023.06687.CEECIND/CP2839/CT0002}{10.54499/2023.06687.CEECIND/CP2839/CT0002.}

The Board of Observational and Instrumental Astronomy (NAOS) at the Federal University of Rio Grande do Norte's research activities are supported by continuous grants from the Brazilian funding agency CNPq. This study was partially funded by the Coordena\c{c}\~ao de Aperfei\c{c}oamento de Pessoal de N\'ivel Superior—Brasil (CAPES) — Finance Code 001 and the CAPES-Print program.

BLCM  acknowledge CAPES postdoctoral fellowships. BLCM  acknowledges CNPq research fellowships (Grant No. 305804/2022-7).

NBC  acknowledges support from an NSERC Discovery Grant, a Canada Research Chair, and an Arthur B. McDonald Fellowship, and thanks the Trottier Space Institute for its financial support and dynamic intellectual environment.

JRM  acknowledges CNPq research fellowships (Grant No. 308928/2019-9).

XDu  acknowledges the support from the European Research Council (ERC) under the European Union’s Horizon 2020 research and innovation programme (grant agreement SCORE No 851555) and from the Swiss National Science Foundation under the grant SPECTRE (No 200021\_215200).\\

DE  acknowledge support from the Swiss National Science Foundation for project 200021\_200726. The authors acknowledge the financial support of the SNSF.

ICL  acknowledges CNPq research fellowships (Grant No. 313103/2022-4).

CMo  acknowledges the funding from the Swiss National Science Foundation under grant 200021\_204847 “PlanetsInTime”.

Co-funded by the European Union (ERC, FIERCE, 101052347). Views and opinions expressed are however those of the author(s) only and do not necessarily reflect those of the European Union or the European Research Council. Neither the European Union nor the granting authority can be held responsible for them.

GAW is supported by a Discovery Grant from the Natural Sciences and Engineering Research Council (NSERC) of Canada.

RA  acknowledges the Swiss National Science Foundation (SNSF) support under the Post-Doc Mobility grant P500PT\_222212 and the support of the Institut Trottier de Recherche sur les Exoplan\`etes (IREx).

KAM  acknowledges support from the Swiss National Science Foundation (SNSF) under the Postdoc Mobility grant P500PT\_230225.

KAC acknowledges support from the TESS mission via subaward s3449 from MIT.

ED-M  further acknowledges the support from FCT through Stimulus FCT contract 2021.01294.CEECIND. ED-M  acknowledges the support by the Ram\'on y Cajal contract RyC2022-035854-I funded by MICIU/AEI/10.13039/501100011033 and by ESF+.

AKS  acknowledges financial support from La Caixa Foundation (ID 100010434) under the grant LCF/BQ/DI23/11990071.

We acknowledge the Natural Sciences and Engineering Research Council of Canada (NSERC) in supporting this research.\\

Funding for the TESS mission is provided by NASA's Science Mission Directorate. We acknowledge the use of public TESS data from pipelines at the TESS Science Office and at the TESS Science Processing Operations Center.

This work makes use of observations from the LCOGT network. Part of the LCOGT telescope time was granted by NOIRLab through the Mid-Scale Innovations Program (MSIP). MSIP is funded by NSF.

This research has made use of the Exoplanet Follow-up Observation Program (ExoFOP; DOI: 10.26134/ExoFOP5) website, which is operated by the California Institute of Technology, under contract with the National Aeronautics and Space Administration under the Exoplanet Exploration Program.

This paper includes data collected by the TESS mission that are publicly available from the Mikulski Archive for Space Telescopes (MAST).

This research has made use of the NASA Exoplanet Archive, which is operated by the California Institute of Technology, under contract with the National Aeronautics and Space Administration under the Exoplanet Exploration Program.

This publication makes use of The Data \& Analysis Center for Exoplanets (DACE), which is a facility based at the University of Geneva (CH) dedicated to extrasolar planets data visualisation, exchange and analysis. DACE is a platform of the Swiss National Centre of Competence in Research (NCCR) PlanetS, federating the Swiss expertise in Exoplanet research. The DACE platform is available at \url{https://dace.unige.ch}.\\

This work made use of \texttt{tpfplotter} by J. Lillo-Box (publicly available at \url{www.github.com/jlillo/tpfplotter}), which also made use of the python packages \texttt{astropy}, \texttt{lightkurve}, \texttt{matplotlib} and \texttt{numpy}.

This research also made use of {\tt exoplanet} \citep{Foreman-Mackey2020} and its dependencies \citep{Agol2020,Kumar2019,Astropy2013,Astropy2018,Kipping2013,Luger2018,Salvatier2016,Theano2016}.

\end{acknowledgements}


\bibliographystyle{aa}
\bibliography{bib}

\begin{appendix}
\twocolumn
\section{Photometry}

\begin{figure}[htp]
    \centering
    \begin{subfigure}{0.34\textwidth}
        \centering
        \includegraphics[width=\textwidth]{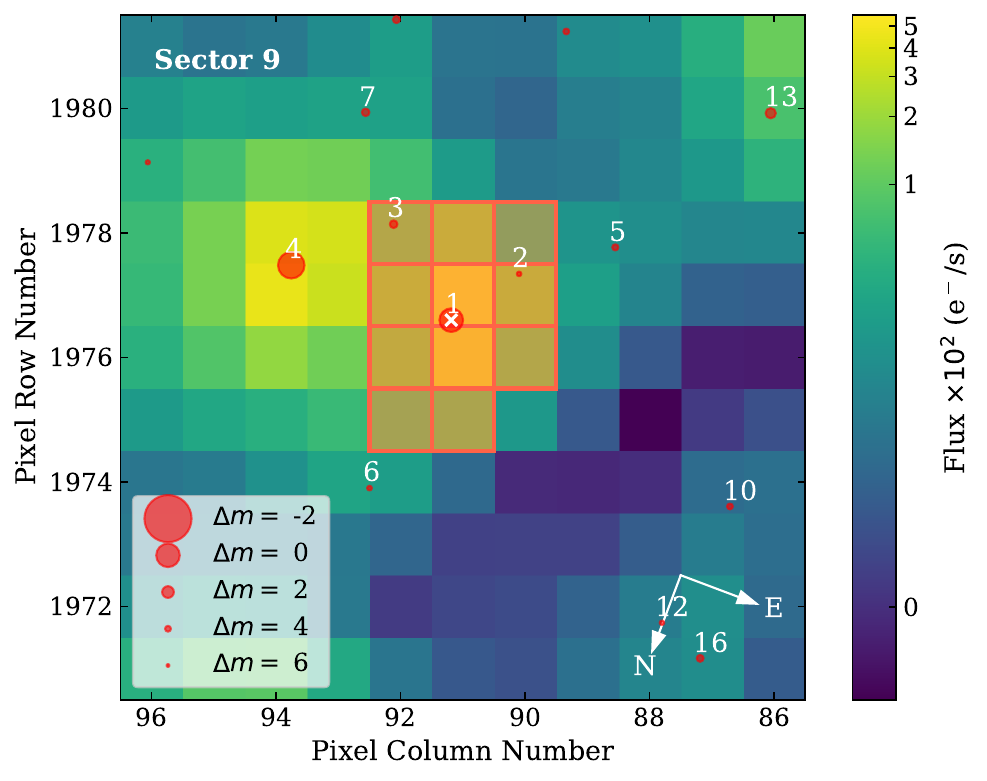}
    \end{subfigure}
    \begin{subfigure}{0.34\textwidth}
        \centering
        \includegraphics[width=\textwidth]{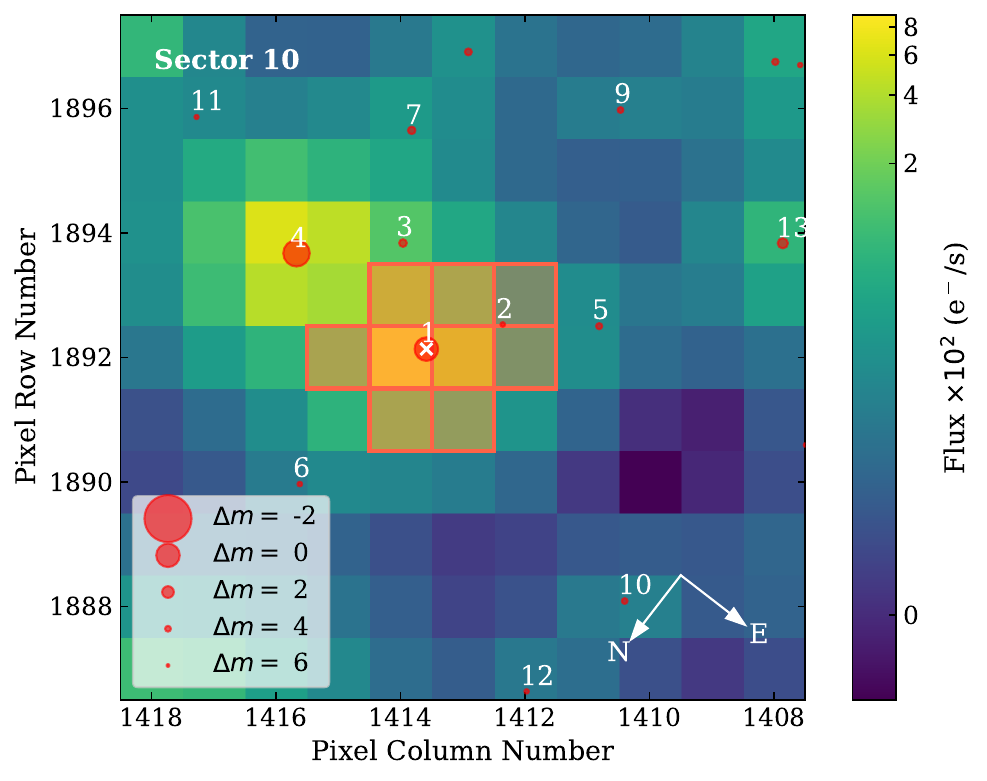}
    \end{subfigure}
    \begin{subfigure}{0.34\textwidth}
        \centering
        \includegraphics[width=\textwidth]{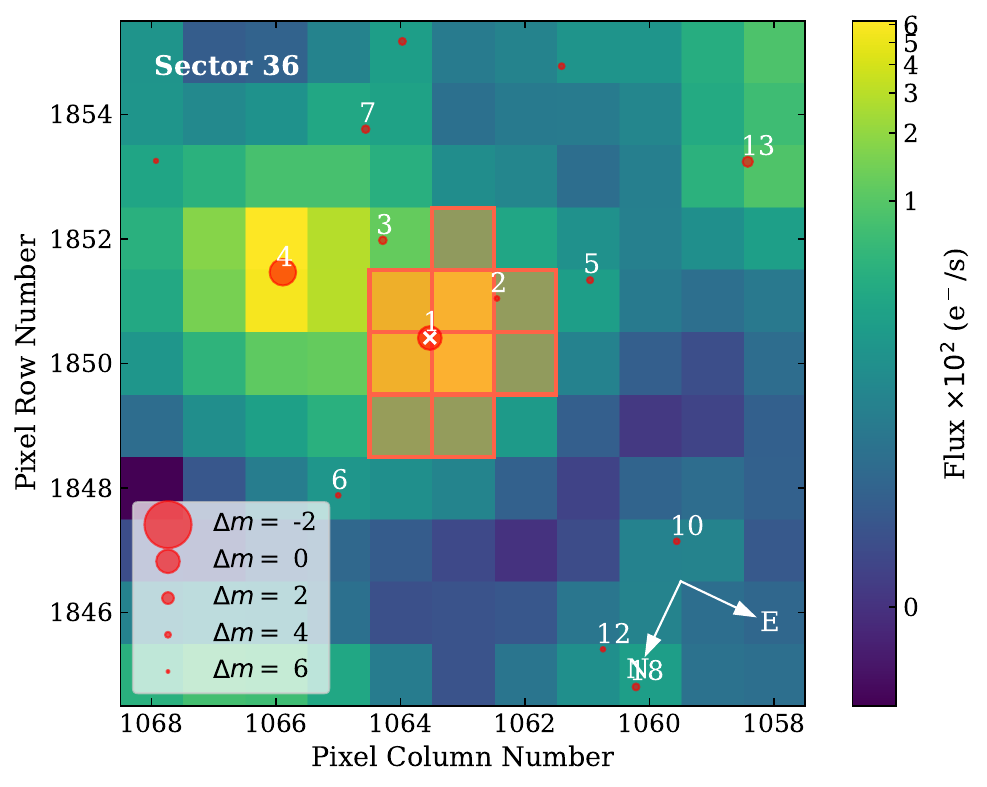}
    \end{subfigure}
    \begin{subfigure}{0.34\textwidth}
        \centering
        \includegraphics[width=\textwidth]{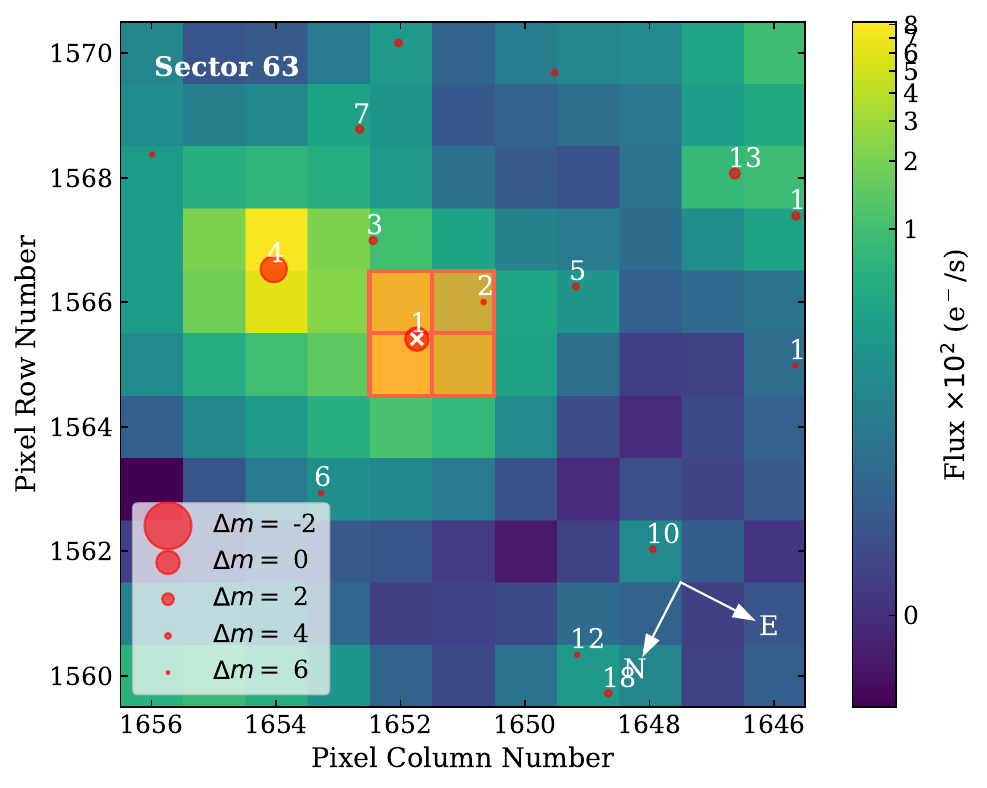}
    \end{subfigure}
\caption{The Target Pixel File (TPF) for TOI-672 (marked as a white cross) from TESS Sectors 9, 10, 36, and 63 \textit{(top to bottom)}. Other Gaia DR3 sources within a limit of 6 Gaia magnitudes difference from TOI-672 are marked as red circles, and are numbered in distance order from TOI-672. The aperture mask is outlined and shaded in red. This figure was created with {\tt tpfplotter} \citep{Aller2020}.}
\label{fig:tpfplotter}
\end{figure}

\begin{figure}[htp]
    \centering
    \includegraphics[width=\columnwidth]{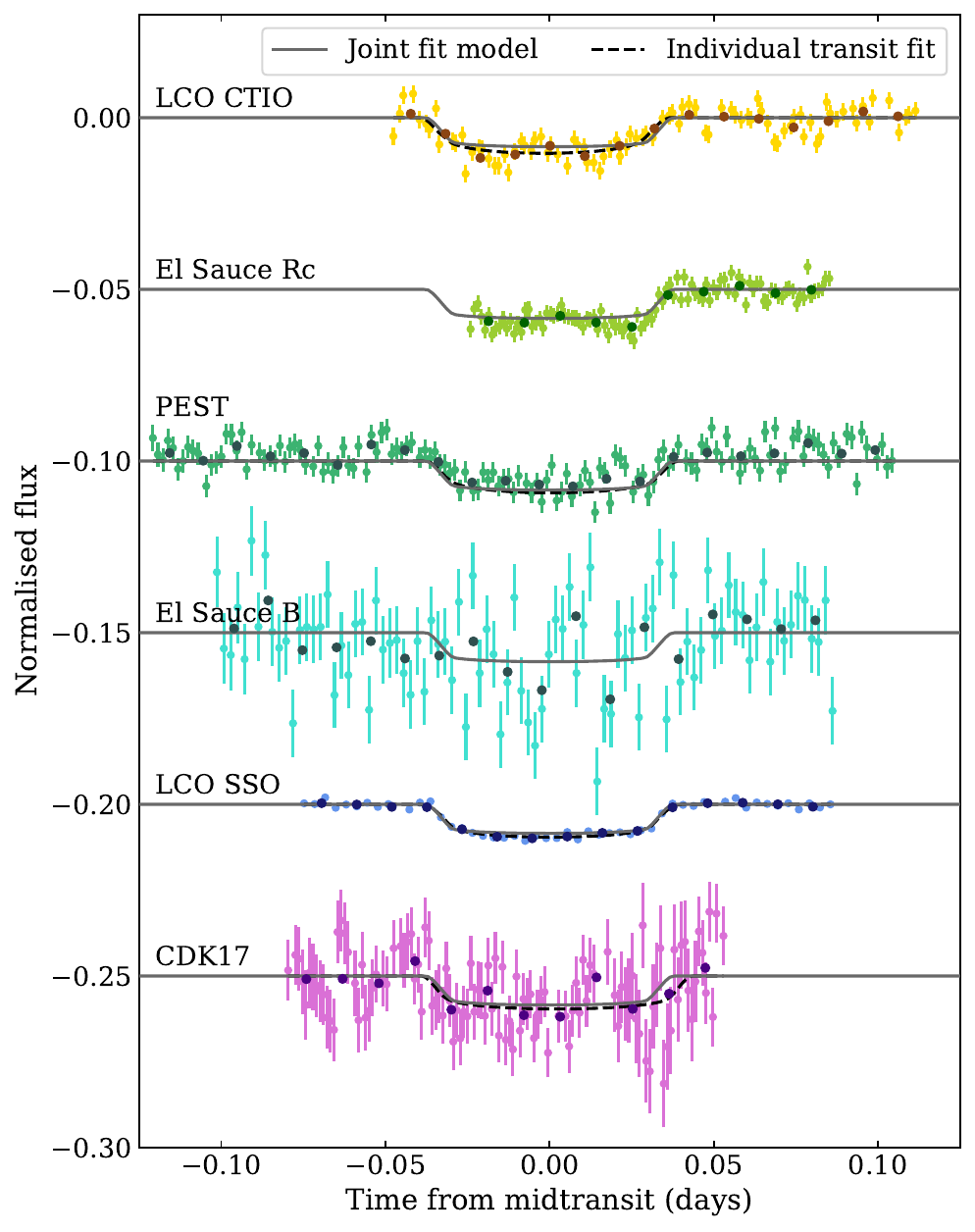}
    \caption{The photometry from various instruments described in Section~\ref{sec:obs_otherphotom}, centered on the time from midtransit predicted by the joint fit model. \textit{From top to bottom, in chronological order:} LCO CTIO (yellow circles); El Sauce Rc (green, offset by -0.05 in flux); PEST (sea green, offset by -0.10); El Sauce B (turquoise, offset by -0.15); LCO SSO (blue, offset by -0.20); and CDK17 (orchid, offset by -0.25). The photometry is binned as darker circles. The planet model from the joint fit using the TESS limb darkening is overlaid (dark grey solid lines), alongside the individual fits to each transit from the TTV search described in Section~\ref{sec:disc_ttvs} for the LCO CTIO, PEST, LCO SSO, and CDK17 photometry (dashed black lines). The different models generally agree well, with a slight difference in transit ``dip'' shape caused by the difference in limb darkening parameters between the fits.}
    \label{fig:sg1}
\end{figure}

\begin{figure*}
    \centering
    \includegraphics[width=\linewidth]{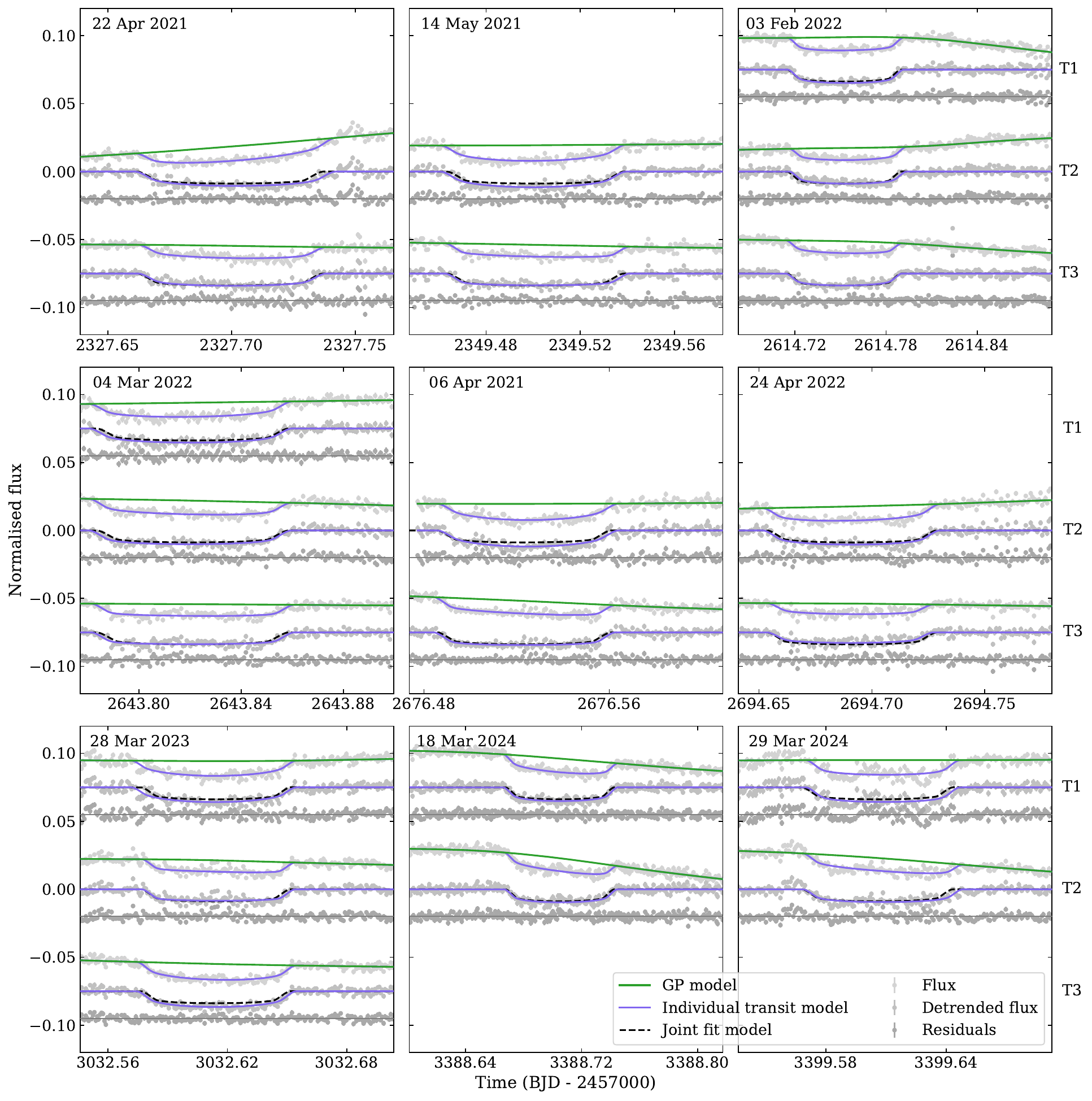}
    \caption{The photometry from ExTrA described in Section~\ref{sec:obs_extra}, where each panel is from a single observation night which is labelled. In each panel, observations can come from Telescope 1 (T1, \textit{top row}, overall offset by +0.075 in flux), Telescope 2 (T2, \textit{middle row}, no overall offset in flux), and/or Telescope 3 (T3, \textit{bottom row}, overall offset by -0.075 in flux). For each telescope, we display the extracted photometry (pale grey dots, offset by +0.02 in flux) which is detrended by a GP (green line), to produce the detrended photometry (medium grey dots, no offset in flux), and then the individual transit model (purple line) is subtracted to show the residuals (dark grey dots, offset by -0.02 in flux). We also show the transit model from the joint fit (dashed black line, using the ExTrA limb darkening), which generally agrees well with the individual transit fits performed for the TTV search (purple line), see Section~\ref{sec:disc_ttvs}. }
    \label{fig:extra_extra}
\end{figure*}

\FloatBarrier

\clearpage
\onecolumn

\begin{table*}[htp]
    \tiny
    \centering
    \caption{All photometric transits of TOI-672 presented in this paper.}
    \begin{threeparttable}
    \begin{tabular}{lllllllll}
    \toprule
    \multicolumn{3}{l}{\textbf{Instrument}} & \textbf{Used in}      & \textbf{Used in}    & \textbf{Mid-transit}  & \textbf{+1 st. } & \textbf{-1 st. } & \textbf{Fig.}\\
    & &                                       & \textbf{joint fit?}   & \textbf{TTV search?}  & \textbf{time (BJD-}   & \textbf{dev.}    & \textbf{dev.}  & \\
    & &                                       &                       &                       & \textbf{2457000)}     &                  & \\
    \midrule
    TESS        & \multicolumn{2}{l}{Sector 9}  & Y & Y & 1546.47947 & 0.00108 & 0.00104 & Fig.~\ref{fig:tess}, \ref{fig:tess_fold} \\
                &           & & \multirow[t]{2}{*}[-0.5em]{$\vdots$} & \multirow[t]{2}{*}[-0.5em]{$\vdots$} & 1550.11438 & 0.00109 & 0.00112 & \multirow[t]{2}{*}[-0.5em]{$\vdots$} \\
                &           & & & & 1553.74764 & 0.00132 & 0.00124 \\
                &           & & & & 1561.01452 & 0.00084 & 0.00086 \\
                &           & & & & 1564.64751 & 0.00102 & 0.00102 \\
                &           & & & & 1568.28141 & 0.00106 & 0.00106 \\
                & \multicolumn{2}{l}{Sector 10} & & & 1575.54835 & 0.00068 & 0.00067 \\
                &           & & & & 1579.18227 & 0.00104 & 0.00104 \\
                &           & & & & 1586.45005 & 0.00090 & 0.00083 \\
                &           & & & & 1590.08250 & 0.00119 & 0.00122 \\
                &           & & & & 1593.71421 & 0.00077 & 0.00077 \\
                & \multicolumn{2}{l}{Sector 36} & & & 2284.09636 & 0.00086 & 0.00082 \\
                &           & & & & 2287.72776 & 0.00111 & 0.00109 \\
                &           & & & & 2291.36441 & 0.00095 & 0.00093 \\
                &           & & & & 2298.63032 & 0.00082 & 0.00080 \\
                &           & & & & 2302.26324 & 0.00088 & 0.00088 \\
                &           & & & & 2305.89829 & 0.00095 & 0.00093 \\
                & \multicolumn{2}{l}{Sector 63} & & & 3014.44672 & 0.00104 & 0.00116 \\
                &           & & & & 3018.08016 & 0.00077 & 0.00078 \\
                &           & & & & 3021.71372 & 0.00080 & 0.00082 \\
                &           & & & & 3025.34665 & 0.00081 & 0.00083 \\
                &           & & & & 3028.98026 & 0.00087 & 0.00086 \\
                &           & & & & 3032.61552 & 0.00125 & 0.00122 \\
                &           & & & & 3036.24692 & 0.00090 & 0.00094 \\
                &           & & & & 3039.88181 & 0.00089 & 0.00085 \\
    \hdashline\noalign{\vskip 0.5ex}
    ExTrA       & 21-04-22 & T2	& N & Y &	2327.70156	&	0.00253	&	0.00248	& Fig. \ref{fig:extra_extra}\\
                &          & T3	& \multirow[t]{2}{*}[-0.5em]{$\vdots$} & \multirow[t]{2}{*}[-0.5em]{$\vdots$} &	2327.70036	&	0.00166	&	0.00137	& \multirow[t]{2}{*}[-0.5em]{$\vdots$}\\
                & 21-05-14 & T2	& & & 2349.50002	&	0.00158	&	0.00162	\\
                &          & T3	& & & 2349.50171	&	0.00123	&	0.00110	\\
                & 22-02-03 & T1	& & &	2614.75339	&	0.00116	&	0.00112	\\
                &          & T2	& & &	2614.75493	&	0.00251	&	0.00211	\\
                &          & T3	& & &	2614.75221	&	0.00124	&	0.00130	\\
                & 22-03-04 & T1	& & &	2643.82024	&	0.00167	&	0.00166	\\
                &          & T2	& & &	2643.82001	&	0.00138	&	0.00150	\\
                &          & T3	& & &	2643.82098	&	0.00127	&	0.00127	\\
                & 22-04-06 & T2	& & &	2676.52495	&	0.00170	&	0.00154	\\
                &          & T3	& & &	2676.52340	&	0.00108	&	0.00110	\\
                & 22-04-24 & T2	& & &	2694.69089	&	0.00182	&	0.00174	\\
                &          & T3	& & &	2694.69036	&	0.00119	&	0.00122	\\
                & 23-02-16 & T1 & & N & - & - & - & -\\
                &          & T2 & & \multirow[t]{2}{*}[-0.5em]{$\vdots$} & - & - & - & - \\
                &          & T3 & & & - & - & - & - \\
                & 23-03-28 & T1	& & Y &	3032.61340	&	0.00235	&	0.00184	& Fig. \ref{fig:extra_extra} \\
                &          & T2	& & \multirow[t]{2}{*}[-0.5em]{$\vdots$} &	3032.61552	&	0.00103	&	0.00104	& \multirow[t]{2}{*}[-0.5em]{$\vdots$} \\
                &          & T3	& & &	3032.61446	&	0.00132	&	0.00137	\\
                & 23-04-08 & T1 & & N & - & - & - & - \\
                &          & T2 & & \multirow[t]{2}{*}[-0.5em]{$\vdots$} & - & - & - & -\\
                &          & T3 & & & - & - & - & -\\
                & 24-03-18 & T1	& & Y &	3388.70552	&	0.00159	&	0.00156	& Fig. \ref{fig:extra_extra} \\
                &          & T2	& & \multirow[t]{2}{*}[-0.5em]{$\vdots$} &	3388.70566	&	0.00220	&	0.00221	& \multirow[t]{2}{*}[-0.5em]{$\vdots$} \\
                & 24-03-29 & T1	& & &	3399.60852	&	0.00177	&	0.00157	\\
                &          & T2	& & &	3399.60841	&	0.00217	&	0.00198	\\
                & 24-04-27 & T2	& Y & &	3428.67508	&	0.00145	&	0.00143	& Fig.~\ref{fig:extra} \\
    \hdashline\noalign{\vskip 0.5ex}
    LCO CTIO    & \multicolumn{2}{l}{19-05-11} & N & Y & 1615.51610 & 0.00098 & 0.00096 & Fig.~\ref{fig:sg1} \\
    \hdashline\noalign{\vskip 0.5ex}
    El Sauce Rc & \multicolumn{2}{l}{19-05-11} & N & N & - & - & - & Fig.~\ref{fig:sg1} \\
    \hdashline\noalign{\vskip 0.5ex}
    PEST        & \multicolumn{2}{l}{19-05-26} & N & Y &  1630.05268 & 0.00111 & 0.00112 & Fig.~\ref{fig:sg1} \\
    \hdashline\noalign{\vskip 0.5ex}
    El Sauce B  & \multicolumn{2}{l}{20-02-01} & N & N & - & - & - & Fig.~\ref{fig:sg1} \\
    \hdashline\noalign{\vskip 0.5ex}
    LCO SSO     & \multicolumn{2}{l}{20-03-19} & N & Y & 1928.00602 & 0.00039 & 0.00038 & Fig.~\ref{fig:sg1} \\
    \hdashline\noalign{\vskip 0.5ex}
    CDK17       & \multicolumn{2}{l}{22-06-14} & N & Y & 2745.56357 & 0.00198 & 0.00185 & Fig.~\ref{fig:sg1} \\
    \bottomrule
    \end{tabular}
    \begin{tablenotes}
    \item Dates are given as YY-MM-DD and are the date of the beginning of the night of the observation. It is indicated which transits are used in the joint fit (Section~\ref{sec:fit}), and which are used to search for TTVs (Section~\ref{sec:disc_ttvs}). Where a transit has been individually fit for the TTV search, the resultant mid-transit time is given. Figure references are given.
    \end{tablenotes}
    \end{threeparttable}
    \label{tab:ttvs}
\end{table*}

\clearpage

\section{Spectroscopic time series}

\begin{figure*}
    \centering
    \includegraphics[width=\linewidth]{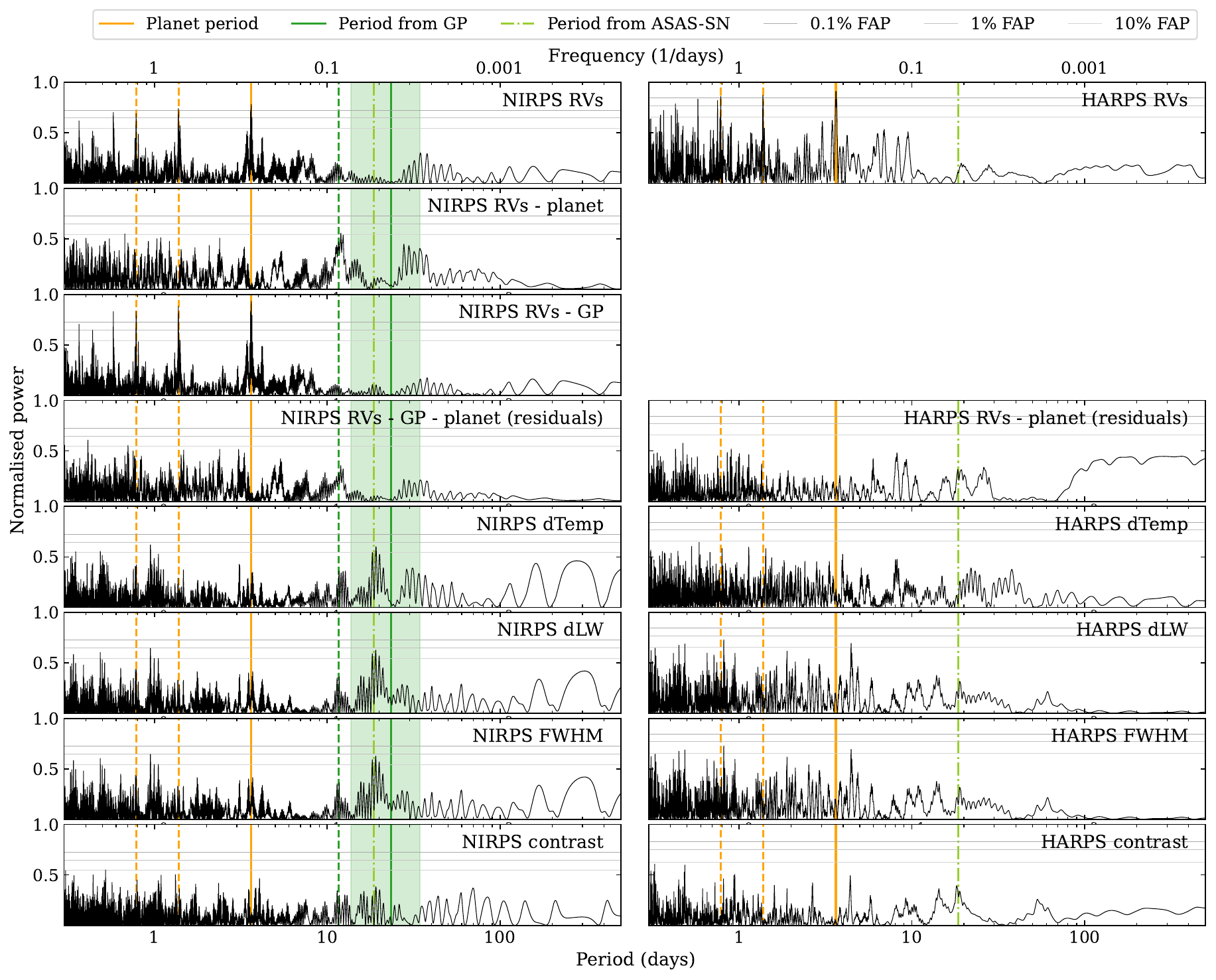}
    \caption{Lomb-Scargle periodograms (black lines) of the NIRPS \textit{(left column)} and HARPS \textit{(right column)} data. The best-fit period of TOI-672\,b is shown (solid vertical orange line), along with its aliases (dashed orange lines). The recurrence timescale fit by the GP used on the NIRPS data (solid vertical green lines) is shown, along with the 1 standard deviation error (shaded green), and half the recurrence timescale (dashed vertical green line). The 0.1, 1, and 10$\%$ False Alarm Probabilities (FAPs, solid horizontal lines from dark to light grey) are calculated using the approximation from \citet{Baluev2008}. We describe the panels from top to bottom. \textit{First row:} the RVs prior to detrending, clearly showing signals above the 0.1\,\% FAP thresholds at the planetary period (and its aliases). \textit{Second row:} the NIRPS RVs with the planet signal removed. \textit{Third row:} the NIRPS RVs detrended with the GP, where the significance of the planet signals has increased. \textit{Fourth row:} the residual RVs after detrending (only in the case of the NIRPS RVs) and subtraction of the planet model, showing no further significant signals. \textit{Fifth to eighth rows:} the dTemp, dLW, FWHM, and contrast activity indicators, respectively (all activity indicators are described in Section~\ref{sec:obs_nirpsharps}). None of the indicators show peaks at the planetary period. The NIRPS dTemp, dLW, and FWHM show a peak around 20\,days of some significance (discussed in Section~\ref{sec:rotation}), and corresponding peaks at 20\,days do also appear in all the HARPS activity indicators, though they are not significant.
}
    \label{fig:activityindicators_ls}
\end{figure*}

\begin{table*}[ht!]
    \tiny
	\caption{NIRPS radial velocities extracted using LBL (version 0.65.006), using NIRPS observations of GJ\,2066 to create the ``friend'' template.}
	\label{tab:nirpsrvs}
    \begin{threeparttable}
	\begin{tabular}{lllllllllll}
	\toprule
	Time & RV & $\sigma_\textrm{RV}$  & d2v & $\sigma_{\rm d2v}$ & DT3500 & $\sigma_{\rm DT3500}$ & FWHM & $\sigma_{\rm FWHM}$ & Contrast & $\sigma_{\rm Contrast}$ \\
	(RJD) & ($ms^{-1}$) & ($ms^{-1}$) & ($ms^{-1}$) & ($ms^{-1}$) & ($ms^{-1}$) & ($ms^{-1}$) & & & &          \\
    \midrule
    60036.61794	&	-4629.14	&	7.84	&	654337.21	&	11831.32	&	139.67	&	1.87	&	9296.24	&	7.38	&	0.8228	&	0.0028	\\
    60042.68622	&	-4575.30	&	5.75	&	594116.31	&	8919.19	    &	108.23	&	1.32	&	9258.67	&	5.56	&	0.8460	&	0.0021	\\
    60044.68910	&	-4639.53	&	5.90	&	616227.86	&	9067.08	    &	113.22	&	1.38	&	9272.47	&	5.66	&	0.8293	&	0.0022	\\
    \vdots & \vdots & \vdots & \vdots & \vdots & \vdots & \vdots & \vdots & \vdots & \vdots & \vdots \\
	\bottomrule
	\end{tabular}
    \begin{tablenotes}
    \item The full NIRPS data products can be found on ExoFOP-TESS at \url{https://exofop.ipac.caltech.edu/tess/target.php?id=151825527}. $\sigma$ is the one standard deviation error on the value. d2v is the differential line width, and DT3500 is the $d{\rm Temp}$ activity indicator, both described in Section~\ref{sec:obs_nirpsharps}.
    \end{tablenotes}
    \end{threeparttable}
\end{table*}

\begin{table*}[ht]
    \tiny
	\caption{HARPS radial velocities reduced using the standard offline HARPS data reduction pipeline (DRS 3.5), using an M2 mask.}
	\label{tab:harpsrvs_drs}
	\begin{threeparttable}
    \begin{tabular}{lllllllllll}
	\toprule
	Time & RV & $\sigma_\textrm{RV}$  & FWHM & $\sigma_{\rm FWHM}$ & Bisector & $\sigma_{\rm Bisector}$ & Contrast & $\sigma_{\rm Contrast}$ & S-index & $\sigma_{\rm S-index}$ \\
	(RJD) & ($ms^{-1}$) & ($ms^{-1}$) & ($ms^{-1}$) & ($ms^{-1}$) & ($ms^{-1}$) & ($ms^{-1}$) & & & &          \\
	\midrule
    60036.61724 & -4507.20 & 10.05 & 3636.44 & 5.14 & 34.13 & 14.21 & 13.885737 & 0.000001 & 1.82 & 0.16 \\
    60044.68818 & -4489.64 & 9.33 & 3687.54 & 5.21 & 32.18 & 13.20 & 14.116411 & 0.000001 & 1.81 & 0.12 \\
    60045.75220 & -4435.54 & 6.73 & 3640.47 & 3.64 & -7.99 & 9.45 & 14.032351 & 0.000001 & 2.60 & 0.30 \\
    \vdots & \vdots & \vdots & \vdots & \vdots & \vdots & \vdots & \vdots & \vdots & \vdots & \vdots \\
	\bottomrule
	\end{tabular}
    \begin{tablenotes}
    \item The full HARPS data products can be found on ExoFOP-TESS at \url{https://exofop.ipac.caltech.edu/tess/target.php?id=151825527}. $\sigma$ is the one standard deviation error on the value.
    \end{tablenotes}
    \end{threeparttable}
\end{table*}

\begin{table*}[ht]
    \tiny
	\caption{HARPS radial velocities extracted using LBL (version 0.65.006), using HARPS observations of TOI-776 to create the ``friend'' template. }
	\label{tab:harpsrvs_lbl}
    \begin{threeparttable}
	\begin{tabular}{lllllllllll}
	\toprule
	Time & RV & $\sigma_\textrm{RV}$  & d2v & $\sigma_{\rm d2v}$ & DT3500 & $\sigma_{\rm DT3500}$ & FWHM & $\sigma_{\rm FWHM}$ & Contrast & $\sigma_{\rm Contrast}$ \\
	(RJD) & ($ms^{-1}$) & ($ms^{-1}$) & ($ms^{-1}$) & ($ms^{-1}$) & ($ms^{-1}$) & ($ms^{-1}$) & & & &          \\
	\midrule
    60036.61724	&	-4690.26	&	3.43	&	93385.95	&	3889.20	&	-45.53	&	1.56	&	7818.31	&	2.78	&	0.999786	&	0.000018	\\
    60044.68818	&	-4688.56	&	3.31	&	110283.75	&	3759.52	&	-45.24	&	1.44	&	7830.40	&	2.69	&	0.999710	&	0.000016	\\
    60045.75217	&	-4632.34	&	2.24	&	86631.91	&	2563.78	&	-47.46	&	1.05	&	7813.48	&	1.83	&	0.999904	&	0.000010	\\
	\vdots & \vdots & \vdots & \vdots & \vdots & \vdots & \vdots & \vdots & \vdots & \vdots & \vdots \\
    \bottomrule
	\end{tabular}
    \begin{tablenotes}
    \item The full HARPS data products can be found on ExoFOP-TESS at \url{https://exofop.ipac.caltech.edu/tess/target.php?id=151825527}. $\sigma$ is the one standard deviation error on the value. d2v is the differential line width, and DT3500 is the $d{\rm Temp}$ activity indicator, both described in Section~\ref{sec:obs_nirpsharps}.
    \end{tablenotes}
    \end{threeparttable}
\end{table*}

\FloatBarrier
\clearpage

\section{Stellar analysis}

\begin{table*}[htbp]
\centering
\tiny
\caption{Lines used for calculating the stellar elemental abundances of TOI-672.}
\label{tab:linelist}
\begin{threeparttable}
\begin{tabular}{p{3cm} p{3cm} p{2.0cm}}
\toprule
\textbf{Element} & \textbf{$\lambda_{\rm cen}$ [nm]} & \textbf{[X/H]} \\
\hline
OH & 1514.58  & -0.302 \\
   & 1514.79 & -0.287 \\
   \noalign{\vskip -1.5ex}
   & $\vdots$ & $\vdots$ \\
   & 1732.22  & -0.259 \\
\hdashline\noalign{\vskip 0.5ex}
$\langle [\mathrm{O/H}] \rangle \pm \sigma_{\rm rand}$ & ($N_{\rm lines} = 33$) & $-0.282\pm 0.002$ \\
\hline
Na 1 & 1074.64 & 0.172 \\
     & 1083.49 & 0.140 \\
     \noalign{\vskip -1.5ex}
     & $\vdots$ & $\vdots$ \\
     & 1476.75 & 0.174 \\
\hdashline\noalign{\vskip 0.5ex}
$\langle [\mathrm{Na/H}] \rangle \pm \sigma_{\rm rand}$ & ($N_{\rm lines} = 5$) & $0.166\pm 0.024$ \\
\bottomrule
\end{tabular}
\begin{tablenotes}
\item Stellar elemental abundances [X/H] are calculated as described in Section~\ref{sec:abundances}. This table is truncated to show the lines used for the first few elements in Table~\ref{tab:Abundance_table}; the full version can be found on the Strasbourg astronomical Data Center (CDS).
\end{tablenotes}
\end{threeparttable}
\end{table*}

\begin{table*}[htbp]
\centering
\tiny
\caption{Stellar elemental abundances of TOI-672.}
\label{tab:Abundance_table}
\begin{threeparttable}
\begin{tabular}{p{1cm} p{1cm} p{1.5cm} p{1.1cm} p{1.1cm} p{1.1cm} p{1.1cm} p{1.1cm} p{1.1cm} p{1.1cm}}
\toprule
\textbf{X} & \textit{\textbf{N}} & \textbf{[X/H]} & \textbf{$\sigma_{rand}$} & \textbf{$\sigma_{T_{\rm eff}}$} & \textbf{$\sigma_{\rm [M/H]}$} & \textbf{$\sigma_{\log g}$} & \textbf{$\sigma_{mac}$} & \textbf{$\sigma_{mic}$} & \textbf{$\sigma_{total}$} \\
\midrule
O       & 33     & -0.282    & 0.002   & 0.005  & 0.063  & 0.004  & 0.008  & 0.005  & 0.064       \\
Na      & 5      & 0.166     & 0.024   & 0.015  & 0.018  & 0.008  & 0.006  & 0.001  & 0.035       \\
Mg      & 7      & -0.204    & 0.072   & 0.087  & 0.029  & 0.011  & 0.001  & 0.005  & 0.117       \\
Al      & 8      & -0.151    & 0.043   & 0.040  & 0.009  & 0.017  & 0.001  & 0.006  & 0.062       \\
Si      & 15     & -0.152    & 0.034   & 0.055  & 0.060  & 0.021  & 0.022  & 0.027  & 0.097       \\
K       & 6      & 0.156     & 0.061   & 0.018  & 0.030  & 0.027  & 0.036  & 0.033  & 0.090       \\
Ca      & 19     & 0.074     & 0.019   & 0.009  & 0.021  & 0.010  & 0.016  & 0.006  & 0.036       \\
Ti      & 54     & -0.096    & 0.015   & 0.087  & 0.036  & 0.009  & 0.012  & 0.017  & 0.098       \\
Cr      & 24     & -0.072    & 0.021   & 0.016  & 0.022  & 0.010  & 0.012  & 0.007  & 0.038       \\
Mn      & 8      & 0.151     & 0.040   & 0.032  & 0.040  & 0.016  & 0.001  & 0.016  & 0.069       \\
Fe      & 63     & -0.138    & 0.022   & 0.039  & 0.008  & 0.017  & 0.041  & 0.019  & 0.066       \\
Ni      & 3      & -0.006    & 0.032   & 0.060  & 0.035  & 0.019  & 0.003  & 0.001  & 0.082       \\
\bottomrule
\end{tabular}
\begin{tablenotes}
\item Stellar elemental abundances [X/H] are calculated as described in Section~\ref{sec:abundances}. This table reports the number of lines $N$ used for each element's abundance calculation and the systematic uncertainties from random errors across multiple lines of the same element (i.e., $\sigma_{[X/H]}/\sqrt{N}$) and from variations in [X/H] with the following stellar parameters: effective temperature $T_{\rm eff}$, overall metallicity [M/H], surface gravity $\log{g}$, macro- and micro-turbulence velocities. The total uncertainty $\sigma_{total}$ on each [X/H] value is computed from the quadrature sum of the individual error terms.
\end{tablenotes}
\end{threeparttable}
\end{table*}

\begin{figure}[htbp]
    \centering
    \includegraphics[width=0.5\linewidth]{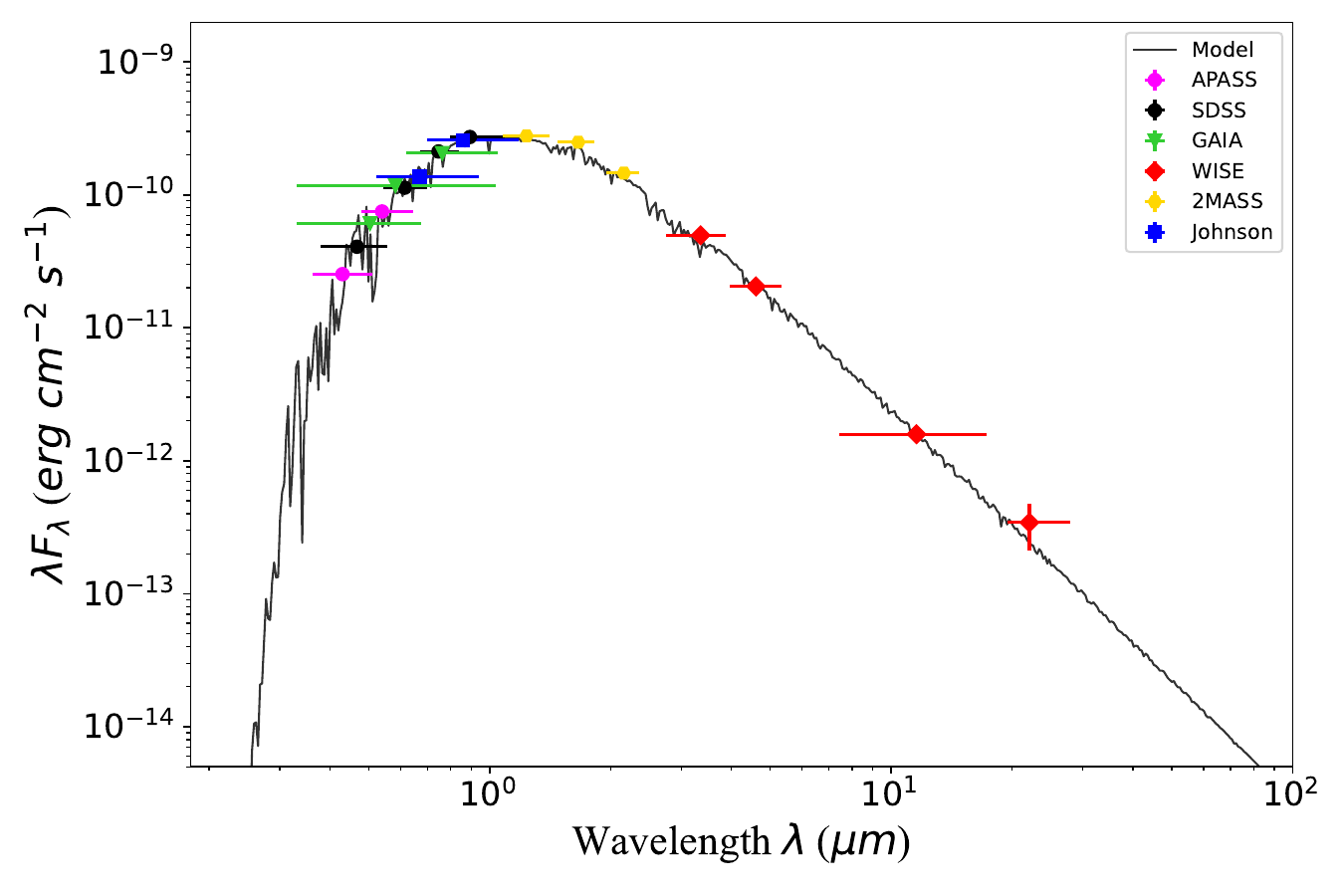}
    \caption{Spectral Energy Distribution (SED) of TOI-672, described in Section~\ref{sec:stellparamsed}, assembled from broadband photometry obtained with APASS (magenta), SDSS (black), Gaia (green), 2MASS (yellow), WISE (red), and the Johnson system (blue). The horizontal bars indicate the effective wavelength ranges of each filter. The SED is fitted with a BT-Settl atmospheric model \citep{Allard2012}.}
    \label{fig:sed}
\end{figure}

\clearpage

\section{Joint fit priors and fit values}

\begin{table*}[htbp!]
    \centering
    \small
    
    \begin{threeparttable}[ht]
    \caption{Prior distributions used in our joint fit model and the resulting fit values.}\label{tab:priors}
    \begin{tabular}{lllll}
    \toprule
    \textbf{Parameter} & & \textbf{(Unit)} & \textbf{Prior distribution} & \textbf{Fit value} \\
	\midrule
    \multicolumn{5}{l}{\textbf{Planet}} \\
    \midrule
    Period                          & $P$           & (days)        & $\mathcal{U}(3.6236, 3.6436)$                     & 3.633581 $\pm$ 0.000001 \\
    Reference time of midtransit    & $t_0$         & (BJD$_{\rm TDB}$-2457000) & $\mathcal{U}(1546.4700, 1546.4900)$   & 1546.4796 $\pm$ 0.0003 \\
    Log of the radius               & $\log{(R_p)}$ & (R$_{\odot}$) & $\mathcal{U}(-3.99, -1.99)^*$    & -3.02$^{+0.04}_{-0.05}$\\
    Radius                          & $R_p$         & (R$_{\oplus}$)& -- (derived)                                      & 5.31$^{+0.24}_{-0.26}$ \\
    Eccentricity                    & $e$           &               & 0 (fixed)                                         & - \\
    Argument of periastron          & $\omega$      & ($^{\circ}$)  & 0 (fixed)                                         & - \\
    \midrule
    \multicolumn{5}{l}{\textbf{Host star}} \\
    \midrule
    Radius  & $R_\star$   & (R$_{\odot}$)   & $\mathcal{N_B}(0.54, 0.02, 0.0, 3.0)$     & 0.54 $\pm$ 0.03\\
    Mass    & $M_\star$   & (M$_{\odot}$)   & $\mathcal{N_B}(0.54, 0.01, 0.0, 3.0)$     & 0.54 $\pm$ 0.03\\
    \midrule
    \multicolumn{5}{l}{\textbf{Photometry}} \\
    \midrule
    \multicolumn{5}{l}{\textbf{TESS}} \\
    Offset                                  &                           & & $\mathcal{N}(0.0, 1.0)$                 & 0.00005 $\pm$ 0.00010 \\
    Log of the jitter                       & $\log{(J_{\rm TESS})}$    & & $\mathcal{N}(-11.58^{\dagger}, 10.0)$   & -11.805 $\pm$ 0.006 \\
    Log of the maximum power                & $\log{(S_0\omega^4_0)}$   & & $\mathcal{N}(-11.58^{\dagger}, 10.0)$   & -10.4 $\pm$ 0.3 \\
    Log of the undamped angular frequency	& $\log{(\omega_0)}$        & & $\mathcal{N}(0.0, 10.0)$                & 1.1 $\pm$ 0.1 \\
    \hdashline\noalign{\vskip 0.5ex}
    \multicolumn{5}{l}{\textbf{ExTrA}} \\
    Offset                                  &                           & & $\mathcal{N}(0.0, 1.0)$                     & 0.0005$^{+0.0246}_{-0.0242}$\\
    Log of the amplitude scale              & $\log{(\sigma)}$          & & $\mathcal{U}(\log{(0.01)}, \log{(0.1)})$    & -3.8$^{+0.7}_{-0.6}$ \\
    Log of the length scale                 & $\log{(\rho)}$            & & $\mathcal{U}(\log{(0.3)}, \log{(3.0)})$     & 0.4$^{+0.5}_{-0.7}$ \\
    \midrule
    \multicolumn{5}{l}{\textbf{RVs}} \\
    \midrule
    Log of the radial velocity semi-amplitude & $\log{(K)}$ &               & $\mathcal{U}(0.0,10.0)$ & 3.47 $\pm$ 0.08\\
    Radial velocity semi-amplitude            & $K$         & (ms$^{-1}$)   & -- (derived)            & 32.0$^{+2.6}_{-2.5}$ \\
    \hdashline\noalign{\vskip 0.5ex}
    \multicolumn{5}{l}{\textbf{NIRPS}} \\[0.5ex]
    Offset                      &                           & (ms$^{-1}$)   & $\mathcal{U}(-4650.0,-4550.0)$                & -4604.3$^{+2.9}_{-2.8}$ \\
    Log of the jitter           & $\log{(J_{\rm NIRPS})}$   & (ms$^{-1}$)   & $\mathcal{N}(2.90^{\ddagger}, 5.0)$           & 1.14$^{+1.06}_{-4.05}$ \\
    Jitter                      & $J_{\rm NIRPS}$           & (ms$^{-1}$)   & -- (derived)                                  & 1.77$^{+1.24}_{-1.54}$ \\
    Amplitude                   & $\eta_{\rm NIRPS}$        & (ms$^{-1}$)   & $\mathcal{HC}(5.0)$                           & 7.5$^{+3.2}_{-4.7}$ \\
    Recurrence timescale        & $\theta_{\rm NIRPS}$      & (days)        & $\mathcal{U}(10.0,40.0)$                      & 23.4$^{+11.0}_{-9.6}$ \\
    Growth and decay timescale  & $\lambda_{\rm NIRPS}$     & (days)        & $\mathcal{U}(\log{(10.0)}, \log{(4000.0)})$   & 165.9$^{+1261.4}_{-143.6}$ \\
    Smoothing parameter         & $\gamma_{\rm NIRPS}$      &               & $\mathcal{N_T}(0.2, 5.0, 0.0, 1.0)$           & 0.32$^{+0.32}_{-0.2}$\\
    \hdashline\noalign{\vskip 0.5ex}
    \multicolumn{5}{l}{\textbf{HARPS}} \\
    Offset                      &                           & (ms$^{-1}$)   & $\mathcal{U}(-4700.0,-4600.0)$                & -4653.5 $\pm$ 2.9 \\
    Log of the jitter           & $\log{(J_{\rm HARPS})}$   & (ms$^{-1}$)   & $\mathcal{N}(1.52^{\ddagger}, 5.0)$           & 2.4$\pm 0.2$\\
    Jitter                      & $J_{\rm HARPS}$           & (ms$^{-1}$)   & -- (derived)                                  & 3.3$^{+0.4}_{-0.3}$ \\
    \bottomrule
    \end{tabular}
    \begin{tablenotes}
    \item The joint fit model is fully described in Section~\ref{sec:fit}. The priors are created using distributions in {\tt PyMC}, with the relevant inputs to each distribution described below. The fit values are given as the median values of the samples, and the uncertainties as the 16th and 84th percentiles. Further (derived) system parameters can be found in Table~\ref{tab:planetparams}.\\
    \item \textbf{Distributions:}
    \item $\mathcal{N}(\mu, \sigma)$: a normal distribution with a mean $\mu$ and a standard deviation $\sigma$;
    \item $\mathcal{N_B}(\mu, \sigma, a, b)$: a bounded normal distribution with a mean $\mu$, a standard deviation $\sigma$, a lower bound $a$, and an upper bound $b$ (bounds optional);
    \item $\mathcal{N_T}(\mu, \sigma, a, b)$: a truncated normal distribution with a mean $\mu$, a standard deviation $\sigma$, a lower bound $a$, and an upper bound $b$ (bounds optional);
    \item $\mathcal{U}(a, b)$: a uniform distribution with a lower bound $a$, and an upper bound $b$;
    \item $\mathcal{HC}(a)$: a Half-Cauchy distribution with scale factor $a$.\\
    \item \textbf{Prior values:}
    \item $^*$ equivalent to $0.5(\log{(D)}) + \log{(R_\star)} \pm 1$ where $D$ is the transit depth (ppm multiplied by $10^{-6}$) and $R_\star$ is the mean of the prior on the stellar radius (\mbox{R$_{\odot}$}), and $-1$ computes the lower bound while $+1$ computes the upper bound;
    \item $^{\dagger}$ equivalent to the log of the variance of the TESS flux;
    \item $^{\ddagger}$ equivalent to 2 multiplied by the log of the minimum error on the NIRPS or HARPS RV data, respectively.
    \end{tablenotes}
    \end{threeparttable}
\end{table*}

\clearpage

\section{Planet sample for Neptunian desert boundary study}

\begin{figure}[ht]
    \centering
    \includegraphics[width=\linewidth]{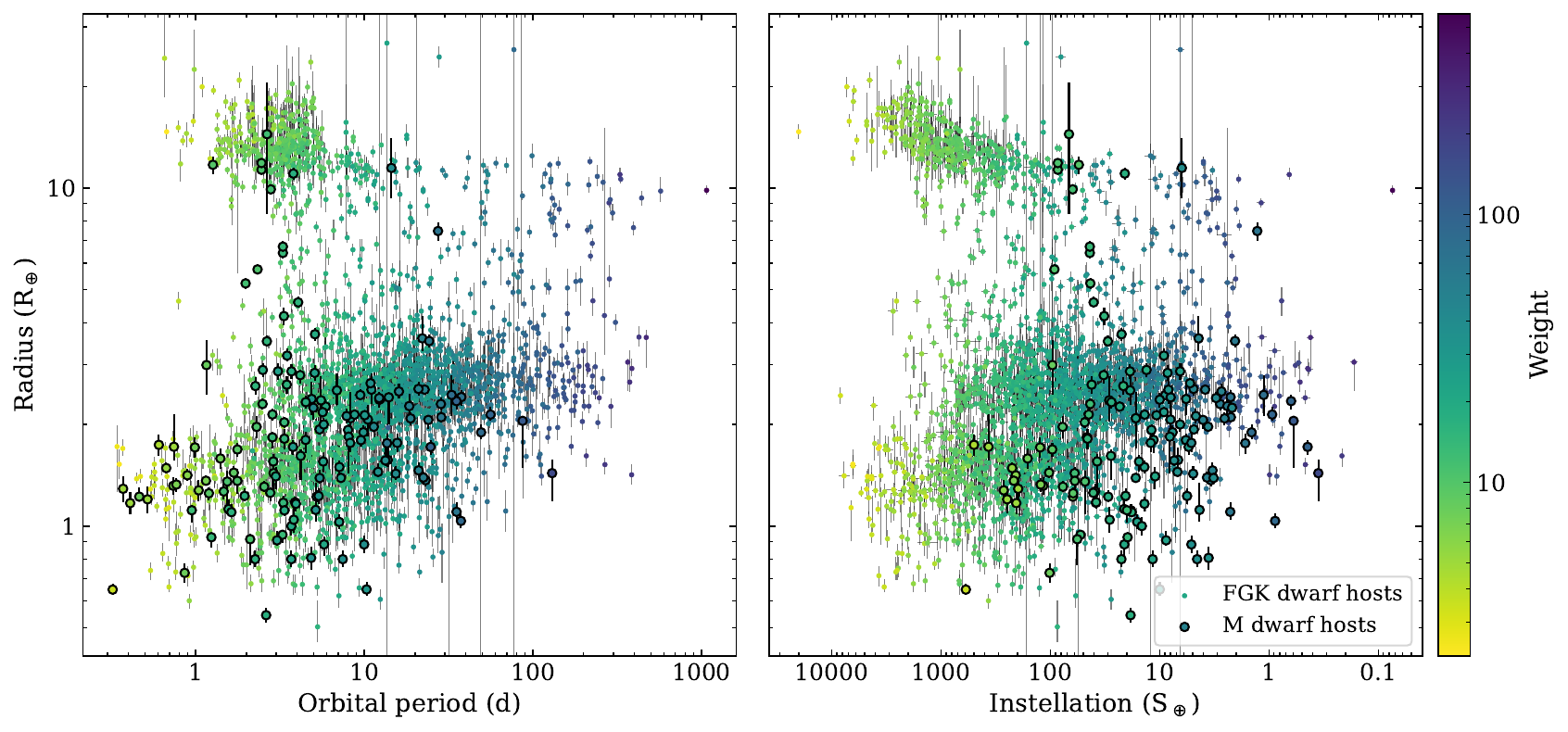}
    \caption{The sample used for the study of the Neptunian desert boundaries, described in Section~\ref{sec:disc_desert} and given in Table~\ref{tab:desert_sample}.}
    \label{fig:desert_sample}
\end{figure}

\begin{table*}
\centering
\tiny
\caption{The sample used for the study of the Neptunian desert boundaries.}
\label{tab:desert_sample}
\begin{threeparttable}
\begin{tabular}{p{1.0cm}p{0.5cm}p{0.4cm}p{0.4cm}p{0.4cm}p{0.4cm}p{0.4cm}p{0.4cm}p{0.5cm}p{0.4cm}p{0.4cm}p{0.7cm}p{0.3cm}p{0.4cm}p{0.7cm}p{0.5cm}p{0.4cm}p{0.4cm}p{0.5cm}}
\toprule
Planet & $a$ & $T_{\rm eff}$ & $R_{\star}$ & $M_{\star}$ & $L$ & $S$ & $P_{orb}$ & $R_{p}/R_{\star}$ & $R_{p}$ & $T_{dur}$ & $\delta$ & $N$ $^{\dagger}$ & $N_{S}$ $^{\ddagger}$ & $\sigma_{\rm CDPP}$ & S/N & $P_{\rm trans}^{-1}$ & $P_{\rm det}^{-1}$ & $w$ \\
       & (au) & (K) & ($R_{\odot}$) & ($M_{\odot}$) & ($L_{\odot}$) & ($S_{\oplus}$) & (days) & & ($R_{\oplus}$) & (hrs) & ($^*$) & & & ($^\S$) \\
\midrule

K00752.01 & 0.086 & 5610 & 0.90 & 0.94 & 0.73 & 99  & 9.49 & 0.022 & 2.20 & 3.59 & 0.00050 & 142 & & 207.29 & 287.63 & 20.43 & 1.03 & 20.99 \\
K00752.02 & 0.275 & 5610 & 0.90 & 0.94 & 0.73 & 10  & 54.42 & 0.028 & 2.76 & 5.36 & 0.00078 & 25 & & 184.96 & 211.70 & 65.46 & 1.03 & 67.27 \\
K00755.01 & 0.036 & 5660 & 0.92 & 0.94 & 0.79 & 630 & 2.53 & 0.024 & 2.42 & 1.74 & 0.00058 & 515 & & 325.80 & 404.22 & 8.28 & 1.03 & 8.51 \\
K00756.01 & 0.098 & 5837 & 1.03 & 1.03 & 1.12 & 116 & 11.09 & 0.037 & 4.14 & 3.66 & 0.00136 & 95 & & 235.39 & 561.33 & 20.53 & 1.03 & 21.10 \\
K00756.02 & 0.051 & 5837 & 1.03 & 1.03 & 1.12 & 431 & 4.13 & 0.026 & 2.94 & 2.04 & 0.00068 & 240 & & 299.06 & 354.53 & 10.63 & 1.03 & 10.93 \\
\multicolumn{17}{c}{\vdots} \\
toi892.01 & 0.100 & 6163 & 1.33 & 1.19 & 2.32 & 231  & 10.63 & 0.079 & 11.48 & 5.42 & 7530.0 & 2 & 1 & 220.61 & 51.60 & 16.19 & 1.03 & 16.64 \\
toi905.01 & 0.046 & 5434 & 0.93 & 0.95 & 0.69 & 319  & 3.74 & 0.131 & 13.30 & 2.04 & 15256.9 & 42 & 6 & 1943.18 & 332.65 & 10.73 & 1.03 & 11.02 \\
toi942.01 & 0.049 & 5043 & 0.83 & 0.83 & 0.40 & 169  & 4.32 & 0.040 & 3.61 & 3.18 & 1480.0 & 6 & 1 & 556.19 & 6.55 & 12.71 & 10.09 & 128.16 \\
toi954.01 & 0.052 & 5932 & 1.78 & 1.35 & 3.62 & 1361 & 3.68 & 0.046 & 8.98 & 4.54 & 2070.0 & 7 & 1 & 178.90 & 30.81 & 6.22 & 1.03 & 6.39 \\
toi966.01 & 0.046 & 6231 & 1.19 & 1.15 & 1.93 & 892  & 3.41 & 0.107 & 13.84 & 3.21 & 16294.4 & 22 & 3 & 1748.80 & 208.27 & 8.43 & 1.03 & 8.66 \\

\bottomrule
\end{tabular}
\begin{tablenotes}
\item The sample is described in Section~\ref{sec:disc_desert} and depicted in Fig.~\ref{fig:desert_sample}. This table is truncated (error columns are omitted and precisions are reduced); the full version can be found on the Strasbourg astronomical Data Center (CDS).
\item $^*$It is important to note that the depth, $\delta$, is reported in units of ppm$\times 10^{-6}$ for the Kepler planets, and ppm for the TESS planets, to match the differing units for the Kepler and TESS RMS CDPP values. 
\item $^{\dagger}$The number of transits, $N$, for the TESS targets is the total number of transits across all sectors; more specifically, we take an average number of transits in a sector based on planet period and the length of each sector, then sum over all sectors. 
\item $^{\ddagger}$For TESS targets, the total number of sectors in which the planet transits is given as $N_{S}$. 
\item $^{\S}$The RMS CDPP computed at the transit duration, $\sigma_{\rm CDPP}$, is reported in units of ppm$\times 10^{-6}$ for Kepler; for TESS it is reported in units of ppm. In this table, the value we give for $\sigma_{\rm CDPP}$ for the TESS targets is rather the denominator of Equ.~\ref{equ:tesssn}, i.e., $\sqrt{\sum_i N_i \sigma^2_{{\rm CDPP}, i}}$.
\end{tablenotes}
\end{threeparttable}
\end{table*}

\end{appendix}

\end{document}